\documentclass[twocolumn,english,prb,aps,amssymb,showpacs,superscriptaddress]{revtex4-1}
\usepackage[T1]{fontenc}
\usepackage[latin9]{inputenc}
\usepackage{amsmath}
\usepackage{amssymb}
\usepackage{graphicx}
\usepackage{multirow}
\usepackage{bm}

\begin{document}

%\preprint{}
%\newcommand{\equref}[1]{Eq.\ (\ref{#1})}
%\newcommand{\figref}[1]{FIG.\ \ref{#1}.}
%\newcommand{\tabref}[1]{TABLE\ \ref{#1}.}
\title{Inter-orbital topological superconductivity in spin-orbit coupled superconductors with inversion symmetry breaking}


\author{Yuri Fukaya}
\affiliation{Department of Applied Physics, Nagoya University, Nagoya 464-8603, Japan}

\author{Shun Tamura}
\affiliation{Department of Applied Physics, Nagoya University, Nagoya 464-8603, Japan}

\author{Keiji Yada}
\affiliation{Department of Applied Physics, Nagoya University, Nagoya 464-8603, Japan}

\author{Yukio Tanaka}
\affiliation{Department of Applied Physics, Nagoya University, Nagoya 464-8603, Japan}

\author{Paola Gentile}
\affiliation{SPIN-CNR, I-84084 Fisciano (Salerno), Italy and
Dipartimento di Fisica ''E. R. Caianiello'', Universit\'a di Salerno, I-84084 Fisciano (Salerno), Italy}

\author{Mario Cuoco}
\affiliation{SPIN-CNR, I-84084 Fisciano (Salerno), Italy and
Dipartimento di Fisica ''E. R. Caianiello'', Universit\'a di Salerno, I-84084 Fisciano (Salerno), Italy}


%\date{\today}

\begin{abstract}
We study the superconducting state of multi-orbital spin-orbit coupled systems in the presence of an orbitally driven inversion asymmetry assuming that the inter-orbital attraction is the dominant pairing channel. 
Although the inversion symmetry is absent, we show that superconducting states that avoid mixing of spin-triplet and spin-singlet configurations are allowed, and remarkably, spin-triplet states that are topologically nontrivial can be stabilized in a large portion of the phase diagram. 
The orbital-dependent spin-triplet pairing generally leads to topological superconductivity with point nodes that are protected by a nonvanishing winding number. 
We demonstrate that the disclosed topological phase can exhibit Lifshitz-type transitions upon different driving mechanisms and interactions, e.g., by tuning the strength of the atomic spin-orbit and inversion asymmetry couplings or by varying the doping and the amplitude of order parameter. 
Such distinctive signatures of the nodal phase manifest through an extraordinary reconstruction of the low-energy excitation spectra both in the bulk and at the edge of the superconductor. 
\end{abstract}

\maketitle

\section{Introduction}

Spin-triplet pairing is at the core of intense investigation especially because of its foundational aspect in unconventional superconductivity \cite{sigrist91,Maeno,Maeno3,Kashiwaya11} and owing to its tight connection with the occurrence of topological phases with zero-energy surface 
Andreev bound states \cite{ABS,ABSb,Hu94,TK95,ABSR1,ABSR2} marked by Majorana edge modes
\cite{Yakovenko,SRFL08,RSFL10,qi11,tanaka12,Flensberg2012,Beenakker13,SatoFujimoto2016,SatoAndo2017}.  
Some of the fundamental essences of topological spin-triplet superconductivity are basically captured by the Kitaev model \cite{Kitaev01} and its generalized versions where non-Abelian states of matter and their employment for topological quantum computation can be demonstrated \cite{Kitaev01,Nayak,wilczek09,alicea12,Ramon2017LRdNC}. 
Another remarkable element of odd-parity superconductivity is given by the potential of having active spin degrees of freedom making such states of matter also appealing for superconducting spintronics applications based on spin control and coherent spin manipulation of Cooper pairs \cite{Bergeret2005RMP,Buzdinrev,Birge,Robinson,Eschrig2003,Eschrig2008,Linder2015}.  
The interplay of magnetism and spin-triplet superconductivity can manifest within different unconventional physical scenarios, such as the case of the emergent spin-orbital interaction between the superconducting order parameter and interface magnetization \cite{Gentile13, Terrade16}, the breakdown of the bulk-boundary correspondence \cite{Mercaldo16}, and the anomalous magnetic\cite{Romano13,Romano16} and spin-charge current\cite{Romano17} effects occurring in the proximity between chiral or helical $p$-wave and spin-singlet superconductors.    
Achieving spin-triplet materials platforms, thus, sets the stage for the development of emergent technologies both in nondissipative spintronics and in the expanding area of quantum devices. 

Although embracing strong promises, spin-triplet superconductivity is quite rare in nature and the mechanisms for electron pairs gluing are not completely settled.  
The search for spin-triplet superconductivity has been performed along different routes. For instance, scientific exploration has been focused on the regions of the materials phase diagram that are in proximity to ferromagnetic quantum phase transitions 
\cite{Fay,Pfleiderer}, as in the case of heavy fermions superconductivity, i.e., $\mathrm{UGe}_2$, URhGe, and $\mathrm{UIr}_2$, or in materials on the verge of a magnetic instability, e.g., ruthenates \cite{Maeno,Maeno2}. 

Another remarkable route to achieve spin-triplet pairing 
relies on the presence of a source of inversion symmetry breaking, 
both at the surface/interface and in the bulk, or alternatively, 
in connection with noncollinear magnetic ordering \cite{Martin2012,Ojanen2014,Pientka2013,Braunecker,Klinovaja2013,Kim2014,Pientka2014,a15,Nakosai2013,heimes,Mendler2015PRB}. 
Paradigmatic examples along these directions are provided by 
quasi-one-dimensional heterostructures whose interplay of inversion 
and time-reversal symmetry breaking or noncollinear magnetism 
have been shown to convert spin-singlet pairs into spin-triplet ones 
and in turn to topological phases\cite{a11,a12,a13,a14,a15}. 
Similar mechanisms and physical scenarios are also encountered at the interface between spin-singlet superconductors and 
inhomogeneous ferromagnets with even and odd-in time spin-triplet pairing that are generally generated \cite{Bergeret2005RMP}. 
Semimetals have also been indicated as fundamental building blocks to generate spin-triplet pairing as theoretically proposed and 
demonstrated in topological insulators interfaced with conventional superconductors or by doping 
Dirac/Weyl phases \cite{Volovikbook}, e.g., in the case of 
Cu-doped $\mathrm{Bi}_2\mathrm{Se}_3$ \cite{Hor10,Wray10,Kriener11,Fu2010PRL,sasaki11,hao11,yamakage12}
in anti-perovskites materials \cite{Oudah}, 
as well as Cd$_{3}$As$_{2}$ \cite{Aggarwal,Wang}.  

Generally, there are two fundamental interactions to take into account in inversion asymmetric 
microscopic environments: i) the Rashba spin-orbit coupling~\cite{Rashba1960} due to inversion symmetry breaking at the 
surface or interface in heterostructures, and ii) the Dresselhaus coupling arising from the inversion asymmetry in the 
bulk of the host material~\cite{Dresselhaus1955}. 
For the present analysis, it is worth noting that typically in multi-orbital materials, 
it is the combination of the atomic spin-orbit interaction with the inversion symmetry-breaking sources that effectively 
generates both Rashba and Dresselhaus emergent interactions within the electronic manifold close to the Fermi level. 
Another general observation is that the lack of inversion symmetry is expected to lead to a parity mixing of spin-singlet and spin-triplet configurations~\cite{Gorkov2001,Frigeri2004,Yada2009PRB} with an ensuing series of unexpected features ranging from anomalous magneto-electric~\cite{Lu2008} 
effects to unconventional surface states~\cite{Vorontsov2008}, 
topological phases~\cite{Tanaka2009,Sato2009,SBMT10}, and non-trivial spatial textures of the spin-triplet pairs~\cite{Ying2017}. 
Such symmetry conditions in intrinsic materials are, however, fundamentally linked to the momentum 
dependent structure of the superconducting order parameter. 
In contrast, when considering multi-orbital systems, 
more channels are possible with emergent unconventional 
paths for electron pairing that are expected to be strongly 
tied to the orbital character of the electron-electron attraction and
of the electronic states close to the Fermi level.  

Orbital degrees of freedom are key players in quantum 
materials when considering the degeneracy of $d$-bands of 
the transition elements not being completely removed by the crystal 
distortions or due to the intrinsic spin-orbital entanglement~\cite{Ole12} triggered 
by the atomic spin-orbit coupling. 
In this context, a competition of different and complex types of order is
ubiquitous in realistic materials, such as transition metal oxides, mainly owing to the frustrated
exchange arising from the active orbital degrees of freedom.
Such scenarios are commonly encountered 
in materials 
where the atomic physics plays a significant role in setting 
the character of the electronic structure close to the Fermi level. 
As the $d$-orbitals have an anisotropic spatial distribution, the nature 
of the electronic states is also strongly dependent on the system's dimensionality. 
Indeed, two-dimensional (2D) confined electron liquids originating at the interface or surface of materials 
generally manifest a rich variety of spin-orbital phenomena \cite{Hwang2012}. 
Along this line, understanding how electron pairing is settled in quantum systems exhibiting a strong interplay between orbital degrees of freedom and inversion symmetry breaking represents a fundamental problem in unconventional superconductivity, and it can be of great relevance for a large class of materials. 

In this study, we investigate the nature of the superconducting phase in spin-orbit coupled systems in the absence of inversion symmetry assuming that the inter-orbital attractive channel is dominant and sets the electrons pairing. 
We demonstrate that the underlying inversion symmetry breaking leads to exotic spin-triplet superconductivity.  
Isotropic spin-triplet pairing configurations, without any mixing with spin-singlet, generally occur among the symmetry allowed solutions and are shown to be the ground-state in a large part of the parameters space. 
We then realize an isotropic spin-triplet superconductor whose orbital character can make it topologically non trivial. 
Remarkably, the topological phase exhibits an unconventional nodal structure with unique tunable features. 
An exotic fingerprint of the topological phases is that the number and $k$-position of nodes can be controlled by doping, orbital polarization, 
%e.g. (i.e. $[n_{xy}-(1/2)(n_{xz}+n_{yz})/n_{tot}]$) 
through the competition between spin-orbit coupling and lattice distortions, and temperature (or equivalently, the amplitude of the order parameter).

The paper is organized as follows. 
In Sec. II, we introduce the model Hamiltonian and present the classification of the inter-orbital pairing configurations with respect to the point-group and time-reversal symmetries. 
Section III is devoted to an analysis of the stability of the various orbital entangled superconducting states and the energetics of the isotropic superconducting states. 
Section IV focuses on the electronic spectra of the energetically most favorable phases and the ensuing topological configurations both in the bulk and at the boundary. 
Finally, in Sec. V, we provide a discussion of the results and few concluding remarks.  

%%%%%%%%%%%%%%%%%%%%%%%%%%%%%%%%%%%%%%%%%%%%%%%%%%%%%%%%%%%%%%%%%%%%%%%%
\section{Model and symmetry classification of superconducting phases with inter-orbital pairing}

\begin{figure}[htbp]
\centering
\includegraphics[width=8cm]{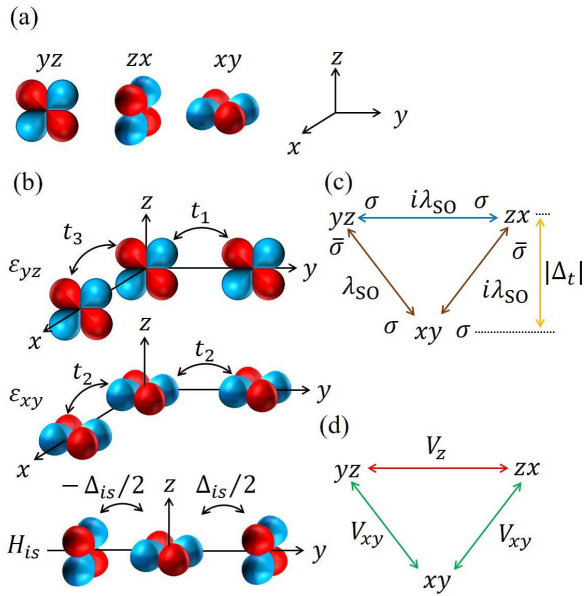}
\caption{(a) $d_{yz}$, $d_{zx}$, and $d_{xy}$-orbitals 
with $L=2$ orbital angular momentum. 
(b) Schematic image of the orbital dependent hopping amplitudes for $\varepsilon_{yz}$, $\varepsilon_{xy}$, and the orbital
connectivity associated with the inversion asymmetry term $\Delta_{is}$.
Here, we do not explicitly indicate the intermediate $p$-orbitals of the oxygen ions surrounding the transition metal 
element that enter the effective $d-d$ hopping processes. 
$\varepsilon_{zx}$ is obtained from 
$\varepsilon_{yz}$ by rotating $\pi/2$ around $z$-axis. 
$\Delta_{is}$ corresponds to the odd-in-space hopping amplitude from 
$d_{xy}$ to $d_{zx}$ along the $y$-direction. Similarly, the odd-in-space hopping amplitude from $d_{xy}$ to $d_{yz}$ along the $x$-direction is obtained by $\pi/2$ rotation around the $z$-axis. 
(c) Sketch of the orbital mixing through the spin-orbit coupling term in the Hamiltonian. 
$\sigma$ denotes the spin state, and $\Bar{\sigma}$ is the opposite spin of $\sigma$. 
$\Delta_{t}$ gives the level splitting between $d_{xy}$-orbital and $d_{yz}/d_{zx}$-orbitals. 
(d) Schematic illustration of inter-orbital interaction. }
\label{Fig1}
\end{figure}

%Hamiltonian
One of the most common crystal structures of transition metal oxides is the perovskite structure, with transition metal (TM) elements surrounded by oxygen (O) in an octahedral environment.
For cubic symmetry, owing to the crystal field potential generated by the oxygen around the TM, the fivefold orbital degeneracy is removed and $d$ orbitals split into two sectors: $t_{2g}$, i.e., $yz$, $zx$, and $xy$, and $e_g$, i.e., $x^2-y^2$ and $3z^2-r^2$.
In the present study, the analysis is focused on two-dimensional (2D) electronic systems with broken out-of-plane inversion symmetry and having only the $t_{2g}$ orbitals (Fig. \ref{Fig1}) close to the Fermi level to set the low energy excitations. For highly symmetric TM-O bonds, the three $t_{2g}$ bands are directional and basically decoupled, e.g., an electron in the $d_{xy}$ orbital can only hop along the $y$ or $x$ direction through the intermediate $p_x$ or $p_y$ orbitals. Similarly, the $d_{yz}$ and $d_{zx}$ bands are quasi-one-dimensional when considering a 2D TM-O bonding network. Furthermore, the atomic spin-orbit interaction (SO) mixes the $t_{2g}$ orbitals thus competing with the quenching of the orbital angular momentum due to the crystal potential. 
Concerning the inversion asymmetry, we consider microscopic couplings that arise from the out-of-plane oxygen displacements around the TM. 
Indeed, by breaking the reflection symmetry with respect to the plane placed in between the TM-O bond~\cite{Khalsa2013PRB}, a mixing of orbitals that are even and odd under such a transformation is generated. 
Such crystal distortions are much more relevant and pronounced in 2D electron gas forming at the interface of insulating polar and nonpolar oxide materials or on their surface and they result in the activation of an effective hybridization, which is odd in space, of $d_{xy}$ and $d_{yz}$ or $d_{zx}$ orbitals along the $y$ or $x$ directions, respectively. Although the polar environment tends to amplify the out-of-plane oxygen displacements with respect to the position of the TM ion, such types of distortions can also occur at the interface of nonpolar oxides and in superlattices~\cite{Autieri2014}. 
  
Thus, the model Hamiltonian, including the $t_{2g}$ hopping connectivity, the atomic spin-orbit coupling, and the inversion symmetry breaking term, reads as
\begin{align}
\mathcal{H}&=\sum_{\bm{k}}\Hat{C}(\bm{k})^{\dagger}H(\bm{k})\Hat{C}(\bm{k}), \\
H(\bm{k})&=H_0(\bm{k})+H_\mathrm{SO}(\bm{k})+H_{is}(\bm{k}),
\end{align}%
where $\Hat{C}^{\dagger}(\bm{k})=\left[ c^{\dagger}_{yz\uparrow \bm{k}}, c^{\dagger}_{zx\uparrow \bm{k}}, c^{\dagger}_{xy\uparrow \bm{k}}, c^{\dagger}_{yz\downarrow \bm{k}}, c^{\dagger}_{zx\downarrow \bm{k}}, c^{\dagger}_{xy\downarrow \bm{k}} \right]$ is a vector whose components are associated with the electron creation operators for a given spin $\sigma$ [$\sigma=(\uparrow,\downarrow)$], orbital $\alpha$ [$\alpha=(xy,yz,zx)$], and momentum $\bm{k}$ in the Brillouin zone. 
In Fig. \ref{Fig1}(a), we report a schematic illustration of the local orbital basis for the $t_{2g}$ states.  
$H_0(\bm{k})$, $H_\mathrm{SO}(\bm{k})$, and $H_{is}(\bm{k})$ indicate the kinetic term, the 
spin-orbit interaction, and the inversion symmetry breaking term, respectively. 
In the spin-orbital basis, $H_0(\bm{k})$ is given by 
\begin{align}
&H_0(\bm{k})=-\mu \left[\Hat{l}_{0}\otimes\Hat{\sigma}_{0} \right]+\Hat{\varepsilon}_{\bm{k}} \otimes \Hat{\sigma}_{0}, \\
&\Hat{\varepsilon}_{\bm{k}}=
\begin{pmatrix}
\varepsilon_{yz} &0 &0 \\
0 & \varepsilon_{zx} &0 \\
0 &0& \varepsilon_{xy}
\end{pmatrix}, \notag \\
&\varepsilon_{yz}=2t_{1}\left( 1-\cos{k_y}\right)+2t_{3}\left(1-\cos{k_x}\right), \notag \\
&\varepsilon_{zx}=2t_{1}\left(1-\cos{k_x}\right)+2t_{3}\left(1-\cos{k_y}\right), \notag \\
&\varepsilon_{xy}=4t_{2}-2t_{2}\cos{k_x}-2t_{2}\cos{k_y}+\Delta_{t}, \notag
\end{align}%
where $\Hat{l}_{0}$ and $\Hat{\sigma}_{0}$ are the unit matrices in orbital and spin space, respectively. 
Here, $\mu$ is the chemical potential, and $t_{1}$, $t_{2}$, and $t_{3}$ are the orbital dependent hopping amplitudes as schematically shown in Fig. \ref{Fig1}(b). 
$\Delta_{t}$ denotes the crystal field potential owing to the symmetry lowering from cubic to tetragonal symmetry. 
The symmetry reduction yields a level splitting between $d_{xy}$ orbital and $d_{yz}/d_{zx}$ orbitals.
$H_\mathrm{SO}(\bm{k})$ denotes the atomic $\bm{L} \cdot \bm{S}$ spin-orbit coupling,
\begin{align}
H_\mathrm{SO}(\bm{k})
=\lambda_{\mathrm{SO}}\left[ \Hat{l}_x \otimes \Hat{\sigma}_x+\Hat{l}_y \otimes \Hat{\sigma}_y+\Hat{l}_z \otimes \Hat{\sigma}_z \right], 
\end{align}%
with $\Hat{\sigma}_{i}(i=x,y,z)$ being the Pauli matrix in spin space.
In order to write down the $\bm{L} \cdot \bm{S}$ interaction, it is convenient to introduce the matrices $\Hat{l}_{x}$, $\Hat{l}_{y}$ and $\Hat{l}_{z}$, which are the projections of the $L=2$ angular momentum operator onto the $t_{2g}$ subspace, i.e.,
\begin{align} 
\Hat{l}_{x}&=
\begin{pmatrix}
0 & 0 & 0 \\
0 & 0 & i \\
0 & -i & 0
\end{pmatrix}, \\
\Hat{l}_{y}&=
\begin{pmatrix}
0 & 0 & -i \\
0 & 0 & 0 \\
i & 0 & 0
\end{pmatrix}, \\
\Hat{l}_{z}&=
\begin{pmatrix}
0 & i & 0 \\
-i & 0 & 0 \\
0 & 0 & 0
\end{pmatrix},
\end{align}%
assuming $\{(d_{yz}, d_{zx}, d_{xy})\}$ as orbital basis.
Finally, as mentioned above, the breaking of the mirror plane in between the TM-O bond, due to the oxygen displacements,
leads to an inversion symmetry breaking term $H_{is}(\bm{k})$ of the type
\begin{align}
H_{is}(\bm{k})
=\Delta_{is} \left[ \Hat{l}_y \otimes \Hat{\sigma}_0 \sin{k_x}-\Hat{l}_x \otimes \Hat{\sigma}_{0} \sin{k_y} \right].
\end{align}%
This contribution gives an inter-orbital process, due to the broken inversion symmetry, that mixes $d_{xy}$ and $d_{yz}$ or $d_{zx}$ along $x$ or $y$ spatial directions [Fig. \ref{Fig1}(b)]. 
$H_{is}$ resembles a Rashba-type Hamiltonian that, however, couples the momentum to the orbital angular momentum rather than the spin. Its origin is due to distortions or other sources of inversion symmetry breaking that lead to local asymmetries deforming the orbital lobes and in turn antisymmetric hopping terms within the orbitals in the $t_{2g}$ sector. In this respect, it is worth pointing out that it is the combination of the local spin-orbit coupling and the antisymmetric inversion symmetry interaction that leads to a nontrivial momentum dependent spin-orbital splitting. While the original Rashba effect \cite{Rashba1960} for the single-band system describes a linear spin splitting and is typically very small, the multi-band character of the model Hamiltonian yields a more complex spin-orbit coupled structure with significant splitting \cite{Winklerbook}. Indeed, near the $\Gamma$ point of the Brillouin zone, one can have a linear spin splitting with respect to the momentum for the lowest energy bands, but a cubiclike splitting in momentum \cite{Zhong2013,Kim2013} for the intermediate ones with enhanced anomalies when the filling is close to the transition from two to four Fermi surfaces. The Rashba-like effects due to the combined atomic spin-orbit coupling and the orbitally driven inversion-symmetry term can be influenced by the application of an external electric field (e.g., via gating) in a dual way. On one hand, the gating directly modifies the filling concentration and, on the other, it can affect the deformation of the orbital lobes by changing the amplitude of the polar distortion \cite{Joshua2015,Steffen2015}. 

In this paper, we set $t_{1}=t_{2}\equiv t$ as a unit of energy for convenience and clarity of presentation. 
The analysis is performed for a representative set of hopping parameters, i.e., $t_{3}/t=0.10$ and $\Delta_{t}/t=-0.50$.
The primary reason for the choice of the electronic parameters is that we aim to model 
superconductivity in transition-metal based layered materials with low electron concentration 
in the $t_{2g}$ sector at the Fermi level both in the presence of atomic spin-orbit and inversion 
symmetry breaking couplings. In this framework, the set of selected parameters is representative of 
a general physical regime where the hierarchy of the energy scales is such that $\Delta_{t}$>$\Delta_{is}$>$\lambda_\mathrm{SO}$ and $\Delta_{t} \sim t$. The choice of this regime is 
also motivated by the fact that this relation can be generally encountered in $3d$ (or $4d$) layered 
oxides or superlattices in the presence of tetragonal distortions with flat octahedra and interface driven 
inversion-symmetry breaking potential. For instance, in the case of the two-dimensional electron gas (2DEG)
forming at the interface of two band insulators [e.g., the $n$-type 2DEG in
LaAlO$_3$/SrTiO$_3$ \cite{Ohtomo2004Nature} (LAO/STO)] or the 2DEG at the surface of a 
band insulator [e.g., in SrTiO$_3$ (STO)], the energy scales for the electronic parameters, as given by $ab$ $initio$ \cite{Zhong2013,Khalsa2013PRB,Zabaleta2016} or spectroscopic studies 
\cite{Salluzzo2009}, 
are such that the bare $\Delta_{t}\sim 50$-$100$ meV, $\Delta_{is} \sim 20$ meV, and $\lambda_\mathrm{SO} \sim 10$ meV, 
while the effective main hopping amplitudes (i.e., $t$) can be in the $200$-$300$ meV range. 
Similar electronic energies can be 
also encountered in $4d$ layered oxides.
Slight variations of these parameters are expected; however, they do not alter the qualitative aspects of the achieved results. We also point out that our analysis is not intended for a specific material case and that variations in the amplitude of the electronic parameters that keep the indicated hierarchy do not alter the qualitative outcome and do not lead to significant changes in the results.

%%%%%%%%%%%%%%%%%%%%%%%%%%%%%%%%%%%%%%%%%%%%%%%%%%

\begin{figure}[htbp]
\centering
\includegraphics[width=8.7cm]{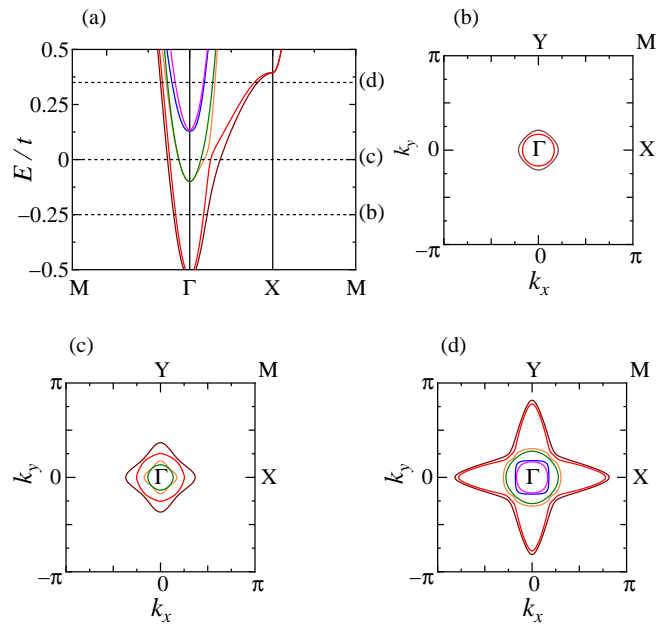}
\caption{(a) Band structure close to the Fermi energy in the normal state at $\lambda_\mathrm{SO}/t=0.10$ and $\Delta_{is}/t=0.20$. 
(b)-(d) Fermi surfaces at (b)$\mu/t=-0.25$, (c)$\mu/t=0.0$, and (d)$\mu/t=0.35$. }
%\label{fig3}
\label{Fig2}
\end{figure}%

The electronic structure of the examined model system can be accessed by direct diagonalization of the matrix Hamiltonian. 
Representative dispersions for $\lambda_\mathrm{SO}/t=0.10$ and $\Delta_{is}/t=0.20$ are shown in Fig. \ref{Fig2}(a).
We observe six non degenerate bands due to the presence of both $H_\mathrm{SO}(\bm{k})$ and $H_{is}(\bm{k})$.
Once the dispersions are determined, one can immediately notice that the number of Fermi surfaces and the structure can be varied by tuning the chemical potential $\mu$. 
Indeed, for $\mu/t=-0.25$, $\mu/t=0.0$, and $\mu/t=0.35$ one can single out all the main possible cases with two, four, and six Fermi sheets, as given in Figs. \ref{Fig2}(b), (c), and (d), respectively. 
For the explored regimes of low doping, all the Fermi surfaces are made of electron-like pockets centered around origin of the Brillouin zone ($\Gamma$). 
The dispersion of the lowest occupied band has weak anisotropy as it has a dominant $d_{xy}$ character (Fig. \ref{Fig2}(a)); moreover, moving to higher electron concentrations, the outer Fermi sheets exhibit a highly anisotropic profile that becomes more pronounced when the chemical potential crosses the bands mainly arising from the $d_{yz}$ and $d_{zx}$-orbitals.

After having considered the normal state properties, we concentrate on the possible superconducting states that can be realized, their energetics and their topological behavior.
The analysis is based on the assumption that the inter-orbital local attractive interaction is the only relevant pairing channel that contributes to the formation of Cooper pairs. Then, the intra-orbital pairing coupling is negligible. Such a hypothesis can be physically applicable in multi-orbital systems because the intra-band Coulomb interaction is typically larger than the inter-band one. 
Indeed, in the $t_{2g}$ restricted sector the Coulomb 
interaction matrix elements of low-energy lattice Hamiltonian can be 
evaluated by employing the Hubbard-Kanamori parametrization 
\cite{Suganobook} in terms of $U$, $U^{'}$ and $J_{H}$, after 
symmetrizing the Slater-integrals \cite{Slaterbook} within 
the $t_{2g}$ shell assuming a cubic splitting of the $t_{2g}$ and 
$e_{g}$ orbitals. $U$ corresponds to the intra-orbital Coulomb repulsion, 
whereas $U^{'}$ (with $U^{'}=U - 2 J_{H}$ in a cubic symmetry) is the 
inter-orbital interaction which is reduced by Hund exchange, $J_{H}$.   
Hence, one has that the inter-orbital Coulomb repulsion is generally always 
smaller than the intra-orbital one. Estimates for transition metal oxide materials 
in $d^1$, $d^2$ or $d^3$ configurations, being relevant for the $t_{2g}$ shell and thus for our work, 
indicate that $U \sim 3.5$ eV and $U^{'}\sim 2.5$ eV \cite{Vaugier2012}.
Thus, it is plausible to expect that the Coulomb repulsion tends to further suppress the electron pairing that occurs within the same band.
In addition, in the case of having the electron-phonon coupling as a source of electrons attraction, it is shown that the effective inter- and intra-orbital attractive interaction can be of the same magnitude (see Appendix for more details).

In this framework, we point out that topological superconductivity is
proposed to occur, owing to inter-orbital pairing, in 
%$\mathrm{Cu}_{x}\mathrm{Bi}_2\mathrm{Se}_3$ 
Cu-doped $\mathrm{Bi}_2\mathrm{Se}_3$ 
for an
inversion symmetric crystal structure \cite{Fu2010PRL}.
Here, although similar inter-orbital pairing conditions are considered, we pursue the superconductivity 
in low-dimensional configurations, e.g., at the interface of oxides, with the important constraint of having 
a broken inversion symmetry. 
Concerning the orbital structure of the pairing interaction, owing to the tetragonal crystalline symmetry, the coupling between the $d_{xy}$-orbital and $d_{yz}$/$d_{zx}$-orbital is equivalent, and thus one can assume that only two independent channels of attraction are allowed, as shown in Fig. \ref{Fig1}(d).
Indeed, $V_{xy}$ denotes the interaction between the $d_{xy}$ and $d_{yz}$/$d_{zx}$-orbitals, while $V_{z}$ refers to the coupling between the $d_{yz}$ and $d_{zx}$-orbitals. 
Then, the pairing interaction is given by 
\begin{align}
H_\mathrm{I}&=V_{xy}\sum_{i}\left[n_{xy,i}n_{yz,i}+n_{xy,i}n_{zx,i} \right] \notag \\
&+V_{z}\sum_{i}n_{yz,i}n_{zx,i}, \\
n_{\alpha, i}&=c^{\dagger}_{\alpha \uparrow i}c_{\alpha \uparrow i}+c^{\dagger}_{\alpha \downarrow i}c_{\alpha \downarrow i},
\end{align}%
where $i$ denotes the lattice site.

\subsection{Irreducible representation and symmetry classification}

In this subsection, we classify the inter-orbital superconducting states according to the point group symmetry.
The system upon examination has a tetragonal symmetry associated with the point group $\mathrm{C}_{4v}$, marked by four-fold rotational symmetry $C_4$ and mirror symmetries $M_{yz}$ and $M_{zx}$.
Based on the rotational and reflection symmetry transformations, all the possible inter-orbital isotropic pairings can be classified into
five irreducible representations of the $\mathrm{C}_{4v}$ point group
as summarized in Table \ref{irreducible1}. 
\begin{table}[htbp]
\caption{Irreducible representation of the inter-orbital isotropic superconducting states for
the tetragonal group $\mathrm{C}_{4v}$. In the columns, we report the sign of the order parameter upon a four-fold rotational symmetry transformation, $C_{4}$, 
and the reflection mirror symmetry $M_{yz}$, as well as the explicit spin and orbital structure of the gap function. 
In the E representation, $+$ and $-$ of the subscript mean the doubly degenerate mirror-even ($+$) and mirror-odd ($-$) solutions, respectively. } 
\label{irreducible1}
  \begin{center}
    \begin{tabular}{c|cc|c|c} 
$\mathrm{C}_{4v}$ &$C_{4}$&$M_{yz}$&orbital & basis function \\ \hline \hline
 & & & $(d_{xy}, d_{yz})$ & $d^{(xy,yz)}_{y}$ \\
$\mathrm{A}_{1}$ & $+$ & $+$ & $(d_{xy}, d_{zx})$ & $d^{(xy,zx)}_{x}=-d^{(xy,yz)}_{y}$ \\
 & & & $(d_{yz}, d_{zx})$ & $d^{(yz,zx)}_{z}$ \\ \hline
\multirow{2}{*}{$\mathrm{A}_{2}$} & \multirow{2}{*}{$+$} & \multirow{2}{*}{$-$} & $(d_{xy}, d_{yz})$ & $d^{(xy,yz)}_{x}$ \\
 & & & $(d_{xy}, d_{zx})$  & $d^{(xy,zx)}_{y}=d^{(xy,yz)}_{x}$ \\ \hline
\multirow{2}{*}{$\mathrm{B}_{1}$} & \multirow{2}{*}{$-$} & \multirow{2}{*}{$+$} & $(d_{xy}, d_{yz})$  & $d^{(xy,yz)}_{y}$\\
 & & & $(d_{xy}, d_{zx})$  & $d^{(xy,zx)}_{x}=d^{(xy,yz)}_{y}$ \\ \hline
 & & & $(d_{xy}, d_{yz})$ & $d^{(xy,yz)}_{x}$ \\
$\mathrm{B}_{2}$ & $-$ & $-$ & $(d_{xy}, d_{zx})$ & $d^{(xy,zx)}_{y}=-d^{(xy,yz)}_{x}$ \\
 & & & $(d_{yz}, d_{zx})$ & $\psi^{(yz, zx)}$ \\ \hline
\multirow{4}{*}{E} & \multirow{4}{*}{$\pm i$} & \multirow{4}{*}{$\pm$} & $(d_{xy}, d_{yz})$ & $\psi^{(xy, yz)}$, $d^{(xy,yz)}_{z}$ \\
 & & & \multirow{2}{*}{$(d_{xy}, d_{zx})$} & $\psi^{(xy, zx)}_{+}=\mp id^{(xy,yz)}_{z+}$ \\
 & & & & $d^{(xy,zx)}_{z-}=\mp i\psi^{(xy,yz)}_{-}$ \\
 & & & $(d_{yz}, d_{zx})$ & $d^{(yz,zx)}_{x}$, $d^{(yz,zx)}_{y}$ \\ \hline
    \end{tabular}
  \end{center}
\end{table}% 
For our purposes, only solutions 
that do not break the time-reversal symmetry are considered and are reported in Table \ref{irreducible1}. 
Then, the superconducting order parameter associated to bands $\alpha$ and $\beta$ can be classified as 
an isotropic ($s$-wave) spin-triplet/orbital-singlet  $\bm{d}^{(\alpha, \beta)}$-vector and $s$-wave spin-singlet/orbital-triplet with amplitude $\psi^{(\alpha, \beta)}$ or as a mixing of both configurations.
With these assumptions, one can generally describe the isotropic order parameter with spin-singlet and triplet components as 
\begin{align}
\Hat{\Delta}_{\alpha, \beta}=i\Hat{\sigma}_{y}\left[ \psi^{(\alpha, \beta)}+ \Hat{\bm{\sigma}}\cdot \bm{d}^{(\alpha, \beta)} \right], 
\end{align}%
with $\alpha$ and $\beta$ standing for the orbital index, and having for each channel three possible orbital flavors.
Furthermore, owing to the selected tetragonal crystal symmetry, one can achieve three different types of inter-orbital pairings. 
The spin-singlet configurations are orbital triplets and can be described by a symmetric superposition of opposite spin states in different orbitals. 
On the other hand, spin-triplet components can be expressed by means of the following $\bf{d}$-vectors: 
\begin{align*}
\bm{d}^{(xy, yz)}&=\left(d^{(xy, yz)}_{x}, d^{(xy, yz)}_{y}, d^{(xy, yz)}_{z}\right), \\
\bm{d}^{(xy, zx)}&=\left(d^{(xy, zx)}_{x}, d^{(xy, zx)}_{y}, d^{(xy, zx)}_{z}\right), \\
\bm{d}^{(yz, zx)}&=\left(d^{(yz, zx)}_{x}, d^{(yz, zx)}_{y}, d^{(yz, zx)}_{z}\right),
\end{align*}%
with $\bm{d}^{(\alpha, \beta)}$ indicating the spin-triplet configuration built with $\alpha$ and $\beta$-orbitals. 
In general, independently of the orbital mixing, spin-triplet pairing can be expressed in a matrix form as
\begin{eqnarray}
\Delta_T=
\left(\begin{array}{cc}
  \Delta_{\uparrow\uparrow} & \Delta_{\uparrow\downarrow}\\
  \Delta_{\downarrow\uparrow} & \Delta_{\downarrow\downarrow}
\end{array}\right)
= \left(\begin{array}{cc}
  -d_x+i d_y & d_z \\
  d_z & d_x+id_y
\end{array}\right), 
\label{DeltaT}
\end{eqnarray}
\noindent where the $\bf{d}$-vector components are related to the
pairing order parameter with zero spin projection
along the corresponding symmetry axis. The three components
$d_x=\frac{1}{2}(-\Delta_{\uparrow\uparrow}+\Delta_{\downarrow\downarrow})$,
$d_y=\frac{1}{2 i}(\Delta_{\uparrow\uparrow}+\Delta_{\downarrow\downarrow})$
and $d_z=\Delta_{\uparrow\downarrow}$ are expressed in terms of
the equal spin $\Delta_{\uparrow\uparrow} ~\mathrm{and}~
\Delta_{\downarrow\downarrow}$, and the anti-aligned spin
$\Delta_{\uparrow\downarrow}$ gap functions.
As the components of the $\bf{d}$-vector are associated with the zero spin projection of spin-triplet configuration, if the $\bf{d}$-vector points along a given direction, the 
parallel spin configurations lie in the plane perpendicular to the $\bf{d}$-vector orientation.
In the presence of time-reversal symmetry, the superconducting order parameter should satisfy the following relations: 
\begin{align} 
\Delta^{\downarrow \downarrow}_{\alpha, \beta}&=\left[\Delta^{\uparrow \uparrow}_{\alpha, \beta} \right]^{*}, \\
\Delta^{\uparrow \downarrow}_{\alpha,\beta}&=-\left[ \Delta^{\downarrow \uparrow}_{\alpha,\beta} \right]^{*},  
\end{align}%
with the appropriate choice of the U(1) gauge. 
In addition, the pairing order parameter has four-fold rotational symmetry and 
mirror reflection symmetry with respect to the $yz$ and $zx$ planes as dictated by the point group $\mathrm{C}_{4v}$. 
Thus, it has
to be transformed according to the following relations:
\begin{align*}
C_{4}\Hat{\Delta}C^{t}_{4}&=e^{i\frac{n \pi}{2}}\Hat{\Delta}, \\
M_{yz}\Hat{\Delta}M^{t}_{yz}&=\pm \Hat{\Delta}, 
\end{align*}%
where $n$ equals to $0$ for A representation, $2$ for B representation, $1$, and $3$ for E representation. 
Such properties are very important to distinguish the symmetry of the solutions obtained by the Bogoliubov-de Gennes equation. 
%one has to require that these two symmetries hold. 
The energy gap functions are then explicitly constructed by taking into account the corresponding 
irreducible representations. 
For the one-dimensional representations, the $\mathrm{A}_1$ state is given by 
\begin{align*}
d^{(xy, zx)}_{x}&=-d^{(xy, yz)}_{y}, \\
\Delta^{\uparrow \uparrow}_{xy,yz}&=\Delta^{\downarrow \downarrow}_{xy,yz}=id^{(xy, yz)}_{y}, \\
\Delta^{\uparrow \uparrow}_{xy,zx}&=-\Delta^{\downarrow \downarrow}_{xy,zx}=-d^{(xy, zx)}_{x}, \\
\Delta^{\uparrow \downarrow}_{yz, zx}&=\Delta^{\downarrow \uparrow}_{yz, zx}=d^{(yz, zx)}_{z}, 
\end{align*}%
while for the $\mathrm{A}_2$ representation, 
\begin{align*}
d^{(xy, zx)}_{y}&=d^{(xy, yz)}_{x}, \\
\Delta^{\uparrow \uparrow}_{xy,yz}&=-\Delta^{\downarrow \downarrow}_{xy,yz}=-d^{(xy, yz)}_{x}, \\
\Delta^{\uparrow \uparrow}_{xy,zx}&=\Delta^{\downarrow \downarrow}_{xy,zx}=id^{(xy, zx)}_{y},
\end{align*}%
the $\mathrm{B}_1$ representation, 
\begin{align*}
d^{(xy, zx)}_{x}&=d^{(xy, yz)}_{y}, \\
\Delta^{\uparrow \uparrow}_{xy,yz}&=\Delta^{\downarrow \downarrow}_{xy,yz}=id^{(xy, yz)}_{y}, \\
\Delta^{\uparrow \uparrow}_{xy,zx}&=-\Delta^{\downarrow \downarrow}_{xy,zx}=-d^{(xy, zx)}_{x},
\end{align*}%
and the $\mathrm{B}_2$ representation, 
\begin{align*}
d^{(xy, zx)}_{y}&=-d^{(xy, yz)}_{x}, \\
\Delta^{\uparrow \uparrow}_{xy,yz}&=-\Delta^{\downarrow \downarrow}_{xy,yz}=-d^{(xy, yz)}_{x}, \\
\Delta^{\uparrow \uparrow}_{xy,zx}&=\Delta^{\downarrow \downarrow}_{xy,zx}=id^{(xy, zx)}_{y}, \\
\Delta^{\uparrow \downarrow}_{yz, zx}&=-\Delta^{\downarrow \uparrow}_{yz, zx}=\psi^{(yz, zx)}. 
\end{align*}%
Finally, for the E representation, there are doubly degenerate mirror-even $(+)$ and mirror-odd $(-)$ solutions:
\begin{align*}
\psi^{(xy, zx)}_{+}&=\mp id^{(xy,yz)}_{z+}, \\
d^{(xy, zx)}_{z-}&=\mp i\psi^{(xy,yz)}_{-}, \\
\Delta^{\uparrow \downarrow}_{xy, yz}&=\alpha_{-}\psi^{(xy, yz)}_{-}+\alpha_{+}d^{(xy, yz)}_{z+}, \\
\Delta^{\downarrow \uparrow}_{xy, yz}&=-\alpha_{-}\psi^{(xy, yz)}_{-}+\alpha_{+}d^{(xy, yz)}_{z+}, \\
\Delta^{\uparrow \downarrow}_{xy, zx}&=\alpha_{+}\psi^{(xy, zx)}_{+}+\alpha_{-}d^{(xy, zx)}_{z-}, \\
\Delta^{\downarrow \uparrow}_{xy, zx}&=-\alpha_{+}\psi^{(xy, zx)}_{+}+\alpha_{-}d^{(xy, zx)}_{z-}, \\
\Delta^{\uparrow \uparrow}_{yz,zx}&=-\alpha_{-}d^{(yz, zx)}_{x-}+i\alpha_{+}d^{(yz, zx)}_{y+}, \\
\Delta^{\downarrow \downarrow}_{yz,zx}&=\alpha_{-}d^{(yz, zx)}_{x-}+i\alpha_{+}d^{(yz, zx)}_{y+}, 
\end{align*}%
where $\alpha_{+}$ and $\alpha_{-}$ denote arbitrary constants for the linear superposition. 
As a consequence of the symmetry constraint and of the inter-orbital structure of the order parameter, 
different types of isotropic spin-triplet and singlet-triplet mixed configurations can be obtained.
Equal spin-triplet and opposite spin-triplet pairings are mixed in the 
$\mathrm{A}_{1}$ representation. 
On the other hand, in the $\mathrm{B}_{2}$ representation, equal spin-triplet 
and spin-singlet pairings are mixed. 
For the $\mathrm{A}_{2}$ and $\mathrm{B}_{1}$ representations, 
only equal spin-triplet pairings are allowed, and all types of pairings can be realized in the E representation. 
It is worth noting that $\mathrm{A}_{1}$, $\mathrm{B}_{2}$, and E representations have pairings between all the orbitals in the $yz$-$zx$ and $xy$-$yz/zx$ channels, 
while $\mathrm{A}_{2}$ and $\mathrm{B}_{1}$ can make electron pairings only 
in the $xy$-$yz/zx$ channel, that is, by mixing the $d_{xy}$ and $d_{yz}$/$d_{zx}$-orbitals as shown in Table \ref{irreducible1}. 
This symmetry constraint is important when 
searching for the ground-state configuration.

%%%%%%%%%%%%%%%%%%%%%%%%%%%%%%%%%
%%%%%%%%%%%%%%%%%%%%%%%%%%%%%%%%%
\section{Energy gap equation and phase diagram}

In order to investigate which of the possible symmetry-allowed solutions is more stable energetically, 
we solve the Eliashberg equation within the mean field approximation 
by taking into account the multi-orbital effects near the transition temperature. 
The linearized Eliashberg equation within the weak coupling approximation is given by 
\begin{align}
\Lambda\Delta^{\sigma \tau}_{\alpha, \gamma}
&=-\frac{k_\mathrm{B}T}{N}V_{\alpha, \gamma}\sum_{\bm{k}',i\varepsilon_m}F_{\alpha \sigma, \gamma \tau}(\bm{k}',i\varepsilon_m), \\
V_{xy, yz}&=V_{xy, zx}=V_{yz, xy}=V_{zx, xy}\equiv V_{xy}, \notag \\ 
V_{yz, zx}&=V_{zx, yz}\equiv V_{z}, \notag
\label{Eliashberg}
\end{align}%
\begin{align}
&F_{\alpha\sigma, \gamma\tau}(\bm{k}', i\varepsilon_m) \\
&=\sum_{\beta, \delta}\sum_{\sigma', \tau'}\Delta^{\sigma' \tau'}_{\beta, \delta}G^{\sigma \sigma'}_{\alpha, \beta}(\bm{k}',i\varepsilon_m)G^{\tau \tau'}_{\gamma, \delta}(-\bm{k}', -i\varepsilon_m),
\label{anomalousG} \notag
\end{align}%
where $\Lambda$ is the eigenvalue of the linearized Eliashberg equation. 
Here, $\sigma$, $\tau$, $\sigma'$, and $\tau'$ denote the spin states and $\alpha$, $\beta$, $\gamma$, and $\delta$ stand for the orbital indices. 
$F_{\alpha\sigma,\gamma\tau}(\bm{k}',i\varepsilon_m)$ is the anomalous Green's function.
As we assume an isotropic Cooper pairing, which is $k$-independent, 
the summation over momentum and Matsubara frequency 
in Eq. (\ref{Eliashberg}) gets simplified. 
Finally, the problem is reduced to the diagonalization of the 
$24 \times 24$ matrix.  
We then study the relative stability of the irreducible representations as listed in 
Table \ref{irreducible1}.
An analysis of the energetically most favorable superconducting states is performed as a function of $V_{z}/V_{xy}$, 
assuming that $V_{xy}/t=-1.0$ and for a given temperature $T/t=5.0\times10^{-5}$. 
When we keep the ratio $V_{z}/V_{xy}$, the eigenvalue $\Lambda$ is  proportional to $V_{xy}$ within the mean-field approximation. 
The choice of the representative coupling $V_{xy}/t=-1.0$ is guided by the fact that one aims to access a physical regime for the superconducting phase that in principle can be compared to realistic superconducting materials in the weak coupling limit. For instance, if one chooses $t\sim 200-300$ meV, which is common in oxides, and considering that the superconducting transition temperature $T_\mathrm{c}$ is obtained when the magnitude of the greatest eigenvalue gets close to 1, then one would find $T_\mathrm{c}$ to be of the order of $100-300$ mK, which is reasonable for the 2DEG superconductivity at the oxide interface. 
 
\begin{figure}[htbp]
\centering
\includegraphics[width=6cm]{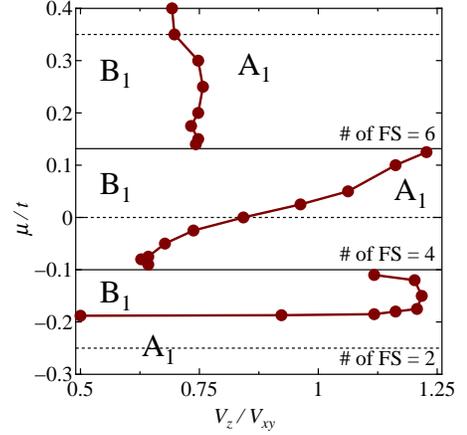}
\caption{Phase diagram as a function of $V_{z}/V_{xy}$ at 
$\lambda_\mathrm{SO}/t=0.10$, $\Delta_{is}/t=0.20$, $T/t=5.0\times10^{-5}$, and $V_{xy}/t=-1.0$. 
The brown solid line is the border between $\mathrm{A}_1$ and $\mathrm{B}_1$ 
states. 
The black solid line indicates the value of the chemical potential for which the number of Fermi surfaces changes.
The black dotted lines correspond to the values of the chemical potentials used in Fig. 2 for the normal state Fermi surfaces. }
%\label{fig4}
\label{Fig3}
\end{figure}%
Figure \ref{Fig3} shows the superconducting phase diagram for representative amplitudes of the spin-orbit coupling, $\lambda_\mathrm{SO}/t=0.10$, and inversion asymmetry interaction, $\Delta_{is}/t=0.20$, 
while varying both the chemical potential and the ratio of the pairing couplings $V_{z}/V_{xy}$. 
Owing to the inequivalent mixing of the orbitals in the paired configurations, it is plausible to expect a significant competition between the various symmetry allowed states and that such an interplay is sensitive not only to the pairing orbital anisotropy, but also to the structure and the number of Fermi surfaces.  
A direct observation is that for $V_{z}$ larger than $V_{xy}$, the $\mathrm{A}_1$ phase is stabilized with respect to the $\mathrm{B}_1$ phase because it contains a $d^{(yz, zx)}_z$ channel of a spin-triplet pairing in the $yz$-$zx$ sector that is absent in $\mathrm{B}_1$ phase. 
However, such a simple deduction does not directly explain why the $\mathrm{A}_1$ phase wins the competition with other superconducting phases, e.g., the $\mathrm{B}_2$ and E phases, which also can gain condensation energy by pairing electrons in the $yz$-$zx$ sector.
As a different type of $\bf{d}$-vector orientation enters into the $\mathrm{A}_1$ and E configurations, yet in the $\mathrm{B}_2$ state, the $yz$-$zx$ channel has a spin-singlet pairing, one can deduce that the interplay between the spin of the Copper pairs and that of the single-electron states close to the Fermi level is relevant to single out the most favorable superconducting phase.

The boundary between the $\mathrm{A}_1$ and $\mathrm{B}_1$ phases exhibits a sudden variation 
when one tunes the chemical potential across the value for which the number of Fermi surfaces changes. Such an abrupt transition 
is, however, plausible when passing through a Lifshitz point in the electronic structure of the normal state because other pairing channels get activated at the Fermi level.
The relation between the modification of the superconducting state and electronic topological or Lifshitz transition\cite{Lifshitz1960} that the Fermi surface can undergo is a subject of general interest. Indeed, there are many theoretical studies and experimental signatures pointing to a subtle interplay of Lifshitz transitions and superconductivity in cuprates \cite{Benhabib2015,Norman2010}, heavy-fermion superconductors \cite{Yelland2011} and more recently in iron-based superconductors \cite{Shi2017,Ren2017,Khan2014,Liu2011}. In those cases, major changes of the superconducting state seem to occur when going through a Lifshitz transition because Fermi pockets can appear or disappear at the Fermi level and in turn lead to different physical effects.  

Here, along this line of investigation, the role of the electron filling is also quite important and sets the competition between the energetically most stable phases. Indeed, one can notice that the $\mathrm{A}_1$ ($\mathrm{B}_1$) phase is stabilized for higher (lower) $V_{z}/V_{xy}$ and lower (higher) $\mu$.
Furthermore, we find that, in the case of two Fermi surfaces, the 
$\mathrm{A}_1$ state is further stabilized by decreasing the chemical potential and moving to a regime of extremely low concentration.
On the other hand, a transition to the $\mathrm{B}_1$ phase is achieved by electron doping. 
In the doping regime of four bands at the Fermi level, 
the $\mathrm{A}_1$-$\mathrm{B}_1$ boundary evolves approximately as a linear function of $V_{z}/V_{xy}$. This implies that the $\mathrm{A}_1$ configuration tends to be 
less stable and a higher ratio $V_{z}/V_{xy}$ is needed to achieve such a configuration at a given chemical potential.   
Finally, approaching the doping regime of six Fermi surfaces, the $\mathrm{A}_1$-$\mathrm{B}_1$ boundary becomes independent of the amplitude of $\mu$.
It is remarkable that the doping can substantially alter the competition between the $\mathrm{A}_1$ and $\mathrm{B}_1$ phases, thus manifesting the 
intricate consequences of the spin-orbital character of the electronic structure close to the Fermi level.  

\begin{figure}[htbp]
\centering
\includegraphics[width=8.5cm]{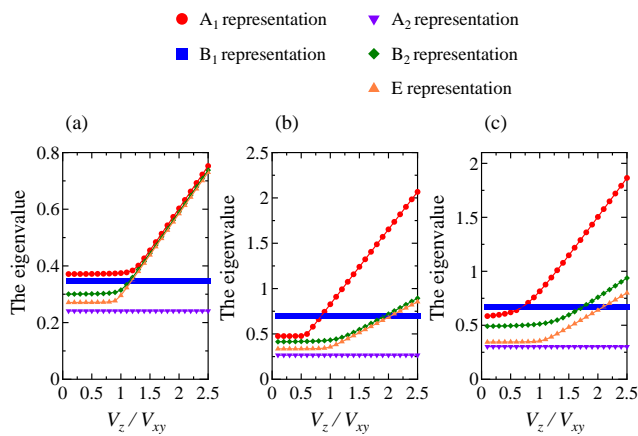}
\caption{Evolution of the eigenvalue of the Eliashberg matrix equation as a function of 
$V_{z}/V_{xy}$ at (a) $\mu /t=-0.25$, (b) $\mu /t=0.0$, and (c) $\mu /t=0.35$ at $\lambda_\mathrm{SO}/t=0.10$, $\Delta_{is}/t=0.20$, $T/t=5.0\times10^{-5}$, and $V_{xy}/t=-1.0$. 
(a) $\mathrm{A}_1$ representation is dominant. 
(b)and(c) $\mathrm{B}_1$ is dominant for small amplitude of the ratio 
$V_{z}/V_{xy}$. }
%\label{fig5}
\label{Fig4}
\end{figure}

To explicitly and quantitatively demonstrate the energy competition among all the symmetry allowed phases, one can follow the behavior of the eigenvalues of the linearized Eliashberg equations as a function of the ratio $V_{z}/V_{xy}$ (Fig. \ref{Fig4}). 
Figures. \ref{Fig4}(a)-(c) show the eigenvalues of the Eliashberg matrix equation for all the irreducible representations 
as a function of $V_{z}/V_{xy}$ when the number of Fermi surfaces is 
(a) two ($\mu/t=-0.25$), (b) four ($\mu/t=0.0$), and (c) six ($\mu/t=0.35$) as indicated by the dotted lines in Fig. \ref{Fig3}. 
With the increase in $V_{z}$, 
the magnitude of the eigenvalues of the irreducible representations 
including the $yz$-$zx$ channel, 
i.e., $\mathrm{A}_{1}$, $\mathrm{B}_{2}$, and E representations, 
increases in all the cases with two, four, and six Fermi surfaces. 
On the other hand, the eigenvalues of the A$_2$ and B$_1$ representations are independent of $V_{z}$, as $V_{z}$ is irrelevant 
for this pairing channel. 
When the number of Fermi surfaces is two, 
the $\mathrm{A}_1$ representation is the most dominant pairing for all $V_{z}$. 
Although the magnitude of the eigenvalues for the  
$\mathrm{B}_2$ and $\mathrm{E}$ representations also increases with $V_{z}$, 
these solutions never become dominant as compared with the $\mathrm{A}_1$ state. 
When the number of Fermi surfaces is four or six, 
the eigenvalue of the $\mathrm{B}_1$ phase is larger
than that of the $\mathrm{A}_1$ representation for lower $V_{z}$. 

Finally, we have investigated the phase diagram by scanning a larger range of temperatures for few representative cases of pairing interaction and filling concentration (see Appendix). The results are not significantly changed except in a region of extremely high temperature, corresponding to an unphysically large amplitude of the pairing interaction. There, although B$_1$ keeps being the most stable state, the largest eigenvalues indicate a competition between the B$_1$ and B$_2$ rather than the B$_1$ and A$_1$ configurations.

%%%%%%%%%%%%%%%%%%%%%%%%%%%%%%%%%
%%%%%%%%%%%%%%%%%%%%%%%%%%%%%%%%%
\section{Topological properties and energy excitation spectrum in the bulk and at the edge}

In the previous section, we confirmed that both the $\mathrm{A}_1$ and $\mathrm{B}_1$ pairings can be energetically stabilized in a large region of the parameter space. 
Thus, it is relevant to further consider the nature of the electronic structure of 
these superconducting phases in order to provide key elements and indications that can be employed for the detection of the most favorable inter-orbital superconductivity. 
The analysis is based on the solution of the Bogoliubov-de Gennes (BdG) equation for the evaluation of the 
low-energy spectral excitations both in the bulk and at the edge of the superconductor
for both the $\mathrm{A}_1$ and $\mathrm{B}_1$ phases. 
The matrix Hamiltonian in momentum space is given by 
\begin{align}
&H_\mathrm{BdG}(\bm{k})=
\begin{pmatrix}
H(\bm{k}) & \hat{\Delta} \\
\hat{\Delta}^\dagger & -H^{*}(-\bm{k})
\end{pmatrix}. 
\end{align}%
with $H(\bm{k})$ being the normal state Hamiltonian. 

\subsection{Bulk energy spectrum and topological superconductivity}

In order to determine the excitation spectrum, we solve the BdG equations for both the $\mathrm{A}_1$ and $\mathrm{B}_1$ configurations. 
For convenience, we introduce the gap amplitude $|\Delta_{0}|$, and we set the components of the $\bf{d}$-vectors to be
\begin{align}
d^{(xy, yz)}_{y}=-d^{(xy, zx)}_{x}=d^{(yz, zx)}_{z}=|\Delta_{0}|,
\end{align}%
for $\mathrm{A}_1$ and  
\begin{align}
d^{(xy, yz)}_{y}=d^{(xy, zx)}_{x}=|\Delta_{0}|,
\end{align}%
for $\mathrm{B}_1$ state. Here, the parameter $|\Delta_{0}|/t=1.0\times10^{-3}$ is set as a scale of energy. 

\begin{figure}[htbp]
\centering
\includegraphics[width=8cm]{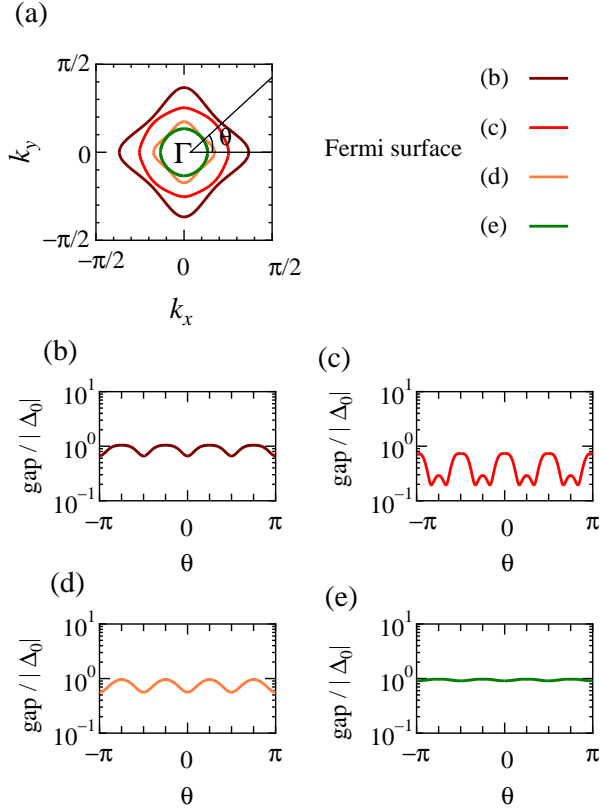}
\caption{(a) Fermi surfaces at $\mu/t=0.0$ in the normal state. 
[(b)-(e)] $\mathrm{A}_1$ quasi-particle energy gap along the Fermi surface as a function of the polar angle $\theta$ as shown in (a) for 
$\lambda_\mathrm{SO}/t=0.10$, $\Delta_{is}/t=0.20$, 
and $|\Delta_{0}|/t=1.0\times10^{-3}$ corresponding to the Fermi surfaces in (a). }
%The line in 
%is the magnitude of energy gap on the 
%Fermi surface shown in curves (b), (c), (d) and (e) in (a). }
%\label{fig7}
\label{Fig5}
\end{figure}
\begin{figure}[htbp]
\centering
\includegraphics[width=8cm]{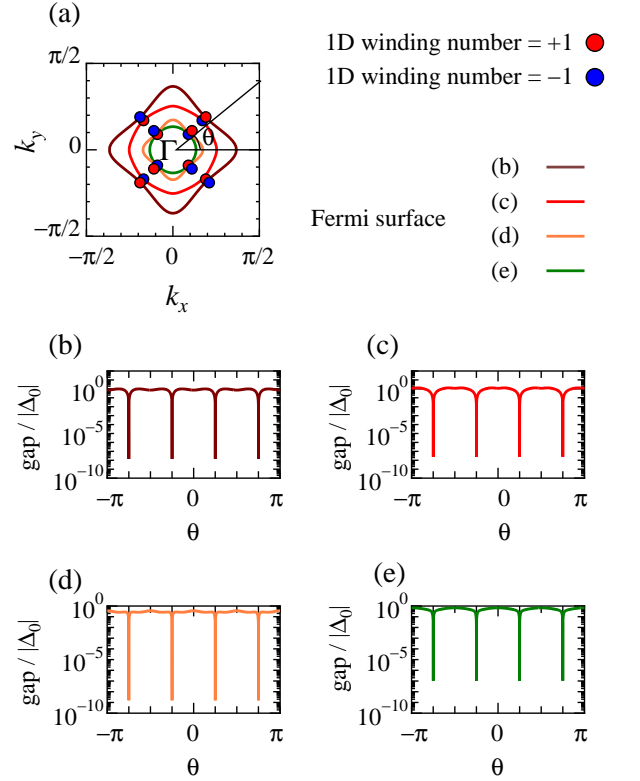}
\caption{(a) Fermi surfaces and 
position of the nodes at $\mu/t=0.0$. We indicate the winding numbers defined at 
each node. 
(b)-(e) indicate the quasi-particle energy spectra for the $\mathrm{B}_1$ state with $\lambda_\mathrm{SO}/t$ = 0.10, $\Delta_{is}/t=0.20$, and $|\Delta_{0}|/t=1.0\times10^{-3}$ at the corresponding Fermi surfaces shown in (a).}
%\label{fig8}
\label{Fig6}
\end{figure}

%
We start focusing on the doping regime of four bands at the Fermi level. 
In this case, the $\mathrm{A}_1$ state has a fully gapped electronic structure 
for all the bands at the Fermi level as demonstrated by the inspection of the in-plane 
angular dependence of the gap magnitude [Figs. \ref{Fig5}(b)-(e)]. 
In particular, we notice that the gap amplitude is not isotropic and orbital dependent when moving from the outer to the inner Fermi surface [Figs. \ref{Fig5}(b)-(e)]. The nodal state (Fig. \ref{Fig6}), on the other hand, exhibits a more regular behavior of the gap amplitude which is basically orbital independent and point nodes occurring only along the diagonal of the Brillouin zone on the various Fermi surfaces.

It is interesting to further investigate the nature of the nodal $\mathrm{B}_1$ phase by determining whether the existence of the nodes is related to a non-vanishing topological invariant. 
As the model Hamiltonian owes particle-hole and 
time-reversal symmetry, 
one can define a chiral operator $\Hat{\Gamma}$ as a product of 
the particle-hole $\Hat{C}$ and 
time-reversal $\Hat{\Theta}$ operators. 
As the chiral symmetry operator anticommutes with $H_\mathrm{BdG}(\bm{k})$, by employing a unitary transformation rotating the basis in the eigenbasis of $\Hat{\Gamma}$, the Hamiltonian can be put in an off-diagonal form with antidiagonal blocks. Hence, the determinant
of each block can be put in a complex polar form and, as long as the eigenvalues are non-zero, it can be used to obtain a winding number by evaluating its trajectory in the complex plane. On a general ground, we point out that the number of singularities in the phase of the determinant is a topological invariant~\cite{tewarisau} because it cannot change without the amplitude going to zero, thus implying a gap closing and a topological phase transition.
For this symmetry class, then, one can associate and determine the winding number around each node by following, for instance, the approach already applied successfully in Refs.
[\onlinecite{YSTY10,STYY11,Brydon2011PRB}]. 
\begin{figure}[htbp]
\centering
\includegraphics[width=8cm]{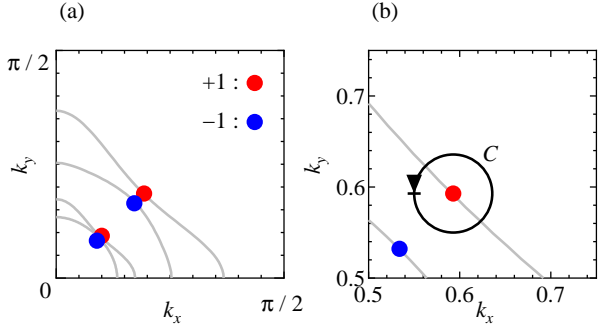}
\caption{(a) Fermi surfaces at $\lambda_\mathrm{SO}/t=0.10$, $\Delta_{is}/t=0.20$, $\mu/t=0.0$, and point nodes position (winding number) at $\Delta_{0}/t=1.0\times10^{-3}$. 
(b) Zoomed view of the plot in (a) and a contour of the integral $C$. }
%\label{fig9}
\label{Fig7}
\end{figure}%
The chiral, particle-hole, and time-reversal operators are expressed as
\begin{align}
\Hat{\Gamma}&=-i\Hat{C}\Hat{\Theta}, \\
\Hat{C}&=
\begin{pmatrix}
0 & \Hat{I}_{6\times6} \\
\Hat{I}_{6\times6} & 0
\end{pmatrix}
=\Hat{l}_{0} \otimes \Hat{\sigma}_{x} \otimes \Hat{\tau}_{0}, \\
\Hat{\Theta}&=\Hat{l}_{0} \otimes i\Hat{\sigma}_{y} \otimes \Hat{\tau}_{0}. 
\end{align}%
Here, $\Hat{I}_{6\times6}$ and $\Hat{\tau}_{0}$ denote the $6\times6$ unit matrix and the identity matrix in the particle-hole space, respectively. 
As we consider time-reversal symmetric pairings, 
the chiral operator anticommutes with the Hamiltonian:  
\begin{align}
\{ H_{\mathrm{BdG}}(\bm{k}), \Hat{\Gamma} \}=0.
%\{ \Hat{H}_{\mathrm{BdG}}, \Hat{\Gamma} \}&=0, \\
%\Hat{\Gamma}&=
%\begin{pmatrix}
%0 & \Hat{l}_{0} \otimes \Hat{\sigma}_{y} \\
%\Hat{l}_{0} \otimes \Hat{\sigma}_{y} & 0
%\end{pmatrix}. 
\end{align}%
One can then introduce a unitary matrix $\Hat{U}_{\Gamma}$ that diagonalizes the chiral operator $\Hat{\Gamma}$:  
\begin{align}
\Hat{U}^{\dagger}_{\Gamma}\Hat{\Gamma}\Hat{U}_{\Gamma}&=
\begin{pmatrix}
\Hat{I}_{6\times6} & 0 \\
0 & -\Hat{I}_{6\times6}
\end{pmatrix}, \\
\Hat{U}^{\dagger}_{\Gamma}&=\Hat{U}_{\Gamma}=\frac{1}{\sqrt{2}}
\begin{pmatrix} 
\Hat{I}_{6\times6} & \Hat{l}_{0} \otimes \Hat{\sigma}_{y} \\
\Hat{l}_{0} \otimes \Hat{\sigma}_{y} & -\Hat{I}_{6\times6}
\end{pmatrix}. 
\end{align}%
In this basis the BdG Hamiltonian is block antidiagonalized by $\Hat{U}_{\Gamma}$,  
\begin{align}
&\Hat{U}^{\dagger}_{\Gamma}H_{\mathrm{BdG}}(\bm{k})\Hat{U}_{\Gamma}=
\begin{pmatrix}
0 & \Hat{q}(\bm{k}) \\
\Hat{q}^{\dagger}(\bm{k}) & 0
\end{pmatrix}, \\
&\Hat{q}=H(\bm{k})\left[ \Hat{l}_0 \otimes \Hat{\sigma}_y \right]-\Hat{\Delta}. 
\end{align}%
Then, the determinant
of the $\Hat{q}(\bm{k})$ matrix block can be put in a complex polar form, and as long as the eigenvalues are nonzero, it can be used to obtain the winding number $W$ by evaluating its trajectory in the complex plane as
\begin{align}
%W(k^{0}_{x})&=\frac{1}{2\pi}\int^{\pi}_{-\pi}\frac{\partial \theta(\bm{k})}{\partial k_y}|_{k_x\rightarrow k^{0}_x}dk_y \\
W&=\frac{1}{2\pi}\oint_{C} d\theta(\bm{k}), \label{winding} \\ 
\theta(\bm{k})&\equiv \arg[\det{\Hat{q}(\bm{k})}] \,. \notag
\label{winding}
\end{align}%
%is obtained. 
$C$ in Eq. (\ref{winding}) is a closed line contour that encloses a given node as schematically shown in Fig. \ref{Fig7}(b). 
From the explicit calculation, 
we find that the amplitude of $W$ is $\pm1$ 
[see Figs. \ref{Fig6}(a) and \ref{Fig7}(a)]. 
\begin{figure}[htbp]
\centering
\includegraphics[width=8.5cm]{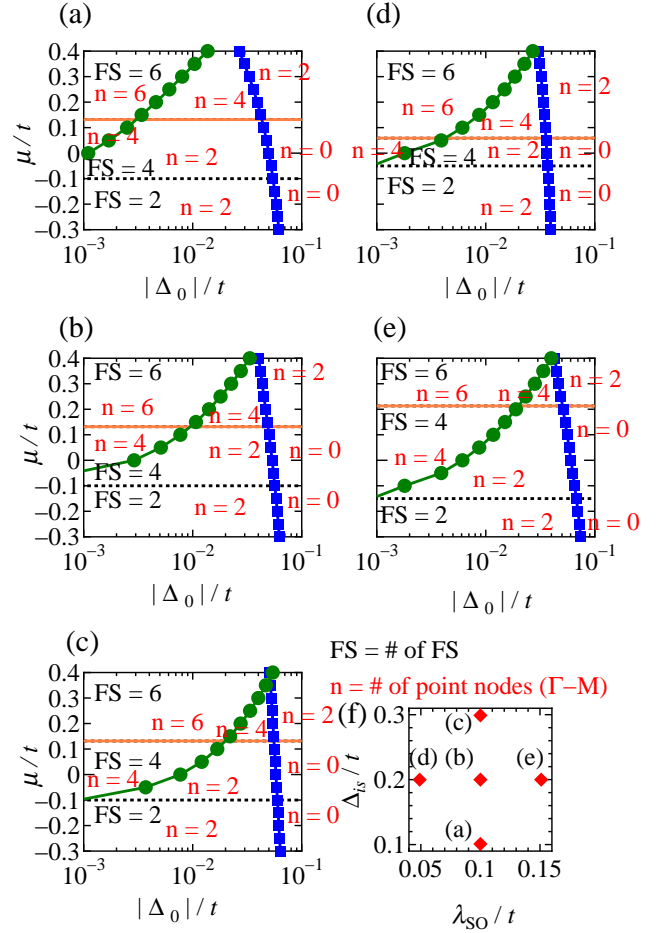}
\caption{Phase diagram of the Lifshitz transitions for the nodal $\mathrm{B}_1$ phase at 
(a) $\lambda_\mathrm{SO}/t=0.10$ and $\Delta_{is}/t=0.10$, (b) $\lambda_\mathrm{SO}/t=0.10$ and $\Delta_{is}/t=0.20$, (c) $\lambda_\mathrm{SO}/t=0.10$ and $\Delta_{is}/t=0.30$, (d)$\lambda_\mathrm{SO}/t=0.05$ and $\Delta_{is}/t=0.20$, and (c) $\lambda_\mathrm{SO}/t=0.15$ and $\Delta_{is}/t=0.20$. (f) Schematic plot of the correspondence between the spin-orbit and the inversion asymmetry couplings and the panels (a)-(e). 
Black and red labels denote the number of Fermi surfaces in the normal state
and the point nodes in the superconducting phase that are located along the diagonal of the Brillouin zone from $\Gamma$ to M, respectively.
The black dotted line and the orange solid line indicate the two-to-four Fermi surface separation and the four-to-six Fermi surface boundary in the normal state. Circles and squares set the transition lines for the nodal superconductor between configurations having different number of nodes in the excitation spectrum.}
%\label{fig11}
\label{Fig8}
\end{figure}%
If the nodes have a nonzero winding number, edge states appear due to the bulk-edge correspondence. 
It is known \cite{STYY11} that the following index theorem is satisfied: for any one-dimensional cut in the Brillouin zone that is indicated by a given momentum $k_{\parallel}$ that is parallel to the edge, 
one has that $w(k_{\parallel})=n_{+}-n_{-}$, with $n_{+}$ and $n_{-}$ being the number of the eigenstates associated to the eigenvalues $+1$ and $-1$ of the chiral operator $\Hat{\Gamma}$, respectively. 
The number of edge states is equal to $|w(k_{\parallel})|$ when considering a boundary configuration with a conserved $k_{\parallel}$.
We can easily show that $W$ which is given in Eq. (\ref{winding}) and $w(k_{\parallel})$ are deeply linked:
$w(k_1)-w(k_2)=-W\mathrm{sgn}(k_1-k_2)$ where $W$ is the total winding number around the nodes between $k_{\parallel}=k_1$ and $k_{\parallel}=k_2$.
Thus, nonzero $W$ on the nodes means nonzero $w(k_{\parallel})$  and the existence of the zero energy edge state with appropriate choice of the crystal plane. 
Such relation sets the main physical connection between the winding number and the properties of the topological superconductor. 

We generally find that two to six point nodes can occur along the $\Gamma$-M
direction, and their number is related to that of the Fermi surfaces.
Interestingly, the position of the point nodes is not fixed and pinned to the lines
of the Fermi surface in the normal state. In general, their position along the diagonal of the Brillouin zone depends on the amplitude $|\Delta_{0}|$ and indirectly on the values of the spin-orbit and inversion asymmetry couplings. 
Thus, two adjacent point nodes with opposite winding numbers can, in principle, be moved until they merge and then disappear by opening a gap in the excitation spectrum. 
This behavior is generally demonstrated in Fig. \ref{Fig8}. 
A phase diagram can be determined in terms of the amplitude $|\Delta_{0}|$ and the chemical potential $\mu$.
The nodal superconductor can undergo different types of Lifshitz transitions, and in general, those occurring in the normal state are not linked to the nodal merging in the superconducting phase. 
Indeed, one of the characteristic features of the nodal superconductor is that, by changing the filling, through $\mu$, one can drive a transition from two to four and six point nodes independently of the number of bands crossing the Fermi level in the normal state. 
It is rather the strength of $|\Delta_{0}|$ that plays an important role in tuning the nodal superconductor. An increase in $|\Delta_{0}|$ tends to reduce the number of nodes until a fully gapped phase appears. 
As the critical lines are sensitive to the spin-orbit $\lambda_\mathrm{SO}$ and inversion asymmetry $\Delta_{is}$ couplings, one can get line crossings that allow for multiple merging of nodes such that the superconductor can undergo a direct transition from six to two at $\mu/t \sim 0.40$ [Figs. \ref{Fig8}(e) and (d)] or from four to zero point nodes, as for instance nearby the crossing between the blue and orange lines at $\mu/t \sim 0.10$ in the Fig. \ref{Fig8}.
As the positions of the point nodes are fixed, each Fermi
surface in the limit of small $|\Delta_{0}|$ and its distance in the Brillouin zone
increases with level splitting by $\Delta_{is}$ and
$\lambda_\mathrm{SO}$, a larger $|\Delta_{0}|$ is required to annihilate the point nodes when both $\Delta_{is}$ and $\lambda_\mathrm{SO}$ grow in amplitude 
as demonstrated by the shift of the green and blue critical lines in Figs. \ref{Fig8}(a)-\ref{Fig8}(c) for different values of $\Delta_{is}$, and Figs. \ref{Fig8}(d), \ref{Fig8}(b), and \ref{Fig8}(e) in terms of $\lambda_\mathrm{SO}$.
When considering these results in the context of two-dimensional superconductors that emerge at the surface or interface of band insulators we observe that the achieved topological transitions can be driven by gate voltage and temperature, as $\mu$ and $\Delta_{is}$ are tunable by electric fields, and the amplitude of $|\Delta_{0}|$ can be controlled by the temperature and the electric field as well.

\subsection{Local density of states at the edge of the superconductor}

Having established that the nodes in the $\mathrm{B}_1$ configuration are protected by a nonvanishing winding number, one can expect that flat zero-energy 
surface Andreev bound states (SABS) occur at the boundary of the superconductor. 

\begin{figure*}[htbp]
\centering
\includegraphics[width=14cm]{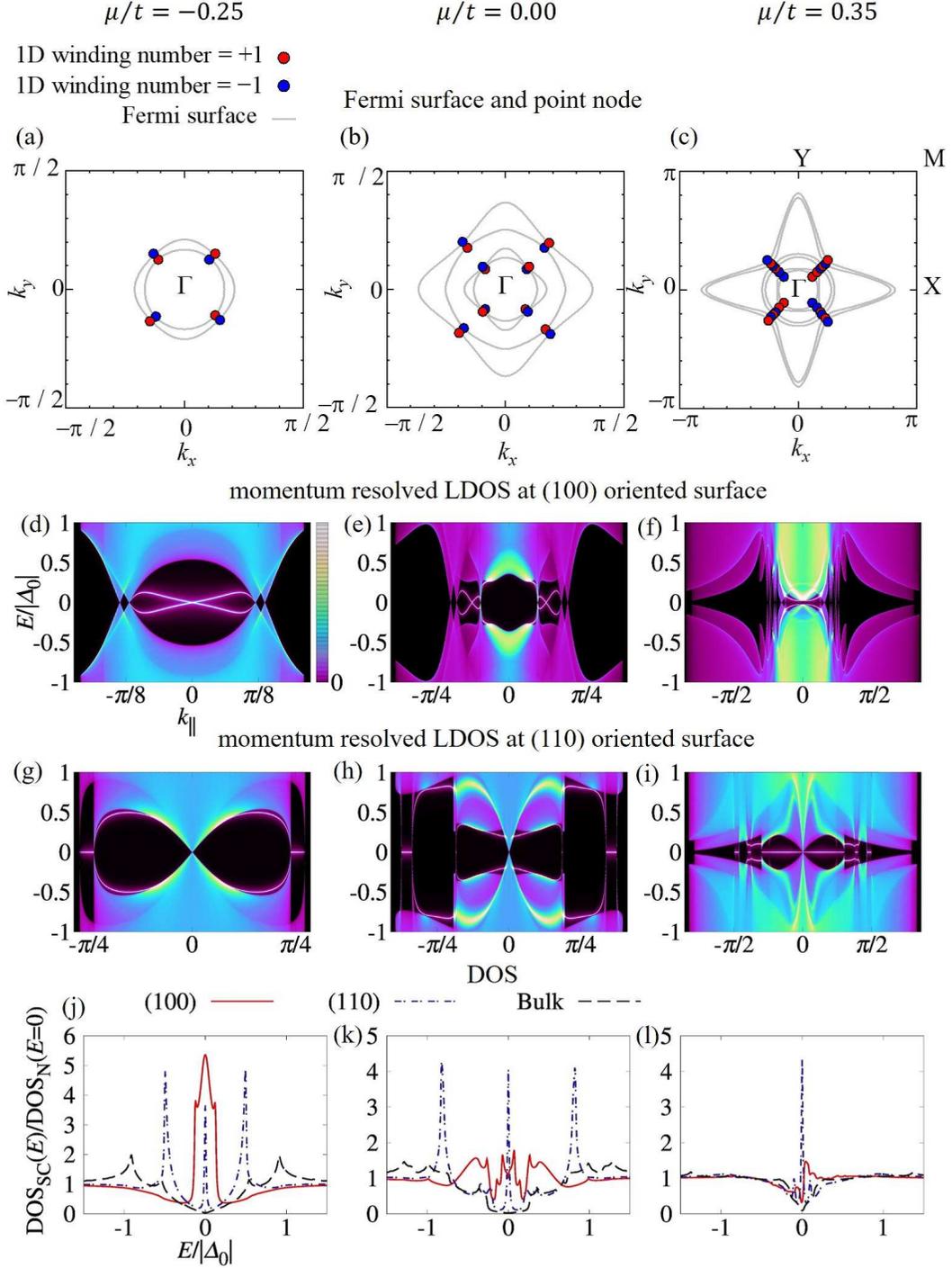}
\caption{Momentum-resolved and angular averaged LDOS for 
B$_{1}$ representation. 
The Fermi surfaces and the position of the point nodes are shown 
for (a) $\mu/t=-0.25$, (b) $\mu/t=0.0$, and (c) $\mu/t=0.35$. 
The momentum ($k_{\parallel}$) resolved LDOS at the (100) oriented surface  
for (d) $\mu/t=-0.25$, (e) $\mu/t=0.0$, and (f) $\mu/t=0.35$. 
%The SABS of (e) are not helical edge states since it does not connect at $E=0$. 
The momentum ($k_{\parallel}$) resolved LDOS at the (110) oriented surface  
for (g) $\mu/t=-0.25$, (h) $\mu/t=0.0$, and (i) $\mu/t=0.35$. 
LDOS normalized by its normal state value at $E=0$ [(DOS$_\mathrm{N}(E=0)$] at the (100) and (110) oriented surfaces, and in the bulk for (j) $\mu/t=-0.25$, (k) $\mu/t=0.0$, and (l) $\mu/t=0.35$. 
The red solid line, blue dash-dotted line, and black dashed line 
denote the LDOS at the (100) oriented surface, (110) oriented surface, 
and in the bulk, respectively.
Other parameters are $\lambda_\mathrm{SO}/t=0.10$, $\Delta_{is}/t=0.20$, and $|\Delta_{0}|/t=1.0\times10^{-3}$. }
%\label{fig10}
\label{Fig9}
\end{figure*}

In this section, we investigate the SABS and the local density of states (LDOS)
for two different terminations of the two-dimensional superconductor, i.e., the (100) and (110) oriented edges.
We start by discussing the LDOS for the (100) and (110) edges at representative values of $\lambda_\mathrm{SO}/t=0.10$, $\Delta_{is}/t=0.20$, and $|\Delta_{0}|/t=1.0\times10^{-3}$, and by varying the chemical potential in order to compare the cases with a different number of point nodes in the bulk energy spectrum at $\mu/t=-0.25$, $\mu/t=0.0$, and $\mu/t=0.35$ as shown in Figs. \ref{Fig9}(a)-\ref{Fig9}(c), respectively.
%%%%%%%%%%%%%%%%%%%%%%%%
%%%%%%%%%%%%%%%%%%

As expected, the momentum-resolved LDOS indicates that zero-energy SABS can be observed but only for specific orientations of the edge. Indeed, as reported in Figs. \ref{Fig9}(d)-\ref{Fig9}(i), one has zero-energy SABS (ZESABS) for the (110) boundary while they are absent
for the (100) edge.
The reason for having inequivalent SABS edge modes is directly related to the presence of a nontrivial winding number that is protecting the point nodes.
For the (110) edge, isolated point nodes exist in the surface Brillouin zone, and
they have winding numbers with opposite sign.
Thus, the ZESABS, which connects the nodes with a positive and negative winding number, emerge in the gap.
On the other hand, when considering the (100) oriented termination, the winding numbers for positive $k_{x}$ and negative
$k_{x}$ are completely opposite in sign, and they cancel each other when projected on
the (100) surface Brillouin zone.
Thus, flat zero-energy states cannot occur for the (100) edge.
Nevertheless, helical edge modes are observed inside the energy gap as demonstrated in
Fig. \ref{Fig9}(d). This is because the Majorana edge modes with positive
and negative chirality can couple, get split, and acquire a dispersion.
The differences in the edge ABS also manifest in the momentum integrated LDOS. 
For the (110) edge,
owing to the presence of the ZESABS, the LDOS normalized by its normal state value at $E=0$ shows pronounced
zero-energy peaks (see dash-dotted line in Figs. \ref{Fig9}(j), (k), and (l)).
On the other hand, for the (100) boundary, they lead to a broad peak or exhibit
many narrow spectral structures reflecting the complex dispersion of the edge states.

\begin{figure}[htbp]
\centering
\includegraphics[width=7cm]{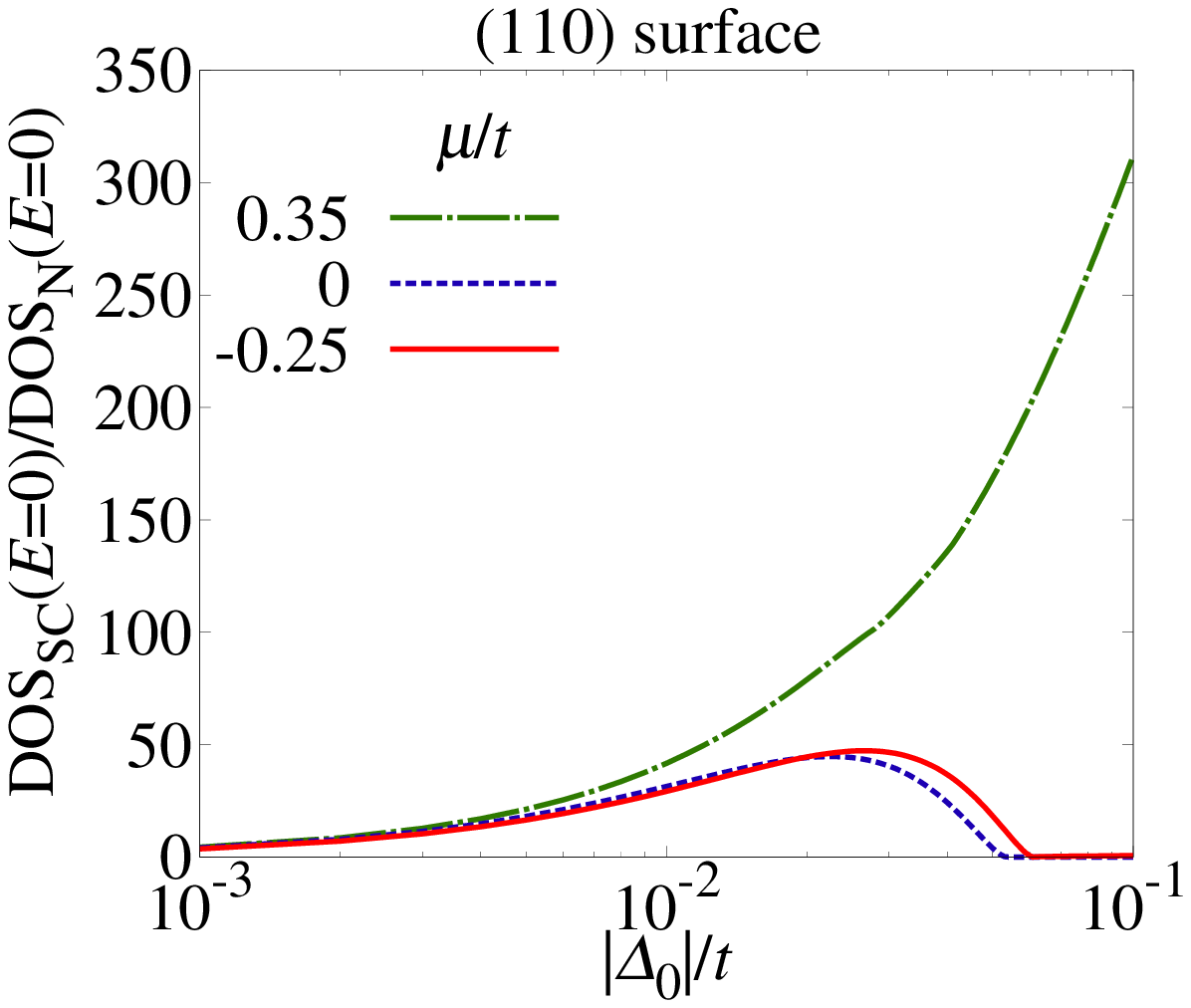}
\caption{The LDOS at $E=0$ for the (110) oriented surface 
for the $\mathrm{B}_1$ representation as a function of 
$|\Delta_{0}|/t$ at $\lambda_\mathrm{SO}/t=0.10$ and 
$\Delta_{is}/t=0.20$. 
The red solid line, blue dotted line, and  
green dashdotted line correspond to 
$\mu/t=-0.25$, $\mu/t=0.0$, and $\mu/t=0.35$. }
%\label{fig12}
\label{Fig10}
\end{figure}%
Finally, we discuss the $|\Delta_{0}|$ dependence of LDOS at zero energy, i.e., $E=0$. 
For the (110) edge, the zero-energy peak mainly originates from the zero-energy flat band.
The height of the zero-energy peak can then be characterized by (i) the
strength of the localization of the edge state and (ii) the total length
of the ZESABS within the surface Brillouin zone.
The strength of the localization is defined by the inverse of the
localization length $1/\xi$ and $1/\xi\propto|\Delta_{0}|$. 
In other words, the peak height generally increases with $|\Delta_{0}|$.
On the other hand, as shown in Fig. \ref{Fig8}, the extension in the momentum space of the zero-energy flat states becomes shorter with increasing $|\Delta_{0}|$.
For simplicity, one can focus on the two Fermi surface configuration. 
In this case, the total length of the zero-energy flat band is roughly
estimated as $\delta {k}(1-|\Delta_{0}|/|\Delta_{0}^{c}|)$ for
$|\Delta_{0}|<|\Delta_{0}^{c}|$ and zero for $|\Delta_{0}|>|\Delta_{0}^{c}|$,
where $\delta {k}$ is the Fermi surface splitting along the $\Gamma$-M
direction and $|\Delta_{0}^{c}|$ is a critical value above which the point-nodes disappear.
Then, the height of the zero-energy peak is proportional to
$|\Delta_{0}|(1-|\Delta_{0}|/|\Delta_{0}^{c}|)$ for $|\Delta_{0}|<|\Delta_{0}^{c}|$
and vanishes for $|\Delta_{0}|>|\Delta_{0}^{c}|$.
This is a nonmonotonic dome-shaped behavior of the ZELDOS as a function of $|\Delta_{0}|$.
The explicit profile can be seen in Fig. \ref{Fig10} at $\mu/t=-0.25$ and $\mu/t=0.0$.
For $\mu/t=0.35$, the point nodes still exist in this parameter regime, and
the height of the zero energy peak develops with $|\Delta_{0}|$.
%ZESABS and gate voltage and temperature
Thus, we have that the ZESABS get strongly renormalized and are tunable by a variation in the electron filling ($\mu$) and amplitude of the order parameter $|\Delta_{0}|$ as shown in Fig. \ref{Fig8}.

%%%%%%%%%%%%%%%%%%%%%%%%%%%%%%%%%
%%%%%%%%%%%%%%%%%%%%%%%%%%%%%%%%%

\section{Discussion and summary}

We investigated and determined the possible superconducting phases arising from inter-orbital pairing in an electronic environment marked by spin-orbit coupling and inversion symmetry breaking while focusing on momentum independent paired configurations. 
One remarkable aspect is that, although the inversion symmetry is absent, one can have symmetry-allowed solutions that avoid mixing of spin-triplet and spin-singlet configurations. 
Importantly, states with only spin-triplet pairings can be stabilized in a large portion of the phase diagram.

Within those spin-triplet superconducting states, we unveiled an unconventional type of topological phase in two-dimensional superconductors that arises from the interplay of spin-orbit coupling and orbitally driven inversion-symmetry breaking. 
For this kind of a model system, atomic physics plays a relevant role and inevitably tends to yield orbital entanglement close to the Fermi level. Thus we assumed that local inter-orbital pairing is the dominant attractive interaction. 
As already mentioned, this type of pairing in the presence of inversion symmetry breaking allows for solutions that do not mix spin-singlet and triplet configurations. 
The orbital-singlet/spin-triplet superconducting phase can have a topological nature with distinctive spin-orbital fingerprints in the low-energy excitations spectra that make it fundamentally different from the topological configuration that is usually obtained in single band noncentrosymmetric superconductors. 
Here, a remarkable finding is that, contrary to the common view that an isotropic pairing structure leads to a fully gapped spectrum, a nodal superconductivity can be achieved when considering an {\it isotropic} spin-triplet pairing. Although in a different context, 
we noticed that akin paths for the generation of an anomalous nodal-line superconductor can also be encountered when local spin-singlet pairing occur in antiferromagnetic semimetals \cite{Brze18}.  

In the present study, for a given symmetry, the superconducting phase can exhibit point nodes that are protected by a nonvanishing winding number. 
The most striking feature of the disclosed topological superconductivity is expressed by its being prone to both topological and Lifshitz-type transitions upon different driving mechanisms and interactions, e.g., when tuning the strength of intrinsic spin-orbit and orbital-momentum couplings or by varying doping and the amplitude of order parameter by, for example, varying the temperature. 
The essence of such a topologically and electronically tunable superconductivity phase is encoded in the fundamental observation of having control of the nodes position in the Brillouin zone. 
Indeed, the location of the point nodes is not determined by the symmetry of the order parameter in the momentum space, as occurs in the single band noncentrosymmetric system, but rather it is a nontrivial consequence of the interplay between spin-triplet pairing and the spin-orbital character of the electronic structure. 
In particular, their position and existence in the Brillouin zone can be manipulated through various types of Lifshitz transitions, if one varies the chemical potential, the amplitude of the spin-triplet order parameter, the inversion symmetry breaking term, and the atomic spin-orbit coupling.  
While electron doping can induce a change in the number of Fermi surfaces, such electronic transition is not always accompanied by a variation in the number of nodes within the superconducting state. 
This behavior allows one to explore different physical scenarios that single out notable experimental paths for the detection of the targeted topological phase. 
Owing to the strong sensitivity of the topological and Lifshitz transitions with respect to the strength of the superconducting order parameter, one can foresee the possibility of observing an extraordinary reconstruction of the superconducting state both in the bulk and at the edge by employing the temperature to drive the pairing order parameter to a vanishing value, i.e., at the critical temperature, starting from a given strength at zero temperature. Then, a substantial thermal reorganization of the superconducting phase can be obtained. 
While a variation in the number of nodes in the low energy excitations spectra cannot be easily extracted by thermodynamic bulk measurements, we find that the electronic structure at the edge of the superconductor generally undergoes a dramatic reconstruction that manifests into a non-monotonous behavior of the zero bias conductance or in an unconventional thermal dependence of the in-gap states. 
Another important detection scheme of the examined spin-triplet superconductivity emerges when considering its sensitivity to the doping or to the strength of the inversion symmetry breaking coupling, which can be accessed by applying an electrostatic gating or pressure. 
Such gate/distortive control can find interesting applications, especially when considering two-dimensional electron gas systems. 

Another interesting feature of the multiple-nodes topological superconducting phase is given by the strong sensitivity of the edge states to the geometric termination, as demonstrated in Fig. \ref{Fig9}. 
This is indeed a consequence of the presence of nodes with an opposite sign winding number within the Brillouin zone. 
Hence, when considering the electronic transport along a profile that is averaging different terminations, it is natural to expect multiple in-gap features.  

Owing to the multi-orbital character of the superconducting state, we expect that non-trivial odd-in-time pair amplitudes are also generated\cite{Black_Schaffer2013PRB,Black_Schaffer2013PRB2,Komendov2015PRB,Asano2015PRB,Komendov2017PRL}. 
In particular, we predict that both local odd-in-time spin-singlet and triplet states can be obtained in the bulk and at the edge. 
The local spin-singlet odd-in-time pair correlations are an exquisite consequence of the multi-orbital superconducting phase. 
Accessing the nature of their competition/cooperation and its connection to the nodal superconducting phase is a general and relevant problem in relation to the generation, manipulation, and control of odd-in-time pair amplitudes. 

It is also relevant to comment on the impact of an intra-orbital pairing on the achieved results. Here, there are few fundamental observations to make. 
Firstly, one may ask whether the topological B$_1$ phase is robust to the adding of an extra pairing component which in the intra-orbital channel is most likely to have a spin-singlet symmetry. For this circumstance, one can start by pointing out that for any intra-orbital pairing component that does not break the chiral symmetry protecting the nodal structure of the superconducting state, the B$_1$ configuration can only undergo a Lifshitz-type transition associated with the merging of nodes having opposite sign in the winding number. Moreover, specifically for the B$_1$ irreducible representation, the intra-orbital spin-singlet component would have a $d_{x^{2}-y^{2}}$-wave symmetry ($\sim \cos{k_x}-\cos{k_y}$) and thus its amplitude would be vanishing along the $\Gamma$-M direction of the Brillouin zone where the nodes of the B$_1$ phase are placed. Hence, the intra-orbital component cannot affect at all the nodal structure of the B$_1$ phase. From this perspective, the B$_1$ phase is remarkably robust to the inclusion of spin-singlet intra-orbital pairing components.
In Appendix, the intra-orbital spin-singlet pairings other than B$_1$ representation ($d_{x^{2}-y^{2}}$-wave) are discussed. 

Concerning the experimental consequences of the topological superconducting phase, one can observe that, apart from the direct spectroscopic access to the temperature dependence of the edge states, the use of a superconductor-normal metal-superconductor (S-N-S) junction can also contribute to design of experiments to directly probe the peculiar behavior of the B$_1$ phase. In particular, by scanning its temperature dependent properties, since the B$_1$ state can undergo a series of Lifshitz transitions within the superconducting phase by gapping out part of the nodes, a dramatic modification of the Andreev spectrum at the S-N boundary is expected to occur. Hence, upon the application of a phase difference between the superconductors in the S-N-S junction, the Josephson current is expected to exhibit an anomalous temperature behavior. In particular, the abrupt changes in the Andreev bound states will drive a rapid variation in the Josephson current through the S-N-S junction when the superconductor undergoes transitions in the number of nodes. 

Finally, we point out that the examined model Hamiltonian is generally applicable to two-dimensional layered materials, in the low/intermediate doping regime, having $t_{2g}$ $d$-bands at the Fermi level and subjected to both atomic spin-orbit coupling and inversion symmetry breaking, for instance owing to lattice distortions and bond bending. 
Many candidate material cases can be encountered in the family of transition metal oxides. 
There, unconventional low-dimensional quantum liquids with low electron density can be obtained by engineering a 2DEG at polar/nonpolar interfaces between two band insulators, on the surface of band insulators (i.e., STO) or by designing single monolayer heterostructures, ultrathin films or superlattices. 
A paradigmatic case of superconducting 2DEG is provided by the LAO/STO heterostructure \cite{Reyren2007Science,Cen2008NatMater,Caviglia2008Nature,Schneider2009PRB}. 
Recent experimental observations by tunneling spectroscopy have pointed out that the superconducting state can be unconventional owing to the occurrence of in-gap states with peaks at zero and finite energies \cite{Kuerten2017PRB}. 
Although these peaks may be associated with a variety of 
concomitant physical mechanisms, e.g., 
surface Andreev bound states \cite{ABS,ABSb,Hu94,TK95,ABSR1,ABSR2}, 
the anomalous proximity effect by 
odd-frequency spin-triplet pairing 
\cite{tanaka12,Sau2015,Huang2015PRB,Burset2015PRB,Lee2017PRB,KashubaPRB2017,Cayao,Proximityp,Proximityp2,Proximityp3,Meissner3,odd1,odd2,odd3}, 
and bound states owing to the presence of magnetic impurities \cite{Ebisu}, 
their nature can provide key information about the 
pairing symmetry of the superconductor. 
Furthermore, the observation of Josephson currents \cite{Tafuri2017PRL} across a constriction 
in the 2DEG confirms a fundamental unconventional 
nature of the superconducting state 
\cite{Josephson1,Josephson2,Josephson3}. 
A common aspect emerging from the two different 
spectroscopic probes is that the superconducting state 
seems to have a multi-component character. 
Although it is not easy to disentangle 
the various contributions that may affect the superconducting phase in the 2DEG, 
we speculate that the proposed topological phase can 
be also included within the possible candidates 
for addressing the puzzling 
properties of the superconductivity of the oxide interface.

\begin{acknowledgments}

This work was supported by a JSPS KAKENHI
(Grants No. JP15H05853, No. JPH06136, and No. JP15H03686),
and the JSPS Core-to-Core program "Oxide Superspin", and the project Quantox of QuantERA ERA-NET Cofund in Quantum Technologies, implemented within the EU H2020 Programme.

\end{acknowledgments}

\section{Appendix}

In this section we address three different issues related to the presented results. Firstly, we investigate how a modification of the pairing interaction affects the phase diagram and the relative competition between the various configurations by scanning a larger range of temperatures at representative cases of filling concentration.
Then, we consider the classification of the irreducible representations of the superconducting phases in the presence of an intra-orbital attractive interaction. Moreover, we demonstrate that the intra-orbital and inter-orbital pairing interactions mediated by phonons have the same amplitude.

\begin{figure}[htbp]
\centering
\includegraphics[width=8cm]{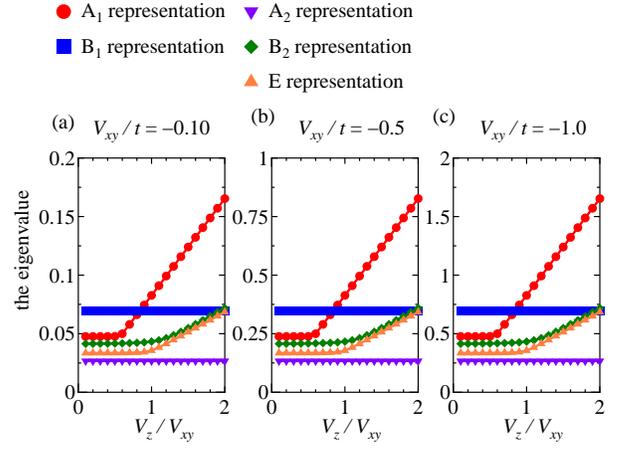}
\caption{The eigenvalues for various irreducible representations 
as a function of $V_{z}/V_{xy}$ 
at (a) $V_{xy}/t=-0.10$, (b) $V_{xy}/t=-0.50$ and (c)$V_{xy}/t=-1.0$, assuming that $T/t=1.0 \times 10^{-5}$, $\lambda_\mathrm{SO}/t=0.10$, $\Delta_{is}/t=0.20$, and $\mu/t=0.0$. }
\label{Appe_1}
\end{figure}

Starting from the impact of the pairing interaction on the phase diagram, in Fig. \ref{Appe_1} we show that at a given temperature the maximal eigenvalue in the various irreducible representations scales with the values of $V_z$ and $V_{xy}$ at $\lambda_\mathrm{SO}/t=0.10$, $\Delta_{is}/t=0.20$ and $\mu/t=0.0$. 
When we keep the ratio $V_{z}/V_{xy}$, the eigenvalue $\Lambda$ is  proportional to $V_{xy}$ within the mean field approximation. 
Hence, the phase diagram is basically determined by the ratio $V_z/V_{xy}$. In addition, since the transition temperature $T_\mathrm{c}$ is achieved when the magnitude of the greatest eigenvalue gets close to 1, then, according to this relation, one can identify the regime of temperatures which is close to the superconducting transition by suitably scaling the pairing interactions. In this way, the corresponding irreducible representation with the largest eigenvalue is the most stable according to the solution of the gap equation. 

\begin{figure}[htbp]
\centering
\includegraphics[width=6cm]{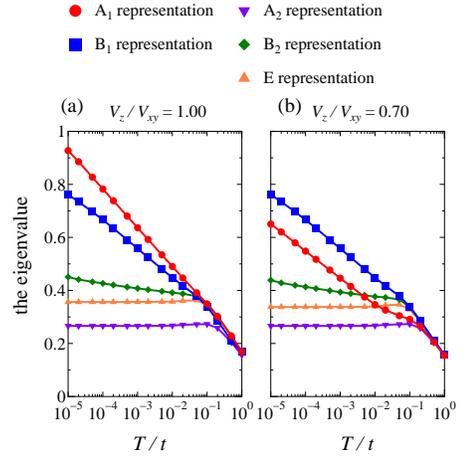}
\caption{The eigenvalues for various irreducible representations 
as a function of $T/t$ 
at (a)$V_{z}/V_{xy}=1.0$ and (b)$V_{z}/V_{xy}=0.70$,
assuming that $V_{xy}/t=-1.0$, $\lambda_\mathrm{SO}/t=0.10$, $\Delta_{is}/t=0.20$, and $\mu/t=0.0$. }
\label{Appe_2}
\end{figure}

In order to understand how a change in the critical temperature can affect the relative stability, in Fig. \ref{Appe_2} we report the eigenvalues for the various irreducible representations at $\lambda_\mathrm{SO}/t=0.10$, $\Delta_{is}/t=0.20$ and $\mu/t=0.0$ as a function of $T/t$ at two different ratio (a) $V_{z}/V_{xy} = 1.0$ and (b) $V_{z}/V_{xy} = 0.70$. We notice that the most stable configuration is not affected by a change in temperature or the strength of the pairing coupling. However, the eigenvalues of the B$_2$ and E representation become larger than that of A$_1$ above $T/t \sim 1.0 \times 10^{-2}$ (see Fig. \ref{Appe_2}(b)), thus affecting the competition between the A$_1$ and B$_1$ configurations. Otherwise, the analysis at different temperatures demonstrate that even for larger values of the pairing interaction the phase diagram is not much affected. 


Concerning the role of the intra-orbital spin-singlet pairing, in Ref. [\onlinecite{sigrist91}], we can classify the possible irreducible representations for the tetragonal group $\mathrm{C}_{4v}$ assuming both isotropic inter-orbital pairing and intra-orbital ones with isotropic and anisotropic structures compatible with the symmetry configuration as shown in Table \ref{irreducible2}.

\begin{table}[htbp]
\caption{Irreducible representation of isotropic inter-orbital superconducting states and intra-orbital spin-singlet ones with isotropic and anisotropic structures for the tetragonal group $\mathrm{C}_{4v}$. 
In the columns, we report the sign of the order parameter upon a four-fold rotational symmetry transformation, $C_{4}$, 
and the reflection mirror symmetry $M_{yz}$, as well as the explicit spin 
and orbital structure of the gap function. 
In the E representation, $+$ and $-$ of the subscript mean the doubly degenerate mirror-even ($+$) and mirror-odd ($-$) solutions, respectively. }
\label{irreducible2}
  \begin{center}
    \begin{tabular}{c|cc|c|c}
$\mathrm{C}_{4v}$ &$C_{4}$&$M_{yz}$&orbital & basis function \\ \hline \hline
\multirow{6}{*}{$\mathrm{A}_{1}$} & \multirow{6}{*}{$+$} & \multirow{6}{*}{$+$} & $(d_{yz}, d_{yz})$ & $\psi^{(yz, yz)}=const.$ \\
 & & & $(d_{zx}, d_{zx})$ & $\psi^{(zx, zx)}$=$\psi^{(yz, yz)}$ \\
 & & & $(d_{xy}, d_{xy})$ & $\psi^{(xy, xy)}=const.$ \\
 & & & $(d_{xy}, d_{yz})$ & $d^{(xy,yz)}_{y}$ \\
 & & & $(d_{xy}, d_{zx})$ & $d^{(xy,zx)}_{x}=-d^{(xy,yz)}_{y}$ \\
 & & & $(d_{yz}, d_{zx})$ & $d^{(yz,zx)}_{z}$ \\ \hline
\multirow{5}{*}{$\mathrm{A}_{2}$} & \multirow{5}{*}{$+$} & \multirow{5}{*}{$-$} & $(d_{yz}, d_{yz})$ & $\psi^{(yz,yz)}=\sin{k_x}\sin{k_y}(\cos{k_x}-\cos{k_y})$ \\
 & & & $(d_{zx}, d_{zx})$ & $\psi^{(zx,zx)}(k_{x},k_{y})=\psi^{(yz, yz)}(k_{y},-k_{x})$ \\
 & & & $(d_{xy}, d_{xy})$ & $\psi^{(xy,xy)}=\sin{k_x}\sin{k_y}(\cos{k_x}-\cos{k_y})$ \\
 & & & $(d_{xy}, d_{yz})$ & $d^{(xy,yz)}_{x}$ \\
 & & & $(d_{xy}, d_{zx})$  & $d^{(xy,zx)}_{y}=d^{(xy,yz)}_{x}$ \\ \hline
\multirow{5}{*}{$\mathrm{B}_{1}$} & \multirow{5}{*}{$-$} & \multirow{5}{*}{$+$} & $(d_{yz}, d_{yz})$ & $\psi^{(yz, yz)}\propto \cos{k_x}-\cos{k_y}$ \\
 & & & $(d_{zx}, d_{zx})$ & $\psi^{(zx, zx)}(k_{x},k_{y})=-\psi^{(yz, yz)}(k_{y},-k_{x})$ \\
 & & & $(d_{xy}, d_{xy})$ & $\psi^{(xy, xy)}\propto \cos{k_x}-\cos{k_y}$ \\
 & & & $(d_{xy}, d_{yz})$ & $d^{(xy,yz)}_{y}$ \\
 & & & $(d_{xy}, d_{zx})$ & $d^{(xy,zx)}_{x}=d^{(xy,yz)}_{y}$ \\ \hline
\multirow{6}{*}{$\mathrm{B}_{2}$} & \multirow{6}{*}{$-$} & \multirow{6}{*}{$-$} & $(d_{yz}, d_{yz})$ & $\psi^{(yz, yz)}\propto \sin{k_x}\sin{k_y}$ \\
 & & & $(d_{zx}, d_{zx})$ & $\psi^{(zx, zx)}(k_{x},k_{y})$=$-\psi^{(yz, yz)}(k_{y},-k_{x})$ \\
 & & & $(d_{xy}, d_{xy})$ & $\psi^{(xy, xy)}\propto \sin{k_x}\sin{k_y}$ \\
 & & & $(d_{xy}, d_{yz})$ & $d^{(xy,yz)}_{x}$ \\
 & & & $(d_{xy}, d_{zx})$ & $d^{(xy,zx)}_{y}=-d^{(xy,yz)}_{x}$ \\
 & & & $(d_{yz}, d_{zx})$ & $\psi^{(yz, zx)}$ \\ \hline
\multirow{4}{*}{E} & \multirow{4}{*}{$\pm i$} & \multirow{4}{*}{$\pm$} & $(d_{xy}, d_{yz})$ & $\psi^{(xy, yz)}$, $d^{(xy,yz)}_{z}$ \\
 & & & \multirow{2}{*}{$(d_{xy}, d_{zx})$} & $\psi^{(xy, zx)}_{+}=\mp id^{(xy,yz)}_{z+}$ \\
 & & & & $d^{(xy,zx)}_{z-}=\mp i\psi^{(xy,yz)}_{-}$ \\
 & & & $(d_{yz}, d_{zx})$ & $d^{(yz,zx)}_{x}$, $d^{(yz,zx)}_{y}$ \\ \hline
    \end{tabular}
  \end{center}
\end{table}%
%%%%%%%%%%%%%%%%%%%%%%%%%%%%%

Finally, we consider the relative strength of the attractive interaction in the inter- and intra-orbital channel as due to electron-phonon coupling in a $t_{2g}$ multi-orbital system.

Consider the electron phonon coupling in $t_{2g}$ system,
\begin{eqnarray}
\mathcal{H}_{ep}&=&\frac{1}{\sqrt N}\sum_{{\bm k},{\bm q},m,l,l'\sigma}\alpha^m_{ll'}({\bm q})c^\dag_{{\bm k}+{\bm q},l,\sigma}c_{{\bm k},l',\sigma},
\end{eqnarray}%
where $m$ denotes the phonon mode, $\alpha^m_{ll'}({\bm q})$ is the electron-phonon coupling constant, and $l$ and $l'$ stands for orbital indices in the basis of $yz$, $zx$, and $xy$.
Here, we consider only the diagonal elements, which are relevant to the attractive interaction.
Off-diagonal ones are relevant to the pair hopping, which enhance the transition temperature.
\begin{eqnarray}
\mathcal{H}_{ep}&=&\frac{1}{\sqrt N}\sum_{{\bm k},{\bm q},m,l,\sigma}\alpha^m_{ll}({\bm q})c^\dag_{{\bm k}+{\bm q},l,\sigma}c_{{\bm k},l,\sigma}.
\end{eqnarray}
The effective interaction in the Eliashberg equation due to this electron-phonon coupling is given by
\begin{eqnarray}
V^m_{ll'}({\bm q},\omega_n)&=&-\alpha^m_{ll}({\bm q})\alpha^m_{l'l'}({\bm q})D^m({\bm q},\omega_n),
\end{eqnarray}
where $D^m({\bm q},\omega_n)$ is the Green's function of phonon
\begin{eqnarray}
D^m({\bm q},\omega_n)=\frac{2\omega_m({\bm q})}{\omega_m^2({\bm q})+\omega^2_n},
\end{eqnarray}
with phonon's frequency $\omega_m({\bm q})$ and bosonic Matsubara frequency $\omega_{n}=2n\pi k_\mathrm{B}T$. Here, we suppose the A$_1$ modes are the most relevant to the interaction.
In A$_1$ modes, 
\begin{eqnarray}
\alpha^m_{yz,yz}({\bm q})=\alpha^m_{zx,zx}({\bm q})
\end{eqnarray}
in the tetragonal symmetry. Then, we have the relation
\begin{eqnarray*}
V^m_{yz,yz}({\bm q},\omega_n)=V^m_{zx,zx}({\bm q},\omega_n)=V^m_{yz,zx}({\bm q},\omega_n)=V^m_{zx,yz}({\bm q},\omega_n).
\end{eqnarray*}
This means that the effective inter and intraorbital attractive interaction in these two orbitals are the same.
We can also include $xy$-orbitals into this relation when $\Delta_t$ and $\Delta_{is}$ is small and the symmetry goes toward the cubic.

\bibliographystyle{apsrev4-1}
\bibliography{BQ13557_v4}

\begin{thebibliography}{133}%
\makeatletter
\providecommand \@ifxundefined [1]{%
 \@ifx{#1\undefined}
}%
\providecommand \@ifnum [1]{%
 \ifnum #1\expandafter \@firstoftwo
 \else \expandafter \@secondoftwo
 \fi
}%
\providecommand \@ifx [1]{%
 \ifx #1\expandafter \@firstoftwo
 \else \expandafter \@secondoftwo
 \fi
}%
\providecommand \natexlab [1]{#1}%
\providecommand \enquote  [1]{``#1''}%
\providecommand \bibnamefont  [1]{#1}%
\providecommand \bibfnamefont [1]{#1}%
\providecommand \citenamefont [1]{#1}%
\providecommand \href@noop [0]{\@secondoftwo}%
\providecommand \href [0]{\begingroup \@sanitize@url \@href}%
\providecommand \@href[1]{\@@startlink{#1}\@@href}%
\providecommand \@@href[1]{\endgroup#1\@@endlink}%
\providecommand \@sanitize@url [0]{\catcode `\\12\catcode `\$12\catcode
  `\&12\catcode `\#12\catcode `\^12\catcode `\_12\catcode `\%12\relax}%
\providecommand \@@startlink[1]{}%
\providecommand \@@endlink[0]{}%
\providecommand \url  [0]{\begingroup\@sanitize@url \@url }%
\providecommand \@url [1]{\endgroup\@href {#1}{\urlprefix }}%
\providecommand \urlprefix  [0]{URL }%
\providecommand \Eprint [0]{\href }%
\providecommand \doibase [0]{http://dx.doi.org/}%
\providecommand \selectlanguage [0]{\@gobble}%
\providecommand \bibinfo  [0]{\@secondoftwo}%
\providecommand \bibfield  [0]{\@secondoftwo}%
\providecommand \translation [1]{[#1]}%
\providecommand \BibitemOpen [0]{}%
\providecommand \bibitemStop [0]{}%
\providecommand \bibitemNoStop [0]{.\EOS\space}%
\providecommand \EOS [0]{\spacefactor3000\relax}%
\providecommand \BibitemShut  [1]{\csname bibitem#1\endcsname}%
\let\auto@bib@innerbib\@empty
%</preamble>
\bibitem [{\citenamefont {Sigrist}\ and\ \citenamefont
  {Ueda}(1991)}]{sigrist91}%
  \BibitemOpen
  \bibfield  {author} {\bibinfo {author} {\bibfnamefont {M.}~\bibnamefont
  {Sigrist}}\ and\ \bibinfo {author} {\bibfnamefont {K.}~\bibnamefont {Ueda}},\
  }\href {\doibase 10.1103/RevModPhys.63.239} {\bibfield  {journal} {\bibinfo
  {journal} {Rev. Mod. Phys.}\ }\textbf {\bibinfo {volume} {63}},\ \bibinfo
  {pages} {239} (\bibinfo {year} {1991})}\BibitemShut {NoStop}%
\bibitem [{\citenamefont {Maeno}\ \emph {et~al.}(1994)\citenamefont {Maeno},
  \citenamefont {Hashimoto}, \citenamefont {Yoshida}, \citenamefont
  {Nishizaki}, \citenamefont {Fujita}, \citenamefont {Bednorz},\ and\
  \citenamefont {Lichtenberg}}]{Maeno}%
  \BibitemOpen
  \bibfield  {author} {\bibinfo {author} {\bibfnamefont {Y.}~\bibnamefont
  {Maeno}}, \bibinfo {author} {\bibfnamefont {H.}~\bibnamefont {Hashimoto}},
  \bibinfo {author} {\bibfnamefont {K.}~\bibnamefont {Yoshida}}, \bibinfo
  {author} {\bibfnamefont {S.}~\bibnamefont {Nishizaki}}, \bibinfo {author}
  {\bibfnamefont {T.}~\bibnamefont {Fujita}}, \bibinfo {author} {\bibfnamefont
  {J.~G.}\ \bibnamefont {Bednorz}}, \ and\ \bibinfo {author} {\bibfnamefont
  {F.}~\bibnamefont {Lichtenberg}},\ }\href@noop {} {\bibfield  {journal}
  {\bibinfo  {journal} {Nature}\ }\textbf {\bibinfo {volume} {372}},\ \bibinfo
  {pages} {532} (\bibinfo {year} {1994})}\BibitemShut {NoStop}%
\bibitem [{\citenamefont {Tou}\ \emph {et~al.}(1998)\citenamefont {Tou},
  \citenamefont {Kitaoka}, \citenamefont {Ishida}, \citenamefont {Asayama},
  \citenamefont {Kimura}, \citenamefont {Onuki}, \citenamefont {Yamamoto},
  \citenamefont {Haga},\ and\ \citenamefont {Maezawa}}]{Maeno3}%
  \BibitemOpen
  \bibfield  {author} {\bibinfo {author} {\bibfnamefont {H.}~\bibnamefont
  {Tou}}, \bibinfo {author} {\bibfnamefont {Y.}~\bibnamefont {Kitaoka}},
  \bibinfo {author} {\bibfnamefont {K.}~\bibnamefont {Ishida}}, \bibinfo
  {author} {\bibfnamefont {K.}~\bibnamefont {Asayama}}, \bibinfo {author}
  {\bibfnamefont {N.}~\bibnamefont {Kimura}}, \bibinfo {author} {\bibfnamefont
  {Y.}~\bibnamefont {Onuki}}, \bibinfo {author} {\bibfnamefont
  {E.}~\bibnamefont {Yamamoto}}, \bibinfo {author} {\bibfnamefont
  {Y.}~\bibnamefont {Haga}}, \ and\ \bibinfo {author} {\bibfnamefont
  {K.}~\bibnamefont {Maezawa}},\ }\href@noop {} {\bibfield  {journal} {\bibinfo
   {journal} {Phys. Rev. Lett.}\ }\textbf {\bibinfo {volume} {80}},\ \bibinfo
  {pages} {3129} (\bibinfo {year} {1998})}\BibitemShut {NoStop}%
\bibitem [{\citenamefont {Kashiwaya}\ \emph {et~al.}(2011)\citenamefont
  {Kashiwaya}, \citenamefont {Kashiwaya}, \citenamefont {Kambara},
  \citenamefont {Furuta}, \citenamefont {Yaguchi}, \citenamefont {Tanaka},\
  and\ \citenamefont {Maeno}}]{Kashiwaya11}%
  \BibitemOpen
  \bibfield  {author} {\bibinfo {author} {\bibfnamefont {S.}~\bibnamefont
  {Kashiwaya}}, \bibinfo {author} {\bibfnamefont {H.}~\bibnamefont
  {Kashiwaya}}, \bibinfo {author} {\bibfnamefont {H.}~\bibnamefont {Kambara}},
  \bibinfo {author} {\bibfnamefont {T.}~\bibnamefont {Furuta}}, \bibinfo
  {author} {\bibfnamefont {H.}~\bibnamefont {Yaguchi}}, \bibinfo {author}
  {\bibfnamefont {Y.}~\bibnamefont {Tanaka}}, \ and\ \bibinfo {author}
  {\bibfnamefont {Y.}~\bibnamefont {Maeno}},\ }\href@noop {} {\bibfield
  {journal} {\bibinfo  {journal} {Phys. Rev. Lett.}\ }\textbf {\bibinfo
  {volume} {107}},\ \bibinfo {pages} {077003} (\bibinfo {year}
  {2011})}\BibitemShut {NoStop}%
\bibitem [{\citenamefont {Buchholtz}\ and\ \citenamefont
  {Zwicknagl}(1981)}]{ABS}%
  \BibitemOpen
  \bibfield  {author} {\bibinfo {author} {\bibfnamefont {L.~J.}\ \bibnamefont
  {Buchholtz}}\ and\ \bibinfo {author} {\bibfnamefont {G.}~\bibnamefont
  {Zwicknagl}},\ }\href@noop {} {\bibfield  {journal} {\bibinfo  {journal}
  {Phys. Rev. B}\ }\textbf {\bibinfo {volume} {23}},\ \bibinfo {pages} {5788}
  (\bibinfo {year} {1981})}\BibitemShut {NoStop}%
\bibitem [{\citenamefont {Hara}\ and\ \citenamefont {Nagai}(1986)}]{ABSb}%
  \BibitemOpen
  \bibfield  {author} {\bibinfo {author} {\bibfnamefont {J.}~\bibnamefont
  {Hara}}\ and\ \bibinfo {author} {\bibfnamefont {K.}~\bibnamefont {Nagai}},\
  }\href@noop {} {\bibfield  {journal} {\bibinfo  {journal} {Prog. Theor.
  Phys.}\ }\textbf {\bibinfo {volume} {76}},\ \bibinfo {pages} {1237} (\bibinfo
  {year} {1986})}\BibitemShut {NoStop}%
\bibitem [{\citenamefont {Hu}(1994)}]{Hu94}%
  \BibitemOpen
  \bibfield  {author} {\bibinfo {author} {\bibfnamefont {C.~R.}\ \bibnamefont
  {Hu}},\ }\href@noop {} {\bibfield  {journal} {\bibinfo  {journal} {Phys. Rev.
  Lett.}\ }\textbf {\bibinfo {volume} {72}},\ \bibinfo {pages} {1526} (\bibinfo
  {year} {1994})}\BibitemShut {NoStop}%
\bibitem [{\citenamefont {Tanaka}\ and\ \citenamefont
  {Kashiwaya}(1995)}]{TK95}%
  \BibitemOpen
  \bibfield  {author} {\bibinfo {author} {\bibfnamefont {Y.}~\bibnamefont
  {Tanaka}}\ and\ \bibinfo {author} {\bibfnamefont {S.}~\bibnamefont
  {Kashiwaya}},\ }\href@noop {} {\bibfield  {journal} {\bibinfo  {journal}
  {Phys. Rev. Lett.}\ }\textbf {\bibinfo {volume} {74}},\ \bibinfo {pages}
  {3451} (\bibinfo {year} {1995})}\BibitemShut {NoStop}%
\bibitem [{\citenamefont {Kashiwaya}\ and\ \citenamefont
  {Tanaka}(2000)}]{ABSR1}%
  \BibitemOpen
  \bibfield  {author} {\bibinfo {author} {\bibfnamefont {S.}~\bibnamefont
  {Kashiwaya}}\ and\ \bibinfo {author} {\bibfnamefont {Y.}~\bibnamefont
  {Tanaka}},\ }\href@noop {} {\bibfield  {journal} {\bibinfo  {journal} {Rep.
  Prog. Phys.}\ }\textbf {\bibinfo {volume} {63}},\ \bibinfo {pages} {1641}
  (\bibinfo {year} {2000})}\BibitemShut {NoStop}%
\bibitem [{\citenamefont {L{\"o}fwander}\ \emph {et~al.}(2001)\citenamefont
  {L{\"o}fwander}, \citenamefont {Shumeiko},\ and\ \citenamefont
  {Wendin}}]{ABSR2}%
  \BibitemOpen
  \bibfield  {author} {\bibinfo {author} {\bibfnamefont {T.}~\bibnamefont
  {L{\"o}fwander}}, \bibinfo {author} {\bibfnamefont {V.~S.}\ \bibnamefont
  {Shumeiko}}, \ and\ \bibinfo {author} {\bibfnamefont {G.}~\bibnamefont
  {Wendin}},\ }\href@noop {} {\bibfield  {journal} {\bibinfo  {journal}
  {Supercond. Sci. Technol.}\ }\textbf {\bibinfo {volume} {14}},\ \bibinfo
  {pages} {R53} (\bibinfo {year} {2001})}\BibitemShut {NoStop}%
\bibitem [{\citenamefont {Kwon}\ \emph {et~al.}(2004)\citenamefont {Kwon},
  \citenamefont {Sengupta},\ and\ \citenamefont {Yakovenko}}]{Yakovenko}%
  \BibitemOpen
  \bibfield  {author} {\bibinfo {author} {\bibfnamefont {H.}~\bibnamefont
  {Kwon}}, \bibinfo {author} {\bibfnamefont {K.}~\bibnamefont {Sengupta}}, \
  and\ \bibinfo {author} {\bibfnamefont {V.}~\bibnamefont {Yakovenko}},\
  }\href@noop {} {\bibfield  {journal} {\bibinfo  {journal} {Eur. Phys. J. B}\
  }\textbf {\bibinfo {volume} {37}},\ \bibinfo {pages} {349} (\bibinfo {year}
  {2004})}\BibitemShut {NoStop}%
\bibitem [{\citenamefont {Schnyder}\ \emph {et~al.}(2008)\citenamefont
  {Schnyder}, \citenamefont {Ryu}, \citenamefont {Furusaki},\ and\
  \citenamefont {Ludwig}}]{SRFL08}%
  \BibitemOpen
  \bibfield  {author} {\bibinfo {author} {\bibfnamefont {A.~P.}\ \bibnamefont
  {Schnyder}}, \bibinfo {author} {\bibfnamefont {S.}~\bibnamefont {Ryu}},
  \bibinfo {author} {\bibfnamefont {A.}~\bibnamefont {Furusaki}}, \ and\
  \bibinfo {author} {\bibfnamefont {A.~W.~W.}\ \bibnamefont {Ludwig}},\
  }\href@noop {} {\bibfield  {journal} {\bibinfo  {journal} {Phys. Rev. B}\
  }\textbf {\bibinfo {volume} {78}},\ \bibinfo {pages} {195125} (\bibinfo
  {year} {2008})}\BibitemShut {NoStop}%
\bibitem [{\citenamefont {Ryu}\ \emph {et~al.}(2010)\citenamefont {Ryu},
  \citenamefont {Schnyder}, \citenamefont {Furusaki},\ and\ \citenamefont
  {Ludwig}}]{RSFL10}%
  \BibitemOpen
  \bibfield  {author} {\bibinfo {author} {\bibfnamefont {S.}~\bibnamefont
  {Ryu}}, \bibinfo {author} {\bibfnamefont {A.~P.}\ \bibnamefont {Schnyder}},
  \bibinfo {author} {\bibfnamefont {A.}~\bibnamefont {Furusaki}}, \ and\
  \bibinfo {author} {\bibfnamefont {A.}~\bibnamefont {Ludwig}},\ }\href@noop {}
  {\bibfield  {journal} {\bibinfo  {journal} {New J. Phys.}\ }\textbf {\bibinfo
  {volume} {12}},\ \bibinfo {pages} {065010} (\bibinfo {year}
  {2010})}\BibitemShut {NoStop}%
\bibitem [{\citenamefont {Qi}\ and\ \citenamefont {Zhang}(2011)}]{qi11}%
  \BibitemOpen
  \bibfield  {author} {\bibinfo {author} {\bibfnamefont {X.-L.}\ \bibnamefont
  {Qi}}\ and\ \bibinfo {author} {\bibfnamefont {S.-C.}\ \bibnamefont {Zhang}},\
  }\href@noop {} {\bibfield  {journal} {\bibinfo  {journal} {Rev. Mod. Phys.}\
  }\textbf {\bibinfo {volume} {83}},\ \bibinfo {pages} {1057} (\bibinfo {year}
  {2011})}\BibitemShut {NoStop}%
\bibitem [{\citenamefont {Tanaka}\ \emph {et~al.}(2012)\citenamefont {Tanaka},
  \citenamefont {Sato},\ and\ \citenamefont {Nagaosa}}]{tanaka12}%
  \BibitemOpen
  \bibfield  {author} {\bibinfo {author} {\bibfnamefont {Y.}~\bibnamefont
  {Tanaka}}, \bibinfo {author} {\bibfnamefont {M.}~\bibnamefont {Sato}}, \ and\
  \bibinfo {author} {\bibfnamefont {N.}~\bibnamefont {Nagaosa}},\ }\href@noop
  {} {\bibfield  {journal} {\bibinfo  {journal} {J. Phys. Soc. Jpn.}\ }\textbf
  {\bibinfo {volume} {81}},\ \bibinfo {pages} {011013} (\bibinfo {year}
  {2012})}\BibitemShut {NoStop}%
\bibitem [{\citenamefont {Leijnse}\ and\ \citenamefont
  {Flensberg}(2012)}]{Flensberg2012}%
  \BibitemOpen
  \bibfield  {author} {\bibinfo {author} {\bibfnamefont {M.}~\bibnamefont
  {Leijnse}}\ and\ \bibinfo {author} {\bibfnamefont {K.}~\bibnamefont
  {Flensberg}},\ }\href@noop {} {\bibfield  {journal} {\bibinfo  {journal}
  {Semiconductor Science and Technology}\ }\textbf {\bibinfo {volume} {27}},\
  \bibinfo {pages} {124003} (\bibinfo {year} {2012})}\BibitemShut {NoStop}%
\bibitem [{\citenamefont {Beenakker}(2013)}]{Beenakker13}%
  \BibitemOpen
  \bibfield  {author} {\bibinfo {author} {\bibfnamefont {C.~W.~J.}\
  \bibnamefont {Beenakker}},\ }\href@noop {} {\bibfield  {journal} {\bibinfo
  {journal} {Annu. Rev. Condens. Matter Phys.}\ }\textbf {\bibinfo {volume}
  {4}},\ \bibinfo {pages} {113} (\bibinfo {year} {2013})}\BibitemShut {NoStop}%
\bibitem [{\citenamefont {Sato}\ and\ \citenamefont
  {Fujimoto}(2016)}]{SatoFujimoto2016}%
  \BibitemOpen
  \bibfield  {author} {\bibinfo {author} {\bibfnamefont {M.}~\bibnamefont
  {Sato}}\ and\ \bibinfo {author} {\bibfnamefont {S.}~\bibnamefont
  {Fujimoto}},\ }\href@noop {} {\bibfield  {journal} {\bibinfo  {journal} {J.
  Phys. Soc. Jpn.}\ }\textbf {\bibinfo {volume} {85}},\ \bibinfo {pages}
  {072001} (\bibinfo {year} {2016})}\BibitemShut {NoStop}%
\bibitem [{\citenamefont {Sato}\ and\ \citenamefont
  {Ando}(2017)}]{SatoAndo2017}%
  \BibitemOpen
  \bibfield  {author} {\bibinfo {author} {\bibfnamefont {M.}~\bibnamefont
  {Sato}}\ and\ \bibinfo {author} {\bibfnamefont {Y.}~\bibnamefont {Ando}},\
  }\href@noop {} {\bibfield  {journal} {\bibinfo  {journal} {Rep. Prog. Phys.}\
  }\textbf {\bibinfo {volume} {80}},\ \bibinfo {pages} {076501} (\bibinfo
  {year} {2017})}\BibitemShut {NoStop}%
\bibitem [{\citenamefont {Kitaev}(2001)}]{Kitaev01}%
  \BibitemOpen
  \bibfield  {author} {\bibinfo {author} {\bibfnamefont {A.~Y.}\ \bibnamefont
  {Kitaev}},\ }\href@noop {} {\bibfield  {journal} {\bibinfo  {journal} {Usp.
  Fiz. Nauk (Suppl.)}\ }\textbf {\bibinfo {volume} {171}},\ \bibinfo {pages}
  {131} (\bibinfo {year} {2001})}\BibitemShut {NoStop}%
\bibitem [{\citenamefont {Nayak}\ \emph {et~al.}(2008)\citenamefont {Nayak},
  \citenamefont {Simon}, \citenamefont {Stern}, \citenamefont {Freedman},\ and\
  \citenamefont {Das~Sarma}}]{Nayak}%
  \BibitemOpen
  \bibfield  {author} {\bibinfo {author} {\bibfnamefont {C.}~\bibnamefont
  {Nayak}}, \bibinfo {author} {\bibfnamefont {S.~H.}\ \bibnamefont {Simon}},
  \bibinfo {author} {\bibfnamefont {A.}~\bibnamefont {Stern}}, \bibinfo
  {author} {\bibfnamefont {M.}~\bibnamefont {Freedman}}, \ and\ \bibinfo
  {author} {\bibfnamefont {S.}~\bibnamefont {Das~Sarma}},\ }\href@noop {}
  {\bibfield  {journal} {\bibinfo  {journal} {Rev. Mod. Phys.}\ }\textbf
  {\bibinfo {volume} {80}},\ \bibinfo {pages} {1083} (\bibinfo {year}
  {2008})}\BibitemShut {NoStop}%
\bibitem [{\citenamefont {Wilczek}(2009)}]{wilczek09}%
  \BibitemOpen
  \bibfield  {author} {\bibinfo {author} {\bibfnamefont {F.}~\bibnamefont
  {Wilczek}},\ }\href@noop {} {\bibfield  {journal} {\bibinfo  {journal} {Nat.
  Phys.}\ }\textbf {\bibinfo {volume} {5}},\ \bibinfo {pages} {614} (\bibinfo
  {year} {2009})}\BibitemShut {NoStop}%
\bibitem [{\citenamefont {Alicea}(2012)}]{alicea12}%
  \BibitemOpen
  \bibfield  {author} {\bibinfo {author} {\bibfnamefont {J.}~\bibnamefont
  {Alicea}},\ }\href@noop {} {\bibfield  {journal} {\bibinfo  {journal} {Rep.
  Prog. Phys.}\ }\textbf {\bibinfo {volume} {75}},\ \bibinfo {pages} {076501}
  (\bibinfo {year} {2012})}\BibitemShut {NoStop}%
\bibitem [{\citenamefont {Aguado}(2017)}]{Ramon2017LRdNC}%
  \BibitemOpen
  \bibfield  {author} {\bibinfo {author} {\bibfnamefont {R.}~\bibnamefont
  {Aguado}},\ }\href {\doibase 10.1393/ncr/i2017-10141-9} {\bibfield  {journal}
  {\bibinfo  {journal} {La Rivista del Nuovo Cimento}\ }\textbf {\bibinfo
  {volume} {40}},\ \bibinfo {pages} {523} (\bibinfo {year} {2017})}\BibitemShut
  {NoStop}%
\bibitem [{\citenamefont {Bergeret}\ \emph {et~al.}(2005)\citenamefont
  {Bergeret}, \citenamefont {Volkov},\ and\ \citenamefont
  {Efetov}}]{Bergeret2005RMP}%
  \BibitemOpen
  \bibfield  {author} {\bibinfo {author} {\bibfnamefont {F.~S.}\ \bibnamefont
  {Bergeret}}, \bibinfo {author} {\bibfnamefont {A.~F.}\ \bibnamefont
  {Volkov}}, \ and\ \bibinfo {author} {\bibfnamefont {K.~B.}\ \bibnamefont
  {Efetov}},\ }\href {\doibase 10.1103/RevModPhys.77.1321} {\bibfield
  {journal} {\bibinfo  {journal} {Rev. Mod. Phys.}\ }\textbf {\bibinfo {volume}
  {77}},\ \bibinfo {pages} {1321} (\bibinfo {year} {2005})}\BibitemShut
  {NoStop}%
\bibitem [{\citenamefont {Buzdin}(2005)}]{Buzdinrev}%
  \BibitemOpen
  \bibfield  {author} {\bibinfo {author} {\bibfnamefont {A.~I.}\ \bibnamefont
  {Buzdin}},\ }\href@noop {} {\bibfield  {journal} {\bibinfo  {journal} {Rev.
  Mod. Phys.}\ }\textbf {\bibinfo {volume} {77}},\ \bibinfo {pages} {935}
  (\bibinfo {year} {2005})}\BibitemShut {NoStop}%
\bibitem [{\citenamefont {Khaire}\ \emph {et~al.}(2010)\citenamefont {Khaire},
  \citenamefont {Khasawneh}, \citenamefont {Pratt},\ and\ \citenamefont
  {Birge}}]{Birge}%
  \BibitemOpen
  \bibfield  {author} {\bibinfo {author} {\bibfnamefont {T.~S.}\ \bibnamefont
  {Khaire}}, \bibinfo {author} {\bibfnamefont {M.~A.}\ \bibnamefont
  {Khasawneh}}, \bibinfo {author} {\bibfnamefont {W.~P.}\ \bibnamefont
  {Pratt}}, \ and\ \bibinfo {author} {\bibfnamefont {N.~O.}\ \bibnamefont
  {Birge}},\ }\href@noop {} {\bibfield  {journal} {\bibinfo  {journal} {Phys.
  Rev. Lett.}\ }\textbf {\bibinfo {volume} {104}},\ \bibinfo {pages} {137002}
  (\bibinfo {year} {2010})}\BibitemShut {NoStop}%
\bibitem [{\citenamefont {Robinson}\ \emph {et~al.}(2010)\citenamefont
  {Robinson}, \citenamefont {Witt},\ and\ \citenamefont {Blamire}}]{Robinson}%
  \BibitemOpen
  \bibfield  {author} {\bibinfo {author} {\bibfnamefont {W.~A.}\ \bibnamefont
  {Robinson}}, \bibinfo {author} {\bibfnamefont {J.~D.~S.}\ \bibnamefont
  {Witt}}, \ and\ \bibinfo {author} {\bibfnamefont {M.~G.}\ \bibnamefont
  {Blamire}},\ }\href@noop {} {\bibfield  {journal} {\bibinfo  {journal}
  {Science}\ }\textbf {\bibinfo {volume} {329}},\ \bibinfo {pages} {59}
  (\bibinfo {year} {2010})}\BibitemShut {NoStop}%
\bibitem [{\citenamefont {Eschrig}\ \emph {et~al.}(2003)\citenamefont
  {Eschrig}, \citenamefont {Kopu}, \citenamefont {Cuevas},\ and\ \citenamefont
  {Sch{\"o}n}}]{Eschrig2003}%
  \BibitemOpen
  \bibfield  {author} {\bibinfo {author} {\bibfnamefont {M.}~\bibnamefont
  {Eschrig}}, \bibinfo {author} {\bibfnamefont {J.}~\bibnamefont {Kopu}},
  \bibinfo {author} {\bibfnamefont {J.~C.}\ \bibnamefont {Cuevas}}, \ and\
  \bibinfo {author} {\bibfnamefont {G.}~\bibnamefont {Sch{\"o}n}},\ }\href@noop
  {} {\bibfield  {journal} {\bibinfo  {journal} {Phys. Rev. Lett.}\ }\textbf
  {\bibinfo {volume} {90}},\ \bibinfo {pages} {137003} (\bibinfo {year}
  {2003})}\BibitemShut {NoStop}%
\bibitem [{\citenamefont {Eschrig}\ and\ \citenamefont
  {L{\"o}fwander}(2008)}]{Eschrig2008}%
  \BibitemOpen
  \bibfield  {author} {\bibinfo {author} {\bibfnamefont {M.}~\bibnamefont
  {Eschrig}}\ and\ \bibinfo {author} {\bibfnamefont {T.}~\bibnamefont
  {L{\"o}fwander}},\ }\href@noop {} {\bibfield  {journal} {\bibinfo  {journal}
  {Nature Physics}\ }\textbf {\bibinfo {volume} {4}},\ \bibinfo {pages} {138}
  (\bibinfo {year} {2008})}\BibitemShut {NoStop}%
\bibitem [{\citenamefont {Linder}\ and\ \citenamefont
  {Robinson}(2015)}]{Linder2015}%
  \BibitemOpen
  \bibfield  {author} {\bibinfo {author} {\bibfnamefont {J.}~\bibnamefont
  {Linder}}\ and\ \bibinfo {author} {\bibfnamefont {J.~W.~A.}\ \bibnamefont
  {Robinson}},\ }\href@noop {} {\bibfield  {journal} {\bibinfo  {journal} {Nat.
  Phys.}\ }\textbf {\bibinfo {volume} {11}},\ \bibinfo {pages} {307} (\bibinfo
  {year} {2015})}\BibitemShut {NoStop}%
\bibitem [{\citenamefont {Gentile}\ \emph {et~al.}(2013)\citenamefont
  {Gentile}, \citenamefont {Cuoco}, \citenamefont {Romano}, \citenamefont
  {Noce}, \citenamefont {Manske},\ and\ \citenamefont {Brydon}}]{Gentile13}%
  \BibitemOpen
  \bibfield  {author} {\bibinfo {author} {\bibfnamefont {P.}~\bibnamefont
  {Gentile}}, \bibinfo {author} {\bibfnamefont {M.}~\bibnamefont {Cuoco}},
  \bibinfo {author} {\bibfnamefont {A.}~\bibnamefont {Romano}}, \bibinfo
  {author} {\bibfnamefont {C.}~\bibnamefont {Noce}}, \bibinfo {author}
  {\bibfnamefont {D.}~\bibnamefont {Manske}}, \ and\ \bibinfo {author}
  {\bibfnamefont {P.~M.~R.}\ \bibnamefont {Brydon}},\ }\href@noop {} {\bibfield
   {journal} {\bibinfo  {journal} {Phys. Rev. Lett.}\ }\textbf {\bibinfo
  {volume} {111}},\ \bibinfo {pages} {097003} (\bibinfo {year}
  {2013})}\BibitemShut {NoStop}%
\bibitem [{\citenamefont {Terrade}\ \emph {et~al.}(2016)\citenamefont
  {Terrade}, \citenamefont {Manske},\ and\ \citenamefont {Cuoco}}]{Terrade16}%
  \BibitemOpen
  \bibfield  {author} {\bibinfo {author} {\bibfnamefont {D.}~\bibnamefont
  {Terrade}}, \bibinfo {author} {\bibfnamefont {D.}~\bibnamefont {Manske}}, \
  and\ \bibinfo {author} {\bibfnamefont {M.}~\bibnamefont {Cuoco}},\
  }\href@noop {} {\bibfield  {journal} {\bibinfo  {journal} {Phys. Rev. B}\
  }\textbf {\bibinfo {volume} {93}},\ \bibinfo {pages} {104523} (\bibinfo
  {year} {2016})}\BibitemShut {NoStop}%
\bibitem [{\citenamefont {Mercaldo}\ \emph {et~al.}(2016)\citenamefont
  {Mercaldo}, \citenamefont {Cuoco},\ and\ \citenamefont
  {Kotetes}}]{Mercaldo16}%
  \BibitemOpen
  \bibfield  {author} {\bibinfo {author} {\bibfnamefont {M.~T.}\ \bibnamefont
  {Mercaldo}}, \bibinfo {author} {\bibfnamefont {M.}~\bibnamefont {Cuoco}}, \
  and\ \bibinfo {author} {\bibfnamefont {P.}~\bibnamefont {Kotetes}},\
  }\href@noop {} {\bibfield  {journal} {\bibinfo  {journal} {Phys. Rev. B}\
  }\textbf {\bibinfo {volume} {94}},\ \bibinfo {pages} {140503(R)} (\bibinfo
  {year} {2016})}\BibitemShut {NoStop}%
\bibitem [{\citenamefont {Romano}\ \emph {et~al.}(2013)\citenamefont {Romano},
  \citenamefont {Gentile}, \citenamefont {Noce}, \citenamefont {Vekhter},\ and\
  \citenamefont {Cuoco}}]{Romano13}%
  \BibitemOpen
  \bibfield  {author} {\bibinfo {author} {\bibfnamefont {A.}~\bibnamefont
  {Romano}}, \bibinfo {author} {\bibfnamefont {P.}~\bibnamefont {Gentile}},
  \bibinfo {author} {\bibfnamefont {C.}~\bibnamefont {Noce}}, \bibinfo {author}
  {\bibfnamefont {I.}~\bibnamefont {Vekhter}}, \ and\ \bibinfo {author}
  {\bibfnamefont {M.}~\bibnamefont {Cuoco}},\ }\href@noop {} {\bibfield
  {journal} {\bibinfo  {journal} {Phys. Rev. Lett.}\ }\textbf {\bibinfo
  {volume} {110}},\ \bibinfo {pages} {267002} (\bibinfo {year}
  {2013})}\BibitemShut {NoStop}%
\bibitem [{\citenamefont {Romano}\ \emph {et~al.}(2016)\citenamefont {Romano},
  \citenamefont {Gentile}, \citenamefont {Noce}, \citenamefont {Vekhter},\ and\
  \citenamefont {Cuoco}}]{Romano16}%
  \BibitemOpen
  \bibfield  {author} {\bibinfo {author} {\bibfnamefont {A.}~\bibnamefont
  {Romano}}, \bibinfo {author} {\bibfnamefont {P.}~\bibnamefont {Gentile}},
  \bibinfo {author} {\bibfnamefont {C.}~\bibnamefont {Noce}}, \bibinfo {author}
  {\bibfnamefont {I.}~\bibnamefont {Vekhter}}, \ and\ \bibinfo {author}
  {\bibfnamefont {M.}~\bibnamefont {Cuoco}},\ }\href@noop {} {\bibfield
  {journal} {\bibinfo  {journal} {Phys. Rev. B}\ }\textbf {\bibinfo {volume}
  {93}},\ \bibinfo {pages} {014510} (\bibinfo {year} {2016})}\BibitemShut
  {NoStop}%
\bibitem [{\citenamefont {Romano}\ \emph {et~al.}(2017)\citenamefont {Romano},
  \citenamefont {Noce}, \citenamefont {Vekhter},\ and\ \citenamefont
  {Cuoco}}]{Romano17}%
  \BibitemOpen
  \bibfield  {author} {\bibinfo {author} {\bibfnamefont {A.}~\bibnamefont
  {Romano}}, \bibinfo {author} {\bibfnamefont {C.}~\bibnamefont {Noce}},
  \bibinfo {author} {\bibfnamefont {I.}~\bibnamefont {Vekhter}}, \ and\
  \bibinfo {author} {\bibfnamefont {M.}~\bibnamefont {Cuoco}},\ }\href@noop {}
  {\bibfield  {journal} {\bibinfo  {journal} {Phys. Rev. B}\ }\textbf {\bibinfo
  {volume} {96}},\ \bibinfo {pages} {054512} (\bibinfo {year}
  {2017})}\BibitemShut {NoStop}%
\bibitem [{\citenamefont {Fay}\ and\ \citenamefont {Appel}(1980)}]{Fay}%
  \BibitemOpen
  \bibfield  {author} {\bibinfo {author} {\bibfnamefont {D.}~\bibnamefont
  {Fay}}\ and\ \bibinfo {author} {\bibfnamefont {J.}~\bibnamefont {Appel}},\
  }\href@noop {} {\bibfield  {journal} {\bibinfo  {journal} {Phys. Rev. B}\
  }\textbf {\bibinfo {volume} {22}},\ \bibinfo {pages} {3173} (\bibinfo {year}
  {1980})}\BibitemShut {NoStop}%
\bibitem [{\citenamefont {Pfleiderer}(2009)}]{Pfleiderer}%
  \BibitemOpen
  \bibfield  {author} {\bibinfo {author} {\bibfnamefont {C.}~\bibnamefont
  {Pfleiderer}},\ }\href@noop {} {\bibfield  {journal} {\bibinfo  {journal}
  {Rev. Mod. Phys.}\ }\textbf {\bibinfo {volume} {81}},\ \bibinfo {pages}
  {1551} (\bibinfo {year} {2009})}\BibitemShut {NoStop}%
\bibitem [{\citenamefont {Mackenzie}\ and\ \citenamefont
  {Maeno}(2003)}]{Maeno2}%
  \BibitemOpen
  \bibfield  {author} {\bibinfo {author} {\bibfnamefont {A.~P.}\ \bibnamefont
  {Mackenzie}}\ and\ \bibinfo {author} {\bibfnamefont {Y.}~\bibnamefont
  {Maeno}},\ }\href@noop {} {\bibfield  {journal} {\bibinfo  {journal} {Rev.
  Mod. Phys.}\ }\textbf {\bibinfo {volume} {75}},\ \bibinfo {pages} {657}
  (\bibinfo {year} {2003})}\BibitemShut {NoStop}%
\bibitem [{\citenamefont {Martin}\ and\ \citenamefont
  {Morpurgo}(2012)}]{Martin2012}%
  \BibitemOpen
  \bibfield  {author} {\bibinfo {author} {\bibfnamefont {I.}~\bibnamefont
  {Martin}}\ and\ \bibinfo {author} {\bibfnamefont {A.~F.}\ \bibnamefont
  {Morpurgo}},\ }\href@noop {} {\bibfield  {journal} {\bibinfo  {journal}
  {Phys. Rev. B}\ }\textbf {\bibinfo {volume} {85}},\ \bibinfo {pages} {144505}
  (\bibinfo {year} {2012})}\BibitemShut {NoStop}%
\bibitem [{\citenamefont {P\"oyh\"onen}\ \emph {et~al.}(2014)\citenamefont
  {P\"oyh\"onen}, \citenamefont {Weststr\"om}, \citenamefont {R\"ontynen},\
  and\ \citenamefont {Ojanen}}]{Ojanen2014}%
  \BibitemOpen
  \bibfield  {author} {\bibinfo {author} {\bibfnamefont {K.}~\bibnamefont
  {P\"oyh\"onen}}, \bibinfo {author} {\bibfnamefont {A.}~\bibnamefont
  {Weststr\"om}}, \bibinfo {author} {\bibfnamefont {J.}~\bibnamefont
  {R\"ontynen}}, \ and\ \bibinfo {author} {\bibfnamefont {T.}~\bibnamefont
  {Ojanen}},\ }\href@noop {} {\bibfield  {journal} {\bibinfo  {journal} {Phys.
  Rev. B}\ }\textbf {\bibinfo {volume} {89}},\ \bibinfo {pages} {115109}
  (\bibinfo {year} {2014})}\BibitemShut {NoStop}%
\bibitem [{\citenamefont {Pientka}\ \emph {et~al.}(2013)\citenamefont
  {Pientka}, \citenamefont {Glazman},\ and\ \citenamefont {von
  Oppen}}]{Pientka2013}%
  \BibitemOpen
  \bibfield  {author} {\bibinfo {author} {\bibfnamefont {F.}~\bibnamefont
  {Pientka}}, \bibinfo {author} {\bibfnamefont {L.~I.}\ \bibnamefont
  {Glazman}}, \ and\ \bibinfo {author} {\bibfnamefont {F.}~\bibnamefont {von
  Oppen}},\ }\href@noop {} {\bibfield  {journal} {\bibinfo  {journal} {Phys.
  Rev. B}\ }\textbf {\bibinfo {volume} {88}},\ \bibinfo {pages} {155420}
  (\bibinfo {year} {2013})}\BibitemShut {NoStop}%
\bibitem [{\citenamefont {Braunecker}\ and\ \citenamefont
  {Simon}(2013)}]{Braunecker}%
  \BibitemOpen
  \bibfield  {author} {\bibinfo {author} {\bibfnamefont {B.}~\bibnamefont
  {Braunecker}}\ and\ \bibinfo {author} {\bibfnamefont {P.}~\bibnamefont
  {Simon}},\ }\href@noop {} {\bibfield  {journal} {\bibinfo  {journal} {Phys.
  Rev. Lett.}\ }\textbf {\bibinfo {volume} {111}},\ \bibinfo {pages} {147202}
  (\bibinfo {year} {2013})}\BibitemShut {NoStop}%
\bibitem [{\citenamefont {Klinovaja}\ \emph {et~al.}(2013)\citenamefont
  {Klinovaja}, \citenamefont {Stano}, \citenamefont {Yazdani},\ and\
  \citenamefont {Loss}}]{Klinovaja2013}%
  \BibitemOpen
  \bibfield  {author} {\bibinfo {author} {\bibfnamefont {J.}~\bibnamefont
  {Klinovaja}}, \bibinfo {author} {\bibfnamefont {P.}~\bibnamefont {Stano}},
  \bibinfo {author} {\bibfnamefont {A.}~\bibnamefont {Yazdani}}, \ and\
  \bibinfo {author} {\bibfnamefont {D.}~\bibnamefont {Loss}},\ }\href@noop {}
  {\bibfield  {journal} {\bibinfo  {journal} {Phys. Rev. Lett.}\ }\textbf
  {\bibinfo {volume} {111}},\ \bibinfo {pages} {186805} (\bibinfo {year}
  {2013})}\BibitemShut {NoStop}%
\bibitem [{\citenamefont {Kim}\ \emph {et~al.}(2014)\citenamefont {Kim},
  \citenamefont {Cheng}, \citenamefont {Bauer}, \citenamefont {Lutchyn},\ and\
  \citenamefont {Das~Sarma}}]{Kim2014}%
  \BibitemOpen
  \bibfield  {author} {\bibinfo {author} {\bibfnamefont {Y.}~\bibnamefont
  {Kim}}, \bibinfo {author} {\bibfnamefont {M.}~\bibnamefont {Cheng}}, \bibinfo
  {author} {\bibfnamefont {B.}~\bibnamefont {Bauer}}, \bibinfo {author}
  {\bibfnamefont {R.~M.}\ \bibnamefont {Lutchyn}}, \ and\ \bibinfo {author}
  {\bibfnamefont {S.}~\bibnamefont {Das~Sarma}},\ }\href@noop {} {\bibfield
  {journal} {\bibinfo  {journal} {Phys. Rev. B}\ }\textbf {\bibinfo {volume}
  {90}},\ \bibinfo {pages} {060401} (\bibinfo {year} {2014})}\BibitemShut
  {NoStop}%
\bibitem [{\citenamefont {Pientka}\ \emph {et~al.}(2014)\citenamefont
  {Pientka}, \citenamefont {Glazman},\ and\ \citenamefont {von
  Oppen}}]{Pientka2014}%
  \BibitemOpen
  \bibfield  {author} {\bibinfo {author} {\bibfnamefont {F.}~\bibnamefont
  {Pientka}}, \bibinfo {author} {\bibfnamefont {L.~I.}\ \bibnamefont
  {Glazman}}, \ and\ \bibinfo {author} {\bibfnamefont {F.}~\bibnamefont {von
  Oppen}},\ }\href@noop {} {\bibfield  {journal} {\bibinfo  {journal} {Phys.
  Rev. B}\ }\textbf {\bibinfo {volume} {89}},\ \bibinfo {pages} {180505}
  (\bibinfo {year} {2014})}\BibitemShut {NoStop}%
\bibitem [{\citenamefont {Nadj-Perge}\ \emph {et~al.}(2013)\citenamefont
  {Nadj-Perge}, \citenamefont {Drozdov}, \citenamefont {Bernevig},\ and\
  \citenamefont {Yazdani}}]{a15}%
  \BibitemOpen
  \bibfield  {author} {\bibinfo {author} {\bibfnamefont {S.}~\bibnamefont
  {Nadj-Perge}}, \bibinfo {author} {\bibfnamefont {I.~K.}\ \bibnamefont
  {Drozdov}}, \bibinfo {author} {\bibfnamefont {B.~A.}\ \bibnamefont
  {Bernevig}}, \ and\ \bibinfo {author} {\bibfnamefont {A.}~\bibnamefont
  {Yazdani}},\ }\href@noop {} {\bibfield  {journal} {\bibinfo  {journal} {Phys.
  Rev. B}\ }\textbf {\bibinfo {volume} {88}},\ \bibinfo {pages} {020407}
  (\bibinfo {year} {2013})}\BibitemShut {NoStop}%
\bibitem [{\citenamefont {Nakosai}\ \emph {et~al.}(2013)\citenamefont
  {Nakosai}, \citenamefont {Tanaka},\ and\ \citenamefont
  {Nagaosa}}]{Nakosai2013}%
  \BibitemOpen
  \bibfield  {author} {\bibinfo {author} {\bibfnamefont {S.}~\bibnamefont
  {Nakosai}}, \bibinfo {author} {\bibfnamefont {Y.}~\bibnamefont {Tanaka}}, \
  and\ \bibinfo {author} {\bibfnamefont {N.}~\bibnamefont {Nagaosa}},\
  }\href@noop {} {\bibfield  {journal} {\bibinfo  {journal} {Phys. Rev. B}\
  }\textbf {\bibinfo {volume} {88}},\ \bibinfo {pages} {180503} (\bibinfo
  {year} {2013})}\BibitemShut {NoStop}%
\bibitem [{\citenamefont {Heimes}\ \emph {et~al.}(2014)\citenamefont {Heimes},
  \citenamefont {Kotetes},\ and\ \citenamefont {Sch\"on}}]{heimes}%
  \BibitemOpen
  \bibfield  {author} {\bibinfo {author} {\bibfnamefont {A.}~\bibnamefont
  {Heimes}}, \bibinfo {author} {\bibfnamefont {P.}~\bibnamefont {Kotetes}}, \
  and\ \bibinfo {author} {\bibfnamefont {G.}~\bibnamefont {Sch\"on}},\
  }\href@noop {} {\bibfield  {journal} {\bibinfo  {journal} {Phys. Rev. B}\
  }\textbf {\bibinfo {volume} {90}},\ \bibinfo {pages} {060507} (\bibinfo
  {year} {2014})}\BibitemShut {NoStop}%
\bibitem [{\citenamefont {Mendler}\ \emph {et~al.}(2015)\citenamefont
  {Mendler}, \citenamefont {Kotetes},\ and\ \citenamefont
  {Sch\"on}}]{Mendler2015PRB}%
  \BibitemOpen
  \bibfield  {author} {\bibinfo {author} {\bibfnamefont {D.}~\bibnamefont
  {Mendler}}, \bibinfo {author} {\bibfnamefont {P.}~\bibnamefont {Kotetes}}, \
  and\ \bibinfo {author} {\bibfnamefont {G.}~\bibnamefont {Sch\"on}},\ }\href
  {\doibase 10.1103/PhysRevB.91.155405} {\bibfield  {journal} {\bibinfo
  {journal} {Phys. Rev. B}\ }\textbf {\bibinfo {volume} {91}},\ \bibinfo
  {pages} {155405} (\bibinfo {year} {2015})}\BibitemShut {NoStop}%
\bibitem [{\citenamefont {Sau}\ \emph {et~al.}(2010)\citenamefont {Sau},
  \citenamefont {Lutchyn}, \citenamefont {Tewari},\ and\ \citenamefont
  {DasSarma}}]{a11}%
  \BibitemOpen
  \bibfield  {author} {\bibinfo {author} {\bibfnamefont {J.~D.}\ \bibnamefont
  {Sau}}, \bibinfo {author} {\bibfnamefont {R.~M.}\ \bibnamefont {Lutchyn}},
  \bibinfo {author} {\bibfnamefont {S.}~\bibnamefont {Tewari}}, \ and\ \bibinfo
  {author} {\bibfnamefont {S.}~\bibnamefont {DasSarma}},\ }\href@noop {}
  {\bibfield  {journal} {\bibinfo  {journal} {Phys. Rev. Lett.}\ }\textbf
  {\bibinfo {volume} {104}},\ \bibinfo {pages} {040502} (\bibinfo {year}
  {2010})}\BibitemShut {NoStop}%
\bibitem [{\citenamefont {Alicea}(2010)}]{a12}%
  \BibitemOpen
  \bibfield  {author} {\bibinfo {author} {\bibfnamefont {J.}~\bibnamefont
  {Alicea}},\ }\href@noop {} {\bibfield  {journal} {\bibinfo  {journal} {Phys.
  Rev. B}\ }\textbf {\bibinfo {volume} {81}},\ \bibinfo {pages} {125318}
  (\bibinfo {year} {2010})}\BibitemShut {NoStop}%
\bibitem [{\citenamefont {Lutchyn}\ \emph {et~al.}(2010)\citenamefont
  {Lutchyn}, \citenamefont {Sau},\ and\ \citenamefont {DasSarma}}]{a13}%
  \BibitemOpen
  \bibfield  {author} {\bibinfo {author} {\bibfnamefont {R.~M.}\ \bibnamefont
  {Lutchyn}}, \bibinfo {author} {\bibfnamefont {J.~D.}\ \bibnamefont {Sau}}, \
  and\ \bibinfo {author} {\bibfnamefont {S.}~\bibnamefont {DasSarma}},\
  }\href@noop {} {\bibfield  {journal} {\bibinfo  {journal} {Phys. Rev. Lett.}\
  }\textbf {\bibinfo {volume} {105}},\ \bibinfo {pages} {077001} (\bibinfo
  {year} {2010})}\BibitemShut {NoStop}%
\bibitem [{\citenamefont {Oreg}\ \emph {et~al.}(2010)\citenamefont {Oreg},
  \citenamefont {Refael},\ and\ \citenamefont {von Oppen}}]{a14}%
  \BibitemOpen
  \bibfield  {author} {\bibinfo {author} {\bibfnamefont {Y.}~\bibnamefont
  {Oreg}}, \bibinfo {author} {\bibfnamefont {G.}~\bibnamefont {Refael}}, \ and\
  \bibinfo {author} {\bibfnamefont {F.}~\bibnamefont {von Oppen}},\ }\href@noop
  {} {\bibfield  {journal} {\bibinfo  {journal} {Phys. Rev. Lett.}\ }\textbf
  {\bibinfo {volume} {105}},\ \bibinfo {pages} {177002} (\bibinfo {year}
  {2010})}\BibitemShut {NoStop}%
\bibitem [{\citenamefont {Volovik}(2003)}]{Volovikbook}%
  \BibitemOpen
  \bibfield  {author} {\bibinfo {author} {\bibfnamefont {G.}~\bibnamefont
  {Volovik}},\ }\href@noop {} {\emph {\bibinfo {title} {The Universe in a
  Helium Droplet}}}\ (\bibinfo  {publisher} {Oxford Science Publications, New
  York},\ \bibinfo {year} {2003})\BibitemShut {NoStop}%
\bibitem [{\citenamefont {Hor}\ \emph {et~al.}(2010)\citenamefont {Hor},
  \citenamefont {Williams}, \citenamefont {Checkelsky}, \citenamefont
  {Roushan}, \citenamefont {Seo}, \citenamefont {Xu}, \citenamefont
  {Zandbergen}, \citenamefont {Yazdani}, \citenamefont {Ong},\ and\
  \citenamefont {Cava}}]{Hor10}%
  \BibitemOpen
  \bibfield  {author} {\bibinfo {author} {\bibfnamefont {Y.~S.}\ \bibnamefont
  {Hor}}, \bibinfo {author} {\bibfnamefont {A.~J.}\ \bibnamefont {Williams}},
  \bibinfo {author} {\bibfnamefont {J.~G.}\ \bibnamefont {Checkelsky}},
  \bibinfo {author} {\bibfnamefont {P.}~\bibnamefont {Roushan}}, \bibinfo
  {author} {\bibfnamefont {J.}~\bibnamefont {Seo}}, \bibinfo {author}
  {\bibfnamefont {Q.}~\bibnamefont {Xu}}, \bibinfo {author} {\bibfnamefont
  {H.~W.}\ \bibnamefont {Zandbergen}}, \bibinfo {author} {\bibfnamefont
  {A.}~\bibnamefont {Yazdani}}, \bibinfo {author} {\bibfnamefont {N.~P.}\
  \bibnamefont {Ong}}, \ and\ \bibinfo {author} {\bibfnamefont {R.~J.}\
  \bibnamefont {Cava}},\ }\href@noop {} {\bibfield  {journal} {\bibinfo
  {journal} {Phys. Rev. Lett}\ }\textbf {\bibinfo {volume} {104}},\ \bibinfo
  {pages} {057001} (\bibinfo {year} {2010})}\BibitemShut {NoStop}%
\bibitem [{\citenamefont {Wray}\ \emph {et~al.}(2010)\citenamefont {Wray},
  \citenamefont {Xu}, \citenamefont {Xia}, \citenamefont {Hor}, \citenamefont
  {Qian}, \citenamefont {Fedorov}, \citenamefont {Lin}, \citenamefont {Bansil},
  \citenamefont {Cava},\ and\ \citenamefont {Hasan}}]{Wray10}%
  \BibitemOpen
  \bibfield  {author} {\bibinfo {author} {\bibfnamefont {L.~A.}\ \bibnamefont
  {Wray}}, \bibinfo {author} {\bibfnamefont {S.~Y.}\ \bibnamefont {Xu}},
  \bibinfo {author} {\bibfnamefont {Y.}~\bibnamefont {Xia}}, \bibinfo {author}
  {\bibfnamefont {Y.~S.}\ \bibnamefont {Hor}}, \bibinfo {author} {\bibfnamefont
  {D.}~\bibnamefont {Qian}}, \bibinfo {author} {\bibfnamefont {A.~V.}\
  \bibnamefont {Fedorov}}, \bibinfo {author} {\bibfnamefont {H.}~\bibnamefont
  {Lin}}, \bibinfo {author} {\bibfnamefont {A.}~\bibnamefont {Bansil}},
  \bibinfo {author} {\bibfnamefont {R.~J.}\ \bibnamefont {Cava}}, \ and\
  \bibinfo {author} {\bibfnamefont {M.~Z.}\ \bibnamefont {Hasan}},\ }\href@noop
  {} {\bibfield  {journal} {\bibinfo  {journal} {Nature Phys.}\ }\textbf
  {\bibinfo {volume} {6}},\ \bibinfo {pages} {855} (\bibinfo {year}
  {2010})}\BibitemShut {NoStop}%
\bibitem [{\citenamefont {Kriener}\ \emph {et~al.}(2011)\citenamefont
  {Kriener}, \citenamefont {Segawa}, \citenamefont {Ren}, \citenamefont
  {Sasaki},\ and\ \citenamefont {Ando}}]{Kriener11}%
  \BibitemOpen
  \bibfield  {author} {\bibinfo {author} {\bibfnamefont {M.}~\bibnamefont
  {Kriener}}, \bibinfo {author} {\bibfnamefont {K.}~\bibnamefont {Segawa}},
  \bibinfo {author} {\bibfnamefont {Z.}~\bibnamefont {Ren}}, \bibinfo {author}
  {\bibfnamefont {S.}~\bibnamefont {Sasaki}}, \ and\ \bibinfo {author}
  {\bibfnamefont {Y.}~\bibnamefont {Ando}},\ }\href@noop {} {\bibfield
  {journal} {\bibinfo  {journal} {Phys.Rev. Lett.}\ }\textbf {\bibinfo {volume}
  {106}},\ \bibinfo {pages} {127004} (\bibinfo {year} {2011})}\BibitemShut
  {NoStop}%
\bibitem [{\citenamefont {Fu}\ and\ \citenamefont {Berg}(2010)}]{Fu2010PRL}%
  \BibitemOpen
  \bibfield  {author} {\bibinfo {author} {\bibfnamefont {L.}~\bibnamefont
  {Fu}}\ and\ \bibinfo {author} {\bibfnamefont {E.}~\bibnamefont {Berg}},\
  }\href {\doibase 10.1103/PhysRevLett.105.097001} {\bibfield  {journal}
  {\bibinfo  {journal} {Phys. Rev. Lett.}\ }\textbf {\bibinfo {volume} {105}},\
  \bibinfo {pages} {097001} (\bibinfo {year} {2010})}\BibitemShut {NoStop}%
\bibitem [{\citenamefont {Sasaki}\ \emph {et~al.}(2011)\citenamefont {Sasaki},
  \citenamefont {Kriener}, \citenamefont {Segawa}, \citenamefont {Yada},
  \citenamefont {Tanaka}, \citenamefont {Sato},\ and\ \citenamefont
  {Ando}}]{sasaki11}%
  \BibitemOpen
  \bibfield  {author} {\bibinfo {author} {\bibfnamefont {S.}~\bibnamefont
  {Sasaki}}, \bibinfo {author} {\bibfnamefont {M.}~\bibnamefont {Kriener}},
  \bibinfo {author} {\bibfnamefont {K.}~\bibnamefont {Segawa}}, \bibinfo
  {author} {\bibfnamefont {K.}~\bibnamefont {Yada}}, \bibinfo {author}
  {\bibfnamefont {Y.}~\bibnamefont {Tanaka}}, \bibinfo {author} {\bibfnamefont
  {M.}~\bibnamefont {Sato}}, \ and\ \bibinfo {author} {\bibfnamefont
  {Y.}~\bibnamefont {Ando}},\ }\href@noop {} {\bibfield  {journal} {\bibinfo
  {journal} {Phys. Rev. Lett.}\ }\textbf {\bibinfo {volume} {107}},\ \bibinfo
  {pages} {217001} (\bibinfo {year} {2011})}\BibitemShut {NoStop}%
\bibitem [{\citenamefont {Hao}\ and\ \citenamefont {Lee}(2011)}]{hao11}%
  \BibitemOpen
  \bibfield  {author} {\bibinfo {author} {\bibfnamefont {L.}~\bibnamefont
  {Hao}}\ and\ \bibinfo {author} {\bibfnamefont {T.~K.}\ \bibnamefont {Lee}},\
  }\href@noop {} {\bibfield  {journal} {\bibinfo  {journal} {Phys. Rev. B}\
  }\textbf {\bibinfo {volume} {83}},\ \bibinfo {pages} {134516} (\bibinfo
  {year} {2011})}\BibitemShut {NoStop}%
\bibitem [{\citenamefont {Yamakage}\ \emph {et~al.}(2012)\citenamefont
  {Yamakage}, \citenamefont {Yada}, \citenamefont {Sato},\ and\ \citenamefont
  {Tanaka}}]{yamakage12}%
  \BibitemOpen
  \bibfield  {author} {\bibinfo {author} {\bibfnamefont {A.}~\bibnamefont
  {Yamakage}}, \bibinfo {author} {\bibfnamefont {K.}~\bibnamefont {Yada}},
  \bibinfo {author} {\bibfnamefont {M.}~\bibnamefont {Sato}}, \ and\ \bibinfo
  {author} {\bibfnamefont {Y.}~\bibnamefont {Tanaka}},\ }\href@noop {}
  {\bibfield  {journal} {\bibinfo  {journal} {Phys. Rev. B}\ }\textbf {\bibinfo
  {volume} {85}},\ \bibinfo {pages} {180509} (\bibinfo {year}
  {2012})}\BibitemShut {NoStop}%
\bibitem [{\citenamefont {Oudah}\ \emph {et~al.}(2016)\citenamefont {Oudah},
  \citenamefont {Ikeda}, \citenamefont {Hausmann}, \citenamefont {Yonezawa},
  \citenamefont {Fukumoto}, \citenamefont {Kobayashi}, \citenamefont {Sato},\
  and\ \citenamefont {Maeno}}]{Oudah}%
  \BibitemOpen
  \bibfield  {author} {\bibinfo {author} {\bibfnamefont {M.}~\bibnamefont
  {Oudah}}, \bibinfo {author} {\bibfnamefont {A.}~\bibnamefont {Ikeda}},
  \bibinfo {author} {\bibfnamefont {J.~N.}\ \bibnamefont {Hausmann}}, \bibinfo
  {author} {\bibfnamefont {S.}~\bibnamefont {Yonezawa}}, \bibinfo {author}
  {\bibfnamefont {T.}~\bibnamefont {Fukumoto}}, \bibinfo {author}
  {\bibfnamefont {S.}~\bibnamefont {Kobayashi}}, \bibinfo {author}
  {\bibfnamefont {M.}~\bibnamefont {Sato}}, \ and\ \bibinfo {author}
  {\bibfnamefont {Y.}~\bibnamefont {Maeno}},\ }\href@noop {} {\bibfield
  {journal} {\bibinfo  {journal} {Nat. Commun.}\ }\textbf {\bibinfo {volume}
  {7}},\ \bibinfo {pages} {13617} (\bibinfo {year} {2016})}\BibitemShut
  {NoStop}%
\bibitem [{\citenamefont {Aggarwal}\ \emph {et~al.}(2015)\citenamefont
  {Aggarwal}, \citenamefont {Gaurav}, \citenamefont {Thakur}, \citenamefont
  {Haque}, \citenamefont {Ganguli},\ and\ \citenamefont {Sheet}}]{Aggarwal}%
  \BibitemOpen
  \bibfield  {author} {\bibinfo {author} {\bibfnamefont {L.}~\bibnamefont
  {Aggarwal}}, \bibinfo {author} {\bibfnamefont {A.}~\bibnamefont {Gaurav}},
  \bibinfo {author} {\bibfnamefont {G.~S.}\ \bibnamefont {Thakur}}, \bibinfo
  {author} {\bibfnamefont {Z.}~\bibnamefont {Haque}}, \bibinfo {author}
  {\bibfnamefont {A.~K.}\ \bibnamefont {Ganguli}}, \ and\ \bibinfo {author}
  {\bibfnamefont {G.}~\bibnamefont {Sheet}},\ }\href@noop {} {\bibfield
  {journal} {\bibinfo  {journal} {Nat. Matter}\ }\textbf {\bibinfo {volume}
  {15}},\ \bibinfo {pages} {32} (\bibinfo {year} {2015})}\BibitemShut {NoStop}%
\bibitem [{\citenamefont {Wang}\ \emph {et~al.}(2015)\citenamefont {Wang},
  \citenamefont {Wang}, \citenamefont {Liu}, \citenamefont {Lu}, \citenamefont
  {Yang}, \citenamefont {Jia}, \citenamefont {Liu}, \citenamefont {Xie},
  \citenamefont {Wei},\ and\ \citenamefont {Wang}}]{Wang}%
  \BibitemOpen
  \bibfield  {author} {\bibinfo {author} {\bibfnamefont {H.}~\bibnamefont
  {Wang}}, \bibinfo {author} {\bibfnamefont {H.}~\bibnamefont {Wang}}, \bibinfo
  {author} {\bibfnamefont {H.}~\bibnamefont {Liu}}, \bibinfo {author}
  {\bibfnamefont {H.}~\bibnamefont {Lu}}, \bibinfo {author} {\bibfnamefont
  {W.}~\bibnamefont {Yang}}, \bibinfo {author} {\bibfnamefont {S.}~\bibnamefont
  {Jia}}, \bibinfo {author} {\bibfnamefont {X.-J.}\ \bibnamefont {Liu}},
  \bibinfo {author} {\bibfnamefont {X.~C.}\ \bibnamefont {Xie}}, \bibinfo
  {author} {\bibfnamefont {J.}~\bibnamefont {Wei}}, \ and\ \bibinfo {author}
  {\bibfnamefont {J.}~\bibnamefont {Wang}},\ }\href@noop {} {\bibfield
  {journal} {\bibinfo  {journal} {Nat. Matter}\ }\textbf {\bibinfo {volume}
  {15}},\ \bibinfo {pages} {38} (\bibinfo {year} {2015})}\BibitemShut {NoStop}%
\bibitem [{\citenamefont {Rashba}(1960)}]{Rashba1960}%
  \BibitemOpen
  \bibfield  {author} {\bibinfo {author} {\bibfnamefont {E.~I.}\ \bibnamefont
  {Rashba}},\ }\href@noop {} {\bibfield  {journal} {\bibinfo  {journal} {Sov.
  Phys. Solid State}\ }\textbf {\bibinfo {volume} {2}},\ \bibinfo {pages}
  {1109} (\bibinfo {year} {1960})}\BibitemShut {NoStop}%
\bibitem [{\citenamefont {Dresselhaus}(1955)}]{Dresselhaus1955}%
  \BibitemOpen
  \bibfield  {author} {\bibinfo {author} {\bibfnamefont {G.}~\bibnamefont
  {Dresselhaus}},\ }\href@noop {} {\bibfield  {journal} {\bibinfo  {journal}
  {Phys. Rev.}\ }\textbf {\bibinfo {volume} {100}},\ \bibinfo {pages} {580}
  (\bibinfo {year} {1955})}\BibitemShut {NoStop}%
\bibitem [{\citenamefont {Gorkov}\ and\ \citenamefont
  {Rashba}(2001)}]{Gorkov2001}%
  \BibitemOpen
  \bibfield  {author} {\bibinfo {author} {\bibfnamefont {L.~P.}\ \bibnamefont
  {Gorkov}}\ and\ \bibinfo {author} {\bibfnamefont {E.~I.}\ \bibnamefont
  {Rashba}},\ }\href@noop {} {\bibfield  {journal} {\bibinfo  {journal} {Phys.
  Rev. Lett.}\ }\textbf {\bibinfo {volume} {87}},\ \bibinfo {pages} {037004}
  (\bibinfo {year} {2001})}\BibitemShut {NoStop}%
\bibitem [{\citenamefont {Frigeri}\ \emph {et~al.}(2004)\citenamefont
  {Frigeri}, \citenamefont {Agterberg}, \citenamefont {Koga},\ and\
  \citenamefont {Sigrist}}]{Frigeri2004}%
  \BibitemOpen
  \bibfield  {author} {\bibinfo {author} {\bibfnamefont {P.~A.}\ \bibnamefont
  {Frigeri}}, \bibinfo {author} {\bibfnamefont {D.~F.}\ \bibnamefont
  {Agterberg}}, \bibinfo {author} {\bibfnamefont {A.}~\bibnamefont {Koga}}, \
  and\ \bibinfo {author} {\bibfnamefont {M.}~\bibnamefont {Sigrist}},\
  }\href@noop {} {\bibfield  {journal} {\bibinfo  {journal} {Phys. Rev. Lett.}\
  }\textbf {\bibinfo {volume} {92}},\ \bibinfo {pages} {097001} (\bibinfo
  {year} {2004})}\BibitemShut {NoStop}%
\bibitem [{\citenamefont {Yada}\ \emph {et~al.}(2009)\citenamefont {Yada},
  \citenamefont {Onari}, \citenamefont {Tanaka},\ and\ \citenamefont
  {Inoue}}]{Yada2009PRB}%
  \BibitemOpen
  \bibfield  {author} {\bibinfo {author} {\bibfnamefont {K.}~\bibnamefont
  {Yada}}, \bibinfo {author} {\bibfnamefont {S.}~\bibnamefont {Onari}},
  \bibinfo {author} {\bibfnamefont {Y.}~\bibnamefont {Tanaka}}, \ and\ \bibinfo
  {author} {\bibfnamefont {J.-i.}\ \bibnamefont {Inoue}},\ }\href {\doibase
  10.1103/PhysRevB.80.140509} {\bibfield  {journal} {\bibinfo  {journal} {Phys.
  Rev. B}\ }\textbf {\bibinfo {volume} {80}},\ \bibinfo {pages} {140509}
  (\bibinfo {year} {2009})}\BibitemShut {NoStop}%
\bibitem [{\citenamefont {Lu}\ and\ \citenamefont {Yip}(2008)}]{Lu2008}%
  \BibitemOpen
  \bibfield  {author} {\bibinfo {author} {\bibfnamefont {C.~K.}\ \bibnamefont
  {Lu}}\ and\ \bibinfo {author} {\bibfnamefont {S.}~\bibnamefont {Yip}},\
  }\href@noop {} {\bibfield  {journal} {\bibinfo  {journal} {Phys. Rev. B}\
  }\textbf {\bibinfo {volume} {78}},\ \bibinfo {pages} {132502} (\bibinfo
  {year} {2008})}\BibitemShut {NoStop}%
\bibitem [{\citenamefont {Vorontsov}\ \emph {et~al.}(2008)\citenamefont
  {Vorontsov}, \citenamefont {Vekhter},\ and\ \citenamefont
  {Eschrig}}]{Vorontsov2008}%
  \BibitemOpen
  \bibfield  {author} {\bibinfo {author} {\bibfnamefont {A.~B.}\ \bibnamefont
  {Vorontsov}}, \bibinfo {author} {\bibfnamefont {I.}~\bibnamefont {Vekhter}},
  \ and\ \bibinfo {author} {\bibfnamefont {M.}~\bibnamefont {Eschrig}},\
  }\href@noop {} {\bibfield  {journal} {\bibinfo  {journal} {Phys. Rev. Lett.}\
  }\textbf {\bibinfo {volume} {101}},\ \bibinfo {pages} {127003} (\bibinfo
  {year} {2008})}\BibitemShut {NoStop}%
\bibitem [{\citenamefont {Tanaka}\ \emph {et~al.}(2009)\citenamefont {Tanaka},
  \citenamefont {Yokoyama}, \citenamefont {Balatsky},\ and\ \citenamefont
  {Nagaosa}}]{Tanaka2009}%
  \BibitemOpen
  \bibfield  {author} {\bibinfo {author} {\bibfnamefont {Y.}~\bibnamefont
  {Tanaka}}, \bibinfo {author} {\bibfnamefont {T.}~\bibnamefont {Yokoyama}},
  \bibinfo {author} {\bibfnamefont {A.~V.}\ \bibnamefont {Balatsky}}, \ and\
  \bibinfo {author} {\bibfnamefont {N.}~\bibnamefont {Nagaosa}},\ }\href@noop
  {} {\bibfield  {journal} {\bibinfo  {journal} {Phys. Rev. B}\ }\textbf
  {\bibinfo {volume} {79}},\ \bibinfo {pages} {060505(R)} (\bibinfo {year}
  {2009})}\BibitemShut {NoStop}%
\bibitem [{\citenamefont {Sato}\ and\ \citenamefont
  {Fujimoto}(2009)}]{Sato2009}%
  \BibitemOpen
  \bibfield  {author} {\bibinfo {author} {\bibfnamefont {M.}~\bibnamefont
  {Sato}}\ and\ \bibinfo {author} {\bibfnamefont {S.}~\bibnamefont
  {Fujimoto}},\ }\href@noop {} {\bibfield  {journal} {\bibinfo  {journal}
  {Phys. Rev. B}\ }\textbf {\bibinfo {volume} {79}},\ \bibinfo {pages} {094504}
  (\bibinfo {year} {2009})}\BibitemShut {NoStop}%
\bibitem [{\citenamefont {Schnyder}\ \emph {et~al.}(2010)\citenamefont
  {Schnyder}, \citenamefont {Brydon}, \citenamefont {Manske},\ and\
  \citenamefont {Timm}}]{SBMT10}%
  \BibitemOpen
  \bibfield  {author} {\bibinfo {author} {\bibfnamefont {A.~P.}\ \bibnamefont
  {Schnyder}}, \bibinfo {author} {\bibfnamefont {P.~M.~R.}\ \bibnamefont
  {Brydon}}, \bibinfo {author} {\bibfnamefont {D.}~\bibnamefont {Manske}}, \
  and\ \bibinfo {author} {\bibfnamefont {C.}~\bibnamefont {Timm}},\ }\href@noop
  {} {\bibfield  {journal} {\bibinfo  {journal} {Phys. Rev. B}\ }\textbf
  {\bibinfo {volume} {82}},\ \bibinfo {pages} {184508} (\bibinfo {year}
  {2010})}\BibitemShut {NoStop}%
\bibitem [{\citenamefont {Ying}\ \emph {et~al.}(2017)\citenamefont {Ying},
  \citenamefont {Cuoco}, \citenamefont {Ortix},\ and\ \citenamefont
  {Gentile}}]{Ying2017}%
  \BibitemOpen
  \bibfield  {author} {\bibinfo {author} {\bibfnamefont {Z.-J.}\ \bibnamefont
  {Ying}}, \bibinfo {author} {\bibfnamefont {M.}~\bibnamefont {Cuoco}},
  \bibinfo {author} {\bibfnamefont {C.}~\bibnamefont {Ortix}}, \ and\ \bibinfo
  {author} {\bibfnamefont {P.}~\bibnamefont {Gentile}},\ }\href@noop {}
  {\bibfield  {journal} {\bibinfo  {journal} {Phys. Rev. B}\ }\textbf {\bibinfo
  {volume} {96}},\ \bibinfo {pages} {100506 (R)} (\bibinfo {year}
  {2017})}\BibitemShut {NoStop}%
\bibitem [{\citenamefont {Ole\'s}(2012)}]{Ole12}%
  \BibitemOpen
  \bibfield  {author} {\bibinfo {author} {\bibfnamefont {A.~M.}\ \bibnamefont
  {Ole\'s}},\ }\href@noop {} {\bibfield  {journal} {\bibinfo  {journal} {J.
  Phys.: Condens. Matter}\ }\textbf {\bibinfo {volume} {24}},\ \bibinfo {pages}
  {313201} (\bibinfo {year} {2012})}\BibitemShut {NoStop}%
\bibitem [{\citenamefont {Hwang}\ \emph {et~al.}(2012)\citenamefont {Hwang},
  \citenamefont {Iwasa}, \citenamefont {Kawasaki}, \citenamefont {Keimer},
  \citenamefont {Nagaosa},\ and\ \citenamefont {Tokura}}]{Hwang2012}%
  \BibitemOpen
  \bibfield  {author} {\bibinfo {author} {\bibfnamefont {H.~Y.}\ \bibnamefont
  {Hwang}}, \bibinfo {author} {\bibfnamefont {Y.}~\bibnamefont {Iwasa}},
  \bibinfo {author} {\bibfnamefont {M.}~\bibnamefont {Kawasaki}}, \bibinfo
  {author} {\bibfnamefont {B.}~\bibnamefont {Keimer}}, \bibinfo {author}
  {\bibfnamefont {N.}~\bibnamefont {Nagaosa}}, \ and\ \bibinfo {author}
  {\bibfnamefont {Y.}~\bibnamefont {Tokura}},\ }\href@noop {} {\bibfield
  {journal} {\bibinfo  {journal} {Nat. Mater.}\ }\textbf {\bibinfo {volume}
  {11}},\ \bibinfo {pages} {103} (\bibinfo {year} {2012})}\BibitemShut
  {NoStop}%
\bibitem [{\citenamefont {Khalsa}\ \emph {et~al.}(2013)\citenamefont {Khalsa},
  \citenamefont {Lee},\ and\ \citenamefont {MacDonald}}]{Khalsa2013PRB}%
  \BibitemOpen
  \bibfield  {author} {\bibinfo {author} {\bibfnamefont {G.}~\bibnamefont
  {Khalsa}}, \bibinfo {author} {\bibfnamefont {B.}~\bibnamefont {Lee}}, \ and\
  \bibinfo {author} {\bibfnamefont {A.~H.}\ \bibnamefont {MacDonald}},\
  }\href@noop {} {\bibfield  {journal} {\bibinfo  {journal} {Phys. Rev. B}\
  }\textbf {\bibinfo {volume} {88}},\ \bibinfo {pages} {041302} (\bibinfo
  {year} {2013})}\BibitemShut {NoStop}%
\bibitem [{\citenamefont {Autieri}\ \emph {et~al.}(2014)\citenamefont
  {Autieri}, \citenamefont {Cuoco},\ and\ \citenamefont {Noce}}]{Autieri2014}%
  \BibitemOpen
  \bibfield  {author} {\bibinfo {author} {\bibfnamefont {C.}~\bibnamefont
  {Autieri}}, \bibinfo {author} {\bibfnamefont {M.}~\bibnamefont {Cuoco}}, \
  and\ \bibinfo {author} {\bibfnamefont {C.}~\bibnamefont {Noce}},\ }\href@noop
  {} {\bibfield  {journal} {\bibinfo  {journal} {Phys. Rev. B}\ }\textbf
  {\bibinfo {volume} {89}},\ \bibinfo {pages} {075102} (\bibinfo {year}
  {2014})}\BibitemShut {NoStop}%
\bibitem [{\citenamefont {Winkler}(2003)}]{Winklerbook}%
  \BibitemOpen
  \bibfield  {author} {\bibinfo {author} {\bibfnamefont {R.}~\bibnamefont
  {Winkler}},\ }\href@noop {} {\emph {\bibinfo {title} {Spin-Orbit Coupling
  Effects in Two-Dimensional Electron and Hole Systems}}}\ (\bibinfo
  {publisher} {Springer Berlin Heidelberg},\ \bibinfo {year}
  {2003})\BibitemShut {NoStop}%
\bibitem [{\citenamefont {Zhong}\ \emph {et~al.}(2013)\citenamefont {Zhong},
  \citenamefont {T\'oth},\ and\ \citenamefont {Held}}]{Zhong2013}%
  \BibitemOpen
  \bibfield  {author} {\bibinfo {author} {\bibfnamefont {Z.}~\bibnamefont
  {Zhong}}, \bibinfo {author} {\bibfnamefont {A.}~\bibnamefont {T\'oth}}, \
  and\ \bibinfo {author} {\bibfnamefont {K.}~\bibnamefont {Held}},\ }\href@noop
  {} {\bibfield  {journal} {\bibinfo  {journal} {Phys. Rev. B}\ }\textbf
  {\bibinfo {volume} {87}},\ \bibinfo {pages} {161102(R)} (\bibinfo {year}
  {2013})}\BibitemShut {NoStop}%
\bibitem [{\citenamefont {Kim}\ \emph {et~al.}(2013)\citenamefont {Kim},
  \citenamefont {Lutchyn},\ and\ \citenamefont {Nayak}}]{Kim2013}%
  \BibitemOpen
  \bibfield  {author} {\bibinfo {author} {\bibfnamefont {Y.}~\bibnamefont
  {Kim}}, \bibinfo {author} {\bibfnamefont {R.~M.}\ \bibnamefont {Lutchyn}}, \
  and\ \bibinfo {author} {\bibfnamefont {C.}~\bibnamefont {Nayak}},\
  }\href@noop {} {\bibfield  {journal} {\bibinfo  {journal} {Phys. Rev. B}\
  }\textbf {\bibinfo {volume} {87}},\ \bibinfo {pages} {245121} (\bibinfo
  {year} {2013})}\BibitemShut {NoStop}%
\bibitem [{\citenamefont {Joshua}\ \emph {et~al.}(2012)\citenamefont {Joshua},
  \citenamefont {Pecker}, \citenamefont {Ruhman}, \citenamefont {Altman},\ and\
  \citenamefont {Ilani}}]{Joshua2015}%
  \BibitemOpen
  \bibfield  {author} {\bibinfo {author} {\bibfnamefont {A.}~\bibnamefont
  {Joshua}}, \bibinfo {author} {\bibfnamefont {S.}~\bibnamefont {Pecker}},
  \bibinfo {author} {\bibfnamefont {J.}~\bibnamefont {Ruhman}}, \bibinfo
  {author} {\bibfnamefont {E.}~\bibnamefont {Altman}}, \ and\ \bibinfo {author}
  {\bibfnamefont {S.}~\bibnamefont {Ilani}},\ }\href@noop {} {\bibfield
  {journal} {\bibinfo  {journal} {Nat. Comm.}\ }\textbf {\bibinfo {volume}
  {3}},\ \bibinfo {pages} {1129} (\bibinfo {year} {2012})}\BibitemShut
  {NoStop}%
\bibitem [{\citenamefont {Steffen}\ \emph {et~al.}(2015)\citenamefont
  {Steffen}, \citenamefont {Loder},\ and\ \citenamefont {Kopp}}]{Steffen2015}%
  \BibitemOpen
  \bibfield  {author} {\bibinfo {author} {\bibfnamefont {K.}~\bibnamefont
  {Steffen}}, \bibinfo {author} {\bibfnamefont {F.}~\bibnamefont {Loder}}, \
  and\ \bibinfo {author} {\bibfnamefont {T.}~\bibnamefont {Kopp}},\ }\href@noop
  {} {\bibfield  {journal} {\bibinfo  {journal} {Phys. Rev. B}\ }\textbf
  {\bibinfo {volume} {91}},\ \bibinfo {pages} {075415} (\bibinfo {year}
  {2015})}\BibitemShut {NoStop}%
\bibitem [{\citenamefont {Ohtomo}\ and\ \citenamefont
  {Hwang}(2004)}]{Ohtomo2004Nature}%
  \BibitemOpen
  \bibfield  {author} {\bibinfo {author} {\bibfnamefont {A.}~\bibnamefont
  {Ohtomo}}\ and\ \bibinfo {author} {\bibfnamefont {H.~Y.}\ \bibnamefont
  {Hwang}},\ }\href@noop {} {\bibfield  {journal} {\bibinfo  {journal} {Nature
  (London)}\ }\textbf {\bibinfo {volume} {427}},\ \bibinfo {pages} {423}
  (\bibinfo {year} {2004})}\BibitemShut {NoStop}%
\bibitem [{\citenamefont {Zabaleta}\ \emph {et~al.}(2016)\citenamefont
  {Zabaleta}, \citenamefont {Borisov}, \citenamefont {Wanke}, \citenamefont
  {Jeschke}, \citenamefont {Parks}, \citenamefont {Baum}, \citenamefont
  {Teker}, \citenamefont {Harada}, \citenamefont {Syassen}, \citenamefont
  {Kopp}, \citenamefont {Pavlenko}, \citenamefont {Valent\'i},\ and\
  \citenamefont {Mannhart}}]{Zabaleta2016}%
  \BibitemOpen
  \bibfield  {author} {\bibinfo {author} {\bibfnamefont {J.}~\bibnamefont
  {Zabaleta}}, \bibinfo {author} {\bibfnamefont {V.~S.}\ \bibnamefont
  {Borisov}}, \bibinfo {author} {\bibfnamefont {R.}~\bibnamefont {Wanke}},
  \bibinfo {author} {\bibfnamefont {H.~O.}\ \bibnamefont {Jeschke}}, \bibinfo
  {author} {\bibfnamefont {S.~C.}\ \bibnamefont {Parks}}, \bibinfo {author}
  {\bibfnamefont {B.}~\bibnamefont {Baum}}, \bibinfo {author} {\bibfnamefont
  {A.}~\bibnamefont {Teker}}, \bibinfo {author} {\bibfnamefont
  {T.}~\bibnamefont {Harada}}, \bibinfo {author} {\bibfnamefont
  {K.}~\bibnamefont {Syassen}}, \bibinfo {author} {\bibfnamefont
  {T.}~\bibnamefont {Kopp}}, \bibinfo {author} {\bibfnamefont {N.}~\bibnamefont
  {Pavlenko}}, \bibinfo {author} {\bibfnamefont {R.}~\bibnamefont {Valent\'i}},
  \ and\ \bibinfo {author} {\bibfnamefont {J.}~\bibnamefont {Mannhart}},\
  }\href@noop {} {\bibfield  {journal} {\bibinfo  {journal} {Phys. Rev. B}\
  }\textbf {\bibinfo {volume} {93}},\ \bibinfo {pages} {235117} (\bibinfo
  {year} {2016})}\BibitemShut {NoStop}%
\bibitem [{\citenamefont {Salluzzo}\ \emph {et~al.}(2009)\citenamefont
  {Salluzzo}, \citenamefont {Cezar}, \citenamefont {Brookes}, \citenamefont
  {Bisogni}, \citenamefont {Luca}, \citenamefont {Richter}, \citenamefont
  {Thiel}, \citenamefont {Mannhart}, \citenamefont {Huijben}, \citenamefont
  {Brinkman}, \citenamefont {Rijnders},\ and\ \citenamefont
  {Ghiringhelli}}]{Salluzzo2009}%
  \BibitemOpen
  \bibfield  {author} {\bibinfo {author} {\bibfnamefont {M.}~\bibnamefont
  {Salluzzo}}, \bibinfo {author} {\bibfnamefont {J.~C.}\ \bibnamefont {Cezar}},
  \bibinfo {author} {\bibfnamefont {N.~B.}\ \bibnamefont {Brookes}}, \bibinfo
  {author} {\bibfnamefont {V.}~\bibnamefont {Bisogni}}, \bibinfo {author}
  {\bibfnamefont {G.~M.~D.}\ \bibnamefont {Luca}}, \bibinfo {author}
  {\bibfnamefont {C.}~\bibnamefont {Richter}}, \bibinfo {author} {\bibfnamefont
  {S.}~\bibnamefont {Thiel}}, \bibinfo {author} {\bibfnamefont
  {J.}~\bibnamefont {Mannhart}}, \bibinfo {author} {\bibfnamefont
  {M.}~\bibnamefont {Huijben}}, \bibinfo {author} {\bibfnamefont
  {A.}~\bibnamefont {Brinkman}}, \bibinfo {author} {\bibfnamefont
  {G.}~\bibnamefont {Rijnders}}, \ and\ \bibinfo {author} {\bibfnamefont
  {G.}~\bibnamefont {Ghiringhelli}},\ }\href@noop {} {\bibfield  {journal}
  {\bibinfo  {journal} {Phys. Rev. Lett.}\ }\textbf {\bibinfo {volume} {102}},\
  \bibinfo {pages} {166804} (\bibinfo {year} {2009})}\BibitemShut {NoStop}%
\bibitem [{\citenamefont {Sugano}\ \emph {et~al.}(1970)\citenamefont {Sugano},
  \citenamefont {Tanabe},\ and\ \citenamefont {Kamimura}}]{Suganobook}%
  \BibitemOpen
  \bibfield  {author} {\bibinfo {author} {\bibnamefont {Sugano}}, \bibinfo
  {author} {\bibfnamefont {Y.}~\bibnamefont {Tanabe}}, \ and\ \bibinfo {author}
  {\bibfnamefont {H.}~\bibnamefont {Kamimura}},\ }\href@noop {} {\emph
  {\bibinfo {title} {Multiplets of transition-metal ions in crystal}}}\
  (\bibinfo  {publisher} {Academic Press, New York London},\ \bibinfo {year}
  {1970})\BibitemShut {NoStop}%
\bibitem [{\citenamefont {Slater}(1960)}]{Slaterbook}%
  \BibitemOpen
  \bibfield  {author} {\bibinfo {author} {\bibfnamefont {J.}~\bibnamefont
  {Slater}},\ }\href@noop {} {\emph {\bibinfo {title} {Quantum Theory of Atomic
  Structure}}}\ (\bibinfo  {publisher} {McGraw-Hill, New York},\ \bibinfo
  {year} {1960})\BibitemShut {NoStop}%
\bibitem [{\citenamefont {Vaugier}\ \emph {et~al.}(2012)\citenamefont
  {Vaugier}, \citenamefont {Jiang},\ and\ \citenamefont
  {Biermann}}]{Vaugier2012}%
  \BibitemOpen
  \bibfield  {author} {\bibinfo {author} {\bibfnamefont {L.}~\bibnamefont
  {Vaugier}}, \bibinfo {author} {\bibfnamefont {H.}~\bibnamefont {Jiang}}, \
  and\ \bibinfo {author} {\bibfnamefont {S.}~\bibnamefont {Biermann}},\
  }\href@noop {} {\bibfield  {journal} {\bibinfo  {journal} {Phys. Rev. B}\
  }\textbf {\bibinfo {volume} {86}},\ \bibinfo {pages} {165105} (\bibinfo
  {year} {2012})}\BibitemShut {NoStop}%
\bibitem [{\citenamefont {Lifshitz}(1960)}]{Lifshitz1960}%
  \BibitemOpen
  \bibfield  {author} {\bibinfo {author} {\bibfnamefont {I.~M.}\ \bibnamefont
  {Lifshitz}},\ }\href@noop {} {\bibfield  {journal} {\bibinfo  {journal} {Sov.
  Phys. JETP}\ }\textbf {\bibinfo {volume} {11}},\ \bibinfo {pages} {1130}
  (\bibinfo {year} {1960})}\BibitemShut {NoStop}%
\bibitem [{\citenamefont {Benhabib}\ \emph {et~al.}(2015)\citenamefont
  {Benhabib}, \citenamefont {Sacuto}, \citenamefont {Civelli}, \citenamefont
  {Paul}, \citenamefont {Cazayous}, \citenamefont {Gallais}, \citenamefont
  {M\'easson}, \citenamefont {Zhong}, \citenamefont {Schneeloch}, \citenamefont
  {Gu}, \citenamefont {Colson},\ and\ \citenamefont {Forget}}]{Benhabib2015}%
  \BibitemOpen
  \bibfield  {author} {\bibinfo {author} {\bibfnamefont {S.}~\bibnamefont
  {Benhabib}}, \bibinfo {author} {\bibfnamefont {A.}~\bibnamefont {Sacuto}},
  \bibinfo {author} {\bibfnamefont {M.}~\bibnamefont {Civelli}}, \bibinfo
  {author} {\bibfnamefont {I.}~\bibnamefont {Paul}}, \bibinfo {author}
  {\bibfnamefont {M.}~\bibnamefont {Cazayous}}, \bibinfo {author}
  {\bibfnamefont {Y.}~\bibnamefont {Gallais}}, \bibinfo {author} {\bibfnamefont
  {M.-A.}\ \bibnamefont {M\'easson}}, \bibinfo {author} {\bibfnamefont {R.~D.}\
  \bibnamefont {Zhong}}, \bibinfo {author} {\bibfnamefont {J.}~\bibnamefont
  {Schneeloch}}, \bibinfo {author} {\bibfnamefont {G.~D.}\ \bibnamefont {Gu}},
  \bibinfo {author} {\bibfnamefont {D.}~\bibnamefont {Colson}}, \ and\ \bibinfo
  {author} {\bibfnamefont {A.}~\bibnamefont {Forget}},\ }\href@noop {}
  {\bibfield  {journal} {\bibinfo  {journal} {Phys. Rev. Lett.}\ }\textbf
  {\bibinfo {volume} {114}},\ \bibinfo {pages} {147001} (\bibinfo {year}
  {2015})}\BibitemShut {NoStop}%
\bibitem [{\citenamefont {Norman}\ \emph {et~al.}(2010)\citenamefont {Norman},
  \citenamefont {Lin},\ and\ \citenamefont {Millis}}]{Norman2010}%
  \BibitemOpen
  \bibfield  {author} {\bibinfo {author} {\bibfnamefont {M.~R.}\ \bibnamefont
  {Norman}}, \bibinfo {author} {\bibfnamefont {J.}~\bibnamefont {Lin}}, \ and\
  \bibinfo {author} {\bibfnamefont {A.~J.}\ \bibnamefont {Millis}},\
  }\href@noop {} {\bibfield  {journal} {\bibinfo  {journal} {Phys. Rev. B}\
  }\textbf {\bibinfo {volume} {81}},\ \bibinfo {pages} {180513(R)} (\bibinfo
  {year} {2010})}\BibitemShut {NoStop}%
\bibitem [{\citenamefont {Yelland}\ \emph {et~al.}(2011)\citenamefont
  {Yelland}, \citenamefont {Barraclough}, \citenamefont {Wang}, \citenamefont
  {Kamenev},\ and\ \citenamefont {Huxley}}]{Yelland2011}%
  \BibitemOpen
  \bibfield  {author} {\bibinfo {author} {\bibfnamefont {E.~A.}\ \bibnamefont
  {Yelland}}, \bibinfo {author} {\bibfnamefont {J.~M.}\ \bibnamefont
  {Barraclough}}, \bibinfo {author} {\bibfnamefont {W.}~\bibnamefont {Wang}},
  \bibinfo {author} {\bibfnamefont {K.~V.}\ \bibnamefont {Kamenev}}, \ and\
  \bibinfo {author} {\bibfnamefont {A.~D.}\ \bibnamefont {Huxley}},\
  }\href@noop {} {\bibfield  {journal} {\bibinfo  {journal} {Nat. Phys.}\
  }\textbf {\bibinfo {volume} {7}},\ \bibinfo {pages} {890} (\bibinfo {year}
  {2011})}\BibitemShut {NoStop}%
\bibitem [{\citenamefont {Shi}\ \emph {et~al.}(2017)\citenamefont {Shi},
  \citenamefont {Han}, \citenamefont {Peng}, \citenamefont {Richard},
  \citenamefont {Qian}, \citenamefont {Wu}, \citenamefont {Qiu}, \citenamefont
  {Wang}, \citenamefont {Hu}, \citenamefont {Sun},\ and\ \citenamefont
  {Ding}}]{Shi2017}%
  \BibitemOpen
  \bibfield  {author} {\bibinfo {author} {\bibfnamefont {X.}~\bibnamefont
  {Shi}}, \bibinfo {author} {\bibfnamefont {Z.-Q.}\ \bibnamefont {Han}},
  \bibinfo {author} {\bibfnamefont {X.-L.}\ \bibnamefont {Peng}}, \bibinfo
  {author} {\bibfnamefont {P.}~\bibnamefont {Richard}}, \bibinfo {author}
  {\bibfnamefont {T.}~\bibnamefont {Qian}}, \bibinfo {author} {\bibfnamefont
  {X.-X.}\ \bibnamefont {Wu}}, \bibinfo {author} {\bibfnamefont {M.-W.}\
  \bibnamefont {Qiu}}, \bibinfo {author} {\bibfnamefont {S.}~\bibnamefont
  {Wang}}, \bibinfo {author} {\bibfnamefont {J.~P.}\ \bibnamefont {Hu}},
  \bibinfo {author} {\bibfnamefont {Y.-J.}\ \bibnamefont {Sun}}, \ and\
  \bibinfo {author} {\bibfnamefont {H.}~\bibnamefont {Ding}},\ }\href@noop {}
  {\bibfield  {journal} {\bibinfo  {journal} {Nat. Comm.}\ }\textbf {\bibinfo
  {volume} {8}},\ \bibinfo {pages} {14988} (\bibinfo {year}
  {2017})}\BibitemShut {NoStop}%
\bibitem [{\citenamefont {Ren}\ \emph {et~al.}(2017)\citenamefont {Ren},
  \citenamefont {Yan}, \citenamefont {Niu}, \citenamefont {Tao}, \citenamefont
  {Hu}, \citenamefont {Peng}, \citenamefont {Xie}, \citenamefont {Zhao},
  \citenamefont {Zhang},\ and\ \citenamefont {Feng}}]{Ren2017}%
  \BibitemOpen
  \bibfield  {author} {\bibinfo {author} {\bibfnamefont {M.}~\bibnamefont
  {Ren}}, \bibinfo {author} {\bibfnamefont {Y.}~\bibnamefont {Yan}}, \bibinfo
  {author} {\bibfnamefont {X.}~\bibnamefont {Niu}}, \bibinfo {author}
  {\bibfnamefont {R.}~\bibnamefont {Tao}}, \bibinfo {author} {\bibfnamefont
  {D.}~\bibnamefont {Hu}}, \bibinfo {author} {\bibfnamefont {R.}~\bibnamefont
  {Peng}}, \bibinfo {author} {\bibfnamefont {B.}~\bibnamefont {Xie}}, \bibinfo
  {author} {\bibfnamefont {J.}~\bibnamefont {Zhao}}, \bibinfo {author}
  {\bibfnamefont {T.}~\bibnamefont {Zhang}}, \ and\ \bibinfo {author}
  {\bibfnamefont {D.-L.}\ \bibnamefont {Feng}},\ }\href@noop {} {\bibfield
  {journal} {\bibinfo  {journal} {Sci. Adv.}\ }\textbf {\bibinfo {volume}
  {3}},\ \bibinfo {pages} {1603238} (\bibinfo {year} {2017})}\BibitemShut
  {NoStop}%
\bibitem [{\citenamefont {Khan}\ and\ \citenamefont
  {Johnson}(2014)}]{Khan2014}%
  \BibitemOpen
  \bibfield  {author} {\bibinfo {author} {\bibfnamefont {S.~N.}\ \bibnamefont
  {Khan}}\ and\ \bibinfo {author} {\bibfnamefont {D.~D.}\ \bibnamefont
  {Johnson}},\ }\href@noop {} {\bibfield  {journal} {\bibinfo  {journal} {Phys.
  Rev. Lett.}\ }\textbf {\bibinfo {volume} {112}},\ \bibinfo {pages} {156401}
  (\bibinfo {year} {2014})}\BibitemShut {NoStop}%
\bibitem [{\citenamefont {Liu}\ \emph {et~al.}(2011)\citenamefont {Liu},
  \citenamefont {Palczewski}, \citenamefont {Dhaka}, \citenamefont {Kondo},
  \citenamefont {Fernandes}, \citenamefont {Mun}, \citenamefont {Hodovanets},
  \citenamefont {Thaler}, \citenamefont {Schmalian}, \citenamefont {Bud'ko},
  \citenamefont {Canfield},\ and\ \citenamefont {Kaminski}}]{Liu2011}%
  \BibitemOpen
  \bibfield  {author} {\bibinfo {author} {\bibfnamefont {C.}~\bibnamefont
  {Liu}}, \bibinfo {author} {\bibfnamefont {A.~D.}\ \bibnamefont {Palczewski}},
  \bibinfo {author} {\bibfnamefont {R.~S.}\ \bibnamefont {Dhaka}}, \bibinfo
  {author} {\bibfnamefont {T.}~\bibnamefont {Kondo}}, \bibinfo {author}
  {\bibfnamefont {R.~M.}\ \bibnamefont {Fernandes}}, \bibinfo {author}
  {\bibfnamefont {E.~D.}\ \bibnamefont {Mun}}, \bibinfo {author} {\bibfnamefont
  {H.}~\bibnamefont {Hodovanets}}, \bibinfo {author} {\bibfnamefont {A.~N.}\
  \bibnamefont {Thaler}}, \bibinfo {author} {\bibfnamefont {J.}~\bibnamefont
  {Schmalian}}, \bibinfo {author} {\bibfnamefont {S.~L.}\ \bibnamefont
  {Bud'ko}}, \bibinfo {author} {\bibfnamefont {P.~C.}\ \bibnamefont
  {Canfield}}, \ and\ \bibinfo {author} {\bibfnamefont {A.}~\bibnamefont
  {Kaminski}},\ }\href@noop {} {\bibfield  {journal} {\bibinfo  {journal}
  {Phys. Rev. B}\ }\textbf {\bibinfo {volume} {84}},\ \bibinfo {pages}
  {020509(R)} (\bibinfo {year} {2011})}\BibitemShut {NoStop}%
\bibitem [{\citenamefont {Tewari}\ and\ \citenamefont {Sau}(2012)}]{tewarisau}%
  \BibitemOpen
  \bibfield  {author} {\bibinfo {author} {\bibfnamefont {S.}~\bibnamefont
  {Tewari}}\ and\ \bibinfo {author} {\bibfnamefont {J.~D.}\ \bibnamefont
  {Sau}},\ }\href@noop {} {\bibfield  {journal} {\bibinfo  {journal} {Phys.
  Rev. Lett.}\ }\textbf {\bibinfo {volume} {109}},\ \bibinfo {pages} {150408}
  (\bibinfo {year} {2012})}\BibitemShut {NoStop}%
\bibitem [{\citenamefont {Yada}\ \emph {et~al.}(2011)\citenamefont {Yada},
  \citenamefont {Sato}, \citenamefont {Tanaka},\ and\ \citenamefont
  {Yokoyama}}]{YSTY10}%
  \BibitemOpen
  \bibfield  {author} {\bibinfo {author} {\bibfnamefont {K.}~\bibnamefont
  {Yada}}, \bibinfo {author} {\bibfnamefont {M.}~\bibnamefont {Sato}}, \bibinfo
  {author} {\bibfnamefont {Y.}~\bibnamefont {Tanaka}}, \ and\ \bibinfo {author}
  {\bibfnamefont {T.}~\bibnamefont {Yokoyama}},\ }\href@noop {} {\bibfield
  {journal} {\bibinfo  {journal} {Phys. Rev. B.}\ }\textbf {\bibinfo {volume}
  {83}},\ \bibinfo {pages} {064505} (\bibinfo {year} {2011})}\BibitemShut
  {NoStop}%
\bibitem [{\citenamefont {Sato}\ \emph {et~al.}(2011)\citenamefont {Sato},
  \citenamefont {Tanaka}, \citenamefont {Yada},\ and\ \citenamefont
  {Yokoyama}}]{STYY11}%
  \BibitemOpen
  \bibfield  {author} {\bibinfo {author} {\bibfnamefont {M.}~\bibnamefont
  {Sato}}, \bibinfo {author} {\bibfnamefont {Y.}~\bibnamefont {Tanaka}},
  \bibinfo {author} {\bibfnamefont {K.}~\bibnamefont {Yada}}, \ and\ \bibinfo
  {author} {\bibfnamefont {T.}~\bibnamefont {Yokoyama}},\ }\href@noop {}
  {\bibfield  {journal} {\bibinfo  {journal} {Phys.\ Rev.\ B}\ }\textbf
  {\bibinfo {volume} {83}},\ \bibinfo {pages} {224511} (\bibinfo {year}
  {2011})}\BibitemShut {NoStop}%
\bibitem [{\citenamefont {Brydon}\ \emph {et~al.}(2011)\citenamefont {Brydon},
  \citenamefont {Schnyder},\ and\ \citenamefont {Timm}}]{Brydon2011PRB}%
  \BibitemOpen
  \bibfield  {author} {\bibinfo {author} {\bibfnamefont {P.~M.~R.}\
  \bibnamefont {Brydon}}, \bibinfo {author} {\bibfnamefont {A.~P.}\
  \bibnamefont {Schnyder}}, \ and\ \bibinfo {author} {\bibfnamefont
  {C.}~\bibnamefont {Timm}},\ }\href {\doibase 10.1103/PhysRevB.84.020501}
  {\bibfield  {journal} {\bibinfo  {journal} {Phys. Rev. B}\ }\textbf {\bibinfo
  {volume} {84}},\ \bibinfo {pages} {020501} (\bibinfo {year}
  {2011})}\BibitemShut {NoStop}%
\bibitem [{\citenamefont {Brzezicki}\ and\ \citenamefont
  {Cuoco}(2018)}]{Brze18}%
  \BibitemOpen
  \bibfield  {author} {\bibinfo {author} {\bibfnamefont {W.}~\bibnamefont
  {Brzezicki}}\ and\ \bibinfo {author} {\bibfnamefont {M.}~\bibnamefont
  {Cuoco}},\ }\href@noop {} {\bibfield  {journal} {\bibinfo  {journal} {Phys.
  Rev. B}\ }\textbf {\bibinfo {volume} {97}},\ \bibinfo {pages} {064513}
  (\bibinfo {year} {2018})}\BibitemShut {NoStop}%
\bibitem [{\citenamefont {Black-Schaffer}\ and\ \citenamefont
  {Balatsky}(2013{\natexlab{a}})}]{Black_Schaffer2013PRB}%
  \BibitemOpen
  \bibfield  {author} {\bibinfo {author} {\bibfnamefont {A.~M.}\ \bibnamefont
  {Black-Schaffer}}\ and\ \bibinfo {author} {\bibfnamefont {A.~V.}\
  \bibnamefont {Balatsky}},\ }\href {\doibase 10.1103/PhysRevB.88.104514}
  {\bibfield  {journal} {\bibinfo  {journal} {Phys. Rev. B}\ }\textbf {\bibinfo
  {volume} {88}},\ \bibinfo {pages} {104514} (\bibinfo {year}
  {2013}{\natexlab{a}})}\BibitemShut {NoStop}%
\bibitem [{\citenamefont {Black-Schaffer}\ and\ \citenamefont
  {Balatsky}(2013{\natexlab{b}})}]{Black_Schaffer2013PRB2}%
  \BibitemOpen
  \bibfield  {author} {\bibinfo {author} {\bibfnamefont {A.~M.}\ \bibnamefont
  {Black-Schaffer}}\ and\ \bibinfo {author} {\bibfnamefont {A.~V.}\
  \bibnamefont {Balatsky}},\ }\href {\doibase 10.1103/PhysRevB.87.220506}
  {\bibfield  {journal} {\bibinfo  {journal} {Phys. Rev. B}\ }\textbf {\bibinfo
  {volume} {87}},\ \bibinfo {pages} {220506} (\bibinfo {year}
  {2013}{\natexlab{b}})}\BibitemShut {NoStop}%
\bibitem [{\citenamefont {Komendov\'a}\ \emph {et~al.}(2015)\citenamefont
  {Komendov\'a}, \citenamefont {Balatsky},\ and\ \citenamefont
  {Black-Schaffer}}]{Komendov2015PRB}%
  \BibitemOpen
  \bibfield  {author} {\bibinfo {author} {\bibfnamefont {L.}~\bibnamefont
  {Komendov\'a}}, \bibinfo {author} {\bibfnamefont {A.~V.}\ \bibnamefont
  {Balatsky}}, \ and\ \bibinfo {author} {\bibfnamefont {A.~M.}\ \bibnamefont
  {Black-Schaffer}},\ }\href {\doibase 10.1103/PhysRevB.92.094517} {\bibfield
  {journal} {\bibinfo  {journal} {Phys. Rev. B}\ }\textbf {\bibinfo {volume}
  {92}},\ \bibinfo {pages} {094517} (\bibinfo {year} {2015})}\BibitemShut
  {NoStop}%
\bibitem [{\citenamefont {Asano}\ and\ \citenamefont
  {Sasaki}(2015)}]{Asano2015PRB}%
  \BibitemOpen
  \bibfield  {author} {\bibinfo {author} {\bibfnamefont {Y.}~\bibnamefont
  {Asano}}\ and\ \bibinfo {author} {\bibfnamefont {A.}~\bibnamefont {Sasaki}},\
  }\href {\doibase 10.1103/PhysRevB.92.224508} {\bibfield  {journal} {\bibinfo
  {journal} {Phys. Rev. B}\ }\textbf {\bibinfo {volume} {92}},\ \bibinfo
  {pages} {224508} (\bibinfo {year} {2015})}\BibitemShut {NoStop}%
\bibitem [{\citenamefont {Komendov\'a}\ and\ \citenamefont
  {Black-Schaffer}(2017)}]{Komendov2017PRL}%
  \BibitemOpen
  \bibfield  {author} {\bibinfo {author} {\bibfnamefont {L.}~\bibnamefont
  {Komendov\'a}}\ and\ \bibinfo {author} {\bibfnamefont {A.~M.}\ \bibnamefont
  {Black-Schaffer}},\ }\href {\doibase 10.1103/PhysRevLett.119.087001}
  {\bibfield  {journal} {\bibinfo  {journal} {Phys. Rev. Lett.}\ }\textbf
  {\bibinfo {volume} {119}},\ \bibinfo {pages} {087001} (\bibinfo {year}
  {2017})}\BibitemShut {NoStop}%
\bibitem [{\citenamefont {Reyren}\ \emph {et~al.}(2007)\citenamefont {Reyren},
  \citenamefont {Thiel}, \citenamefont {Caviglia}, \citenamefont {Kourkoutis},
  \citenamefont {Hammerl}, \citenamefont {Richter}, \citenamefont {Schneider},
  \citenamefont {Kopp}, \citenamefont {Ruetschi}, \citenamefont {Jaccard},
  \citenamefont {Gabay}, \citenamefont {Muller}, \citenamefont {Triscone},
  \citenamefont {Mannhart}, \citenamefont {Kourkoutis},\ and\ \citenamefont
  {Hammerl}}]{Reyren2007Science}%
  \BibitemOpen
  \bibfield  {author} {\bibinfo {author} {\bibfnamefont {N.}~\bibnamefont
  {Reyren}}, \bibinfo {author} {\bibfnamefont {S.}~\bibnamefont {Thiel}},
  \bibinfo {author} {\bibfnamefont {A.~D.}\ \bibnamefont {Caviglia}}, \bibinfo
  {author} {\bibfnamefont {L.~F.}\ \bibnamefont {Kourkoutis}}, \bibinfo
  {author} {\bibfnamefont {G.}~\bibnamefont {Hammerl}}, \bibinfo {author}
  {\bibfnamefont {C.}~\bibnamefont {Richter}}, \bibinfo {author} {\bibfnamefont
  {C.~W.}\ \bibnamefont {Schneider}}, \bibinfo {author} {\bibfnamefont
  {T.}~\bibnamefont {Kopp}}, \bibinfo {author} {\bibfnamefont {A.-S.}\
  \bibnamefont {Ruetschi}}, \bibinfo {author} {\bibfnamefont {D.}~\bibnamefont
  {Jaccard}}, \bibinfo {author} {\bibfnamefont {M.}~\bibnamefont {Gabay}},
  \bibinfo {author} {\bibfnamefont {D.~A.}\ \bibnamefont {Muller}}, \bibinfo
  {author} {\bibfnamefont {J.-M.}\ \bibnamefont {Triscone}}, \bibinfo {author}
  {\bibfnamefont {J.}~\bibnamefont {Mannhart}}, \bibinfo {author}
  {\bibfnamefont {F.~L.}\ \bibnamefont {Kourkoutis}}, \ and\ \bibinfo {author}
  {\bibfnamefont {G.}~\bibnamefont {Hammerl}},\ }\href@noop {} {\bibfield
  {journal} {\bibinfo  {journal} {Science}\ }\textbf {\bibinfo {volume}
  {317}},\ \bibinfo {pages} {1196} (\bibinfo {year} {2007})}\BibitemShut
  {NoStop}%
\bibitem [{\citenamefont {Cen}\ \emph {et~al.}(2008)\citenamefont {Cen},
  \citenamefont {Thiel}, \citenamefont {Hammerl}, \citenamefont {Schneider},
  \citenamefont {Andersen}, \citenamefont {Hellberg}, \citenamefont
  {Mannhart},\ and\ \citenamefont {Levy}}]{Cen2008NatMater}%
  \BibitemOpen
  \bibfield  {author} {\bibinfo {author} {\bibfnamefont {C.}~\bibnamefont
  {Cen}}, \bibinfo {author} {\bibfnamefont {S.}~\bibnamefont {Thiel}}, \bibinfo
  {author} {\bibfnamefont {G.}~\bibnamefont {Hammerl}}, \bibinfo {author}
  {\bibfnamefont {C.~W.}\ \bibnamefont {Schneider}}, \bibinfo {author}
  {\bibfnamefont {K.~E.}\ \bibnamefont {Andersen}}, \bibinfo {author}
  {\bibfnamefont {C.~S.}\ \bibnamefont {Hellberg}}, \bibinfo {author}
  {\bibfnamefont {J.}~\bibnamefont {Mannhart}}, \ and\ \bibinfo {author}
  {\bibfnamefont {J.}~\bibnamefont {Levy}},\ }\href@noop {} {\bibfield
  {journal} {\bibinfo  {journal} {Nat. Mater.}\ }\textbf {\bibinfo {volume}
  {7}},\ \bibinfo {pages} {298} (\bibinfo {year} {2008})}\BibitemShut {NoStop}%
\bibitem [{\citenamefont {Caviglia}\ \emph {et~al.}(2008)\citenamefont
  {Caviglia}, \citenamefont {Gariglio}, \citenamefont {Reyren}, \citenamefont
  {Jaccard}, \citenamefont {Schneider}, \citenamefont {Gabay}, \citenamefont
  {Thiel}, \citenamefont {Hammerl}, \citenamefont {Mannhart},\ and\
  \citenamefont {Triscone}}]{Caviglia2008Nature}%
  \BibitemOpen
  \bibfield  {author} {\bibinfo {author} {\bibfnamefont {A.~D.}\ \bibnamefont
  {Caviglia}}, \bibinfo {author} {\bibfnamefont {S.}~\bibnamefont {Gariglio}},
  \bibinfo {author} {\bibfnamefont {N.}~\bibnamefont {Reyren}}, \bibinfo
  {author} {\bibfnamefont {D.}~\bibnamefont {Jaccard}}, \bibinfo {author}
  {\bibfnamefont {T.}~\bibnamefont {Schneider}}, \bibinfo {author}
  {\bibfnamefont {M.}~\bibnamefont {Gabay}}, \bibinfo {author} {\bibfnamefont
  {S.}~\bibnamefont {Thiel}}, \bibinfo {author} {\bibfnamefont
  {G.}~\bibnamefont {Hammerl}}, \bibinfo {author} {\bibfnamefont
  {J.}~\bibnamefont {Mannhart}}, \ and\ \bibinfo {author} {\bibfnamefont
  {J.-M.}\ \bibnamefont {Triscone}},\ }\href@noop {} {\bibfield  {journal}
  {\bibinfo  {journal} {Nature}\ }\textbf {\bibinfo {volume} {456}},\ \bibinfo
  {pages} {624} (\bibinfo {year} {2008})}\BibitemShut {NoStop}%
\bibitem [{\citenamefont {Schneider}\ \emph {et~al.}(2009)\citenamefont
  {Schneider}, \citenamefont {Caviglia}, \citenamefont {Gariglio},
  \citenamefont {Reyren},\ and\ \citenamefont {Triscone}}]{Schneider2009PRB}%
  \BibitemOpen
  \bibfield  {author} {\bibinfo {author} {\bibfnamefont {T.}~\bibnamefont
  {Schneider}}, \bibinfo {author} {\bibfnamefont {A.~D.}\ \bibnamefont
  {Caviglia}}, \bibinfo {author} {\bibfnamefont {S.}~\bibnamefont {Gariglio}},
  \bibinfo {author} {\bibfnamefont {N.}~\bibnamefont {Reyren}}, \ and\ \bibinfo
  {author} {\bibfnamefont {J.-M.}\ \bibnamefont {Triscone}},\ }\href {\doibase
  10.1103/PhysRevB.79.184502} {\bibfield  {journal} {\bibinfo  {journal} {Phys.
  Rev. B}\ }\textbf {\bibinfo {volume} {79}},\ \bibinfo {pages} {184502}
  (\bibinfo {year} {2009})}\BibitemShut {NoStop}%
\bibitem [{\citenamefont {Kuerten}\ \emph {et~al.}(2017)\citenamefont
  {Kuerten}, \citenamefont {Richter}, \citenamefont {Mohanta}, \citenamefont
  {Kopp}, \citenamefont {Kampf}, \citenamefont {Mannhart},\ and\ \citenamefont
  {Boschker}}]{Kuerten2017PRB}%
  \BibitemOpen
  \bibfield  {author} {\bibinfo {author} {\bibfnamefont {L.}~\bibnamefont
  {Kuerten}}, \bibinfo {author} {\bibfnamefont {C.}~\bibnamefont {Richter}},
  \bibinfo {author} {\bibfnamefont {N.}~\bibnamefont {Mohanta}}, \bibinfo
  {author} {\bibfnamefont {T.}~\bibnamefont {Kopp}}, \bibinfo {author}
  {\bibfnamefont {A.}~\bibnamefont {Kampf}}, \bibinfo {author} {\bibfnamefont
  {J.}~\bibnamefont {Mannhart}}, \ and\ \bibinfo {author} {\bibfnamefont
  {H.}~\bibnamefont {Boschker}},\ }\href {\doibase 10.1103/PhysRevB.96.014513}
  {\bibfield  {journal} {\bibinfo  {journal} {Phys. Rev. B}\ }\textbf {\bibinfo
  {volume} {96}},\ \bibinfo {pages} {014513} (\bibinfo {year}
  {2017})}\BibitemShut {NoStop}%
\bibitem [{\citenamefont {Liu}\ \emph {et~al.}(2015)\citenamefont {Liu},
  \citenamefont {Sau},\ and\ \citenamefont {Das~Sarma}}]{Sau2015}%
  \BibitemOpen
  \bibfield  {author} {\bibinfo {author} {\bibfnamefont {X.}~\bibnamefont
  {Liu}}, \bibinfo {author} {\bibfnamefont {J.~D.}\ \bibnamefont {Sau}}, \ and\
  \bibinfo {author} {\bibfnamefont {S.}~\bibnamefont {Das~Sarma}},\ }\href@noop
  {} {\bibfield  {journal} {\bibinfo  {journal} {Phys. Rev. B}\ }\textbf
  {\bibinfo {volume} {92}},\ \bibinfo {pages} {014513} (\bibinfo {year}
  {2015})}\BibitemShut {NoStop}%
\bibitem [{\citenamefont {Huang}\ \emph {et~al.}(2015)\citenamefont {Huang},
  \citenamefont {Woelfle},\ and\ \citenamefont {Balatsky}}]{Huang2015PRB}%
  \BibitemOpen
  \bibfield  {author} {\bibinfo {author} {\bibfnamefont {Z.}~\bibnamefont
  {Huang}}, \bibinfo {author} {\bibfnamefont {P.}~\bibnamefont {Woelfle}}, \
  and\ \bibinfo {author} {\bibfnamefont {A.~V.}\ \bibnamefont {Balatsky}},\
  }\href@noop {} {\bibfield  {journal} {\bibinfo  {journal} {Phys. Rev. B}\
  }\textbf {\bibinfo {volume} {92}},\ \bibinfo {pages} {121404(R)} (\bibinfo
  {year} {2015})}\BibitemShut {NoStop}%
\bibitem [{\citenamefont {Burset}\ \emph {et~al.}(2015)\citenamefont {Burset},
  \citenamefont {Lu}, \citenamefont {Tkachov}, \citenamefont {Tanaka},
  \citenamefont {Hankiewicz},\ and\ \citenamefont
  {Trauzettel}}]{Burset2015PRB}%
  \BibitemOpen
  \bibfield  {author} {\bibinfo {author} {\bibfnamefont {P.}~\bibnamefont
  {Burset}}, \bibinfo {author} {\bibfnamefont {B.}~\bibnamefont {Lu}}, \bibinfo
  {author} {\bibfnamefont {G.}~\bibnamefont {Tkachov}}, \bibinfo {author}
  {\bibfnamefont {Y.}~\bibnamefont {Tanaka}}, \bibinfo {author} {\bibfnamefont
  {E.}~\bibnamefont {Hankiewicz}}, \ and\ \bibinfo {author} {\bibfnamefont
  {B.}~\bibnamefont {Trauzettel}},\ }\href@noop {} {\bibfield  {journal}
  {\bibinfo  {journal} {Phys. Rev. B}\ }\textbf {\bibinfo {volume} {92}},\
  \bibinfo {pages} {205424} (\bibinfo {year} {2015})}\BibitemShut {NoStop}%
\bibitem [{\citenamefont {Lee}\ \emph {et~al.}(2017)\citenamefont {Lee},
  \citenamefont {Lutchyn},\ and\ \citenamefont {Maciejko}}]{Lee2017PRB}%
  \BibitemOpen
  \bibfield  {author} {\bibinfo {author} {\bibfnamefont {S.-P.}\ \bibnamefont
  {Lee}}, \bibinfo {author} {\bibfnamefont {R.~M.}\ \bibnamefont {Lutchyn}}, \
  and\ \bibinfo {author} {\bibfnamefont {J.}~\bibnamefont {Maciejko}},\ }\href
  {\doibase 10.1103/PhysRevB.95.184506} {\bibfield  {journal} {\bibinfo
  {journal} {Phys. Rev. B}\ }\textbf {\bibinfo {volume} {95}},\ \bibinfo
  {pages} {184506} (\bibinfo {year} {2017})}\BibitemShut {NoStop}%
\bibitem [{\citenamefont {Kashuba}\ \emph {et~al.}(2017)\citenamefont
  {Kashuba}, \citenamefont {Sothmann}, \citenamefont {Burset},\ and\
  \citenamefont {Trauzettel}}]{KashubaPRB2017}%
  \BibitemOpen
  \bibfield  {author} {\bibinfo {author} {\bibfnamefont {O.}~\bibnamefont
  {Kashuba}}, \bibinfo {author} {\bibfnamefont {B.}~\bibnamefont {Sothmann}},
  \bibinfo {author} {\bibfnamefont {P.}~\bibnamefont {Burset}}, \ and\ \bibinfo
  {author} {\bibfnamefont {B.}~\bibnamefont {Trauzettel}},\ }\href@noop {}
  {\bibfield  {journal} {\bibinfo  {journal} {Phys. Rev. B}\ }\textbf {\bibinfo
  {volume} {95}},\ \bibinfo {pages} {174516} (\bibinfo {year}
  {2017})}\BibitemShut {NoStop}%
\bibitem [{\citenamefont {Cayao}\ and\ \citenamefont
  {Black-Schaffer}(2017)}]{Cayao}%
  \BibitemOpen
  \bibfield  {author} {\bibinfo {author} {\bibfnamefont {J.}~\bibnamefont
  {Cayao}}\ and\ \bibinfo {author} {\bibfnamefont {A.~M.}\ \bibnamefont
  {Black-Schaffer}},\ }\href@noop {} {\bibfield  {journal} {\bibinfo  {journal}
  {Phys. Rev. B}\ }\textbf {\bibinfo {volume} {96}},\ \bibinfo {pages} {155426}
  (\bibinfo {year} {2017})}\BibitemShut {NoStop}%
\bibitem [{\citenamefont {Tanaka}\ and\ \citenamefont
  {Kashiwaya}(2004)}]{Proximityp}%
  \BibitemOpen
  \bibfield  {author} {\bibinfo {author} {\bibfnamefont {Y.}~\bibnamefont
  {Tanaka}}\ and\ \bibinfo {author} {\bibfnamefont {S.}~\bibnamefont
  {Kashiwaya}},\ }\href@noop {} {\bibfield  {journal} {\bibinfo  {journal}
  {Phys. Rev. B}\ }\textbf {\bibinfo {volume} {70}},\ \bibinfo {pages} {012507}
  (\bibinfo {year} {2004})}\BibitemShut {NoStop}%
\bibitem [{\citenamefont {Tanaka}\ \emph
  {et~al.}(2005{\natexlab{a}})\citenamefont {Tanaka}, \citenamefont
  {Kashiwaya},\ and\ \citenamefont {Yokoyama}}]{Proximityp2}%
  \BibitemOpen
  \bibfield  {author} {\bibinfo {author} {\bibfnamefont {Y.}~\bibnamefont
  {Tanaka}}, \bibinfo {author} {\bibfnamefont {S.}~\bibnamefont {Kashiwaya}}, \
  and\ \bibinfo {author} {\bibfnamefont {T.}~\bibnamefont {Yokoyama}},\
  }\href@noop {} {\bibfield  {journal} {\bibinfo  {journal} {Phys. Rev. B}\
  }\textbf {\bibinfo {volume} {71}},\ \bibinfo {pages} {094513} (\bibinfo
  {year} {2005}{\natexlab{a}})}\BibitemShut {NoStop}%
\bibitem [{\citenamefont {Asano}\ \emph {et~al.}(2006)\citenamefont {Asano},
  \citenamefont {Tanaka},\ and\ \citenamefont {Kashiwaya}}]{Proximityp3}%
  \BibitemOpen
  \bibfield  {author} {\bibinfo {author} {\bibfnamefont {Y.}~\bibnamefont
  {Asano}}, \bibinfo {author} {\bibfnamefont {Y.}~\bibnamefont {Tanaka}}, \
  and\ \bibinfo {author} {\bibfnamefont {S.}~\bibnamefont {Kashiwaya}},\
  }\href@noop {} {\bibfield  {journal} {\bibinfo  {journal} {Phys. Rev. Lett.}\
  }\textbf {\bibinfo {volume} {96}},\ \bibinfo {pages} {097007} (\bibinfo
  {year} {2006})}\BibitemShut {NoStop}%
\bibitem [{\citenamefont {Tanaka}\ \emph
  {et~al.}(2005{\natexlab{b}})\citenamefont {Tanaka}, \citenamefont {Asano},
  \citenamefont {Golubov},\ and\ \citenamefont {Kashiwaya}}]{Meissner3}%
  \BibitemOpen
  \bibfield  {author} {\bibinfo {author} {\bibfnamefont {Y.}~\bibnamefont
  {Tanaka}}, \bibinfo {author} {\bibfnamefont {Y.}~\bibnamefont {Asano}},
  \bibinfo {author} {\bibfnamefont {A.~A.}\ \bibnamefont {Golubov}}, \ and\
  \bibinfo {author} {\bibfnamefont {S.}~\bibnamefont {Kashiwaya}},\ }\href@noop
  {} {\bibfield  {journal} {\bibinfo  {journal} {Phys. Rev. B}\ }\textbf
  {\bibinfo {volume} {72}},\ \bibinfo {pages} {140503(R)} (\bibinfo {year}
  {2005}{\natexlab{b}})}\BibitemShut {NoStop}%
\bibitem [{\citenamefont {Tanaka}\ and\ \citenamefont {Golubov}(2007)}]{odd1}%
  \BibitemOpen
  \bibfield  {author} {\bibinfo {author} {\bibfnamefont {Y.}~\bibnamefont
  {Tanaka}}\ and\ \bibinfo {author} {\bibfnamefont {A.~A.}\ \bibnamefont
  {Golubov}},\ }\href@noop {} {\bibfield  {journal} {\bibinfo  {journal} {Phys.
  Rev. Lett.}\ }\textbf {\bibinfo {volume} {98}},\ \bibinfo {pages} {037003}
  (\bibinfo {year} {2007})}\BibitemShut {NoStop}%
\bibitem [{\citenamefont {Tanaka}\ \emph
  {et~al.}(2007{\natexlab{a}})\citenamefont {Tanaka}, \citenamefont {Golubov},
  \citenamefont {Kashiwaya},\ and\ \citenamefont {Ueda}}]{odd2}%
  \BibitemOpen
  \bibfield  {author} {\bibinfo {author} {\bibfnamefont {Y.}~\bibnamefont
  {Tanaka}}, \bibinfo {author} {\bibfnamefont {A.~A.}\ \bibnamefont {Golubov}},
  \bibinfo {author} {\bibfnamefont {S.}~\bibnamefont {Kashiwaya}}, \ and\
  \bibinfo {author} {\bibfnamefont {M.}~\bibnamefont {Ueda}},\ }\href@noop {}
  {\bibfield  {journal} {\bibinfo  {journal} {Phys. Rev. Lett.}\ }\textbf
  {\bibinfo {volume} {99}},\ \bibinfo {pages} {037005} (\bibinfo {year}
  {2007}{\natexlab{a}})}\BibitemShut {NoStop}%
\bibitem [{\citenamefont {Tanaka}\ \emph
  {et~al.}(2007{\natexlab{b}})\citenamefont {Tanaka}, \citenamefont {Tanuma},\
  and\ \citenamefont {Golubov}}]{odd3}%
  \BibitemOpen
  \bibfield  {author} {\bibinfo {author} {\bibfnamefont {Y.}~\bibnamefont
  {Tanaka}}, \bibinfo {author} {\bibfnamefont {Y.}~\bibnamefont {Tanuma}}, \
  and\ \bibinfo {author} {\bibfnamefont {A.~A.}\ \bibnamefont {Golubov}},\
  }\href@noop {} {\bibfield  {journal} {\bibinfo  {journal} {Phys. Rev. B}\
  }\textbf {\bibinfo {volume} {76}},\ \bibinfo {pages} {054522} (\bibinfo
  {year} {2007}{\natexlab{b}})}\BibitemShut {NoStop}%
\bibitem [{\citenamefont {Ebisu}\ \emph {et~al.}(2015)\citenamefont {Ebisu},
  \citenamefont {Yada}, \citenamefont {Kasai},\ and\ \citenamefont
  {Tanaka}}]{Ebisu}%
  \BibitemOpen
  \bibfield  {author} {\bibinfo {author} {\bibfnamefont {H.}~\bibnamefont
  {Ebisu}}, \bibinfo {author} {\bibfnamefont {K.}~\bibnamefont {Yada}},
  \bibinfo {author} {\bibfnamefont {H.}~\bibnamefont {Kasai}}, \ and\ \bibinfo
  {author} {\bibfnamefont {Y.}~\bibnamefont {Tanaka}},\ }\href@noop {}
  {\bibfield  {journal} {\bibinfo  {journal} {Phys. Rev. B}\ }\textbf {\bibinfo
  {volume} {91}},\ \bibinfo {pages} {054518} (\bibinfo {year}
  {2015})}\BibitemShut {NoStop}%
\bibitem [{\citenamefont {Stornaiuolo}\ \emph {et~al.}(2017)\citenamefont
  {Stornaiuolo}, \citenamefont {Massarotti}, \citenamefont {Di~Capua},
  \citenamefont {Lucignano}, \citenamefont {Pepe}, \citenamefont {Salluzzo},\
  and\ \citenamefont {Tafuri}}]{Tafuri2017PRL}%
  \BibitemOpen
  \bibfield  {author} {\bibinfo {author} {\bibfnamefont {D.}~\bibnamefont
  {Stornaiuolo}}, \bibinfo {author} {\bibfnamefont {D.}~\bibnamefont
  {Massarotti}}, \bibinfo {author} {\bibfnamefont {R.}~\bibnamefont
  {Di~Capua}}, \bibinfo {author} {\bibfnamefont {P.}~\bibnamefont {Lucignano}},
  \bibinfo {author} {\bibfnamefont {G.~P.}\ \bibnamefont {Pepe}}, \bibinfo
  {author} {\bibfnamefont {M.}~\bibnamefont {Salluzzo}}, \ and\ \bibinfo
  {author} {\bibfnamefont {F.}~\bibnamefont {Tafuri}},\ }\href {\doibase
  10.1103/PhysRevB.95.140502} {\bibfield  {journal} {\bibinfo  {journal} {Phys.
  Rev. B}\ }\textbf {\bibinfo {volume} {95}},\ \bibinfo {pages} {140502}
  (\bibinfo {year} {2017})}\BibitemShut {NoStop}%
\bibitem [{\citenamefont {Tanaka}\ and\ \citenamefont
  {Kashiwaya}(1996)}]{Josephson1}%
  \BibitemOpen
  \bibfield  {author} {\bibinfo {author} {\bibfnamefont {Y.}~\bibnamefont
  {Tanaka}}\ and\ \bibinfo {author} {\bibfnamefont {S.}~\bibnamefont
  {Kashiwaya}},\ }\href@noop {} {\bibfield  {journal} {\bibinfo  {journal}
  {Phys. Rev. B}\ }\textbf {\bibinfo {volume} {53}},\ \bibinfo {pages} {R11957}
  (\bibinfo {year} {1996})}\BibitemShut {NoStop}%
\bibitem [{\citenamefont {Barash}\ \emph {et~al.}(1996)\citenamefont {Barash},
  \citenamefont {Burkhardt},\ and\ \citenamefont {Rainer}}]{Josephson2}%
  \BibitemOpen
  \bibfield  {author} {\bibinfo {author} {\bibfnamefont {Y.~S.}\ \bibnamefont
  {Barash}}, \bibinfo {author} {\bibfnamefont {H.}~\bibnamefont {Burkhardt}}, \
  and\ \bibinfo {author} {\bibfnamefont {D.}~\bibnamefont {Rainer}},\
  }\href@noop {} {\bibfield  {journal} {\bibinfo  {journal} {Phys. Rev. Lett.}\
  }\textbf {\bibinfo {volume} {77}},\ \bibinfo {pages} {4070} (\bibinfo {year}
  {1996})}\BibitemShut {NoStop}%
\bibitem [{\citenamefont {Tanaka}\ and\ \citenamefont
  {Kashiwaya}(1997)}]{Josephson3}%
  \BibitemOpen
  \bibfield  {author} {\bibinfo {author} {\bibfnamefont {Y.}~\bibnamefont
  {Tanaka}}\ and\ \bibinfo {author} {\bibfnamefont {S.}~\bibnamefont
  {Kashiwaya}},\ }\href@noop {} {\bibfield  {journal} {\bibinfo  {journal}
  {Phys. Rev. B}\ }\textbf {\bibinfo {volume} {56}},\ \bibinfo {pages} {892}
  (\bibinfo {year} {1997})}\BibitemShut {NoStop}%
\end{thebibliography}%

\end{document}